# Broken Inversion Symmetry via a Magic Methyl Effect

Calum J. Gibb[1], Jordan Hobbs[2], Caitlin O'Brien[1,2], Kate Hille[2], Benji Maher[2] , Christopher M. Pask [1], Richard. J. Mandle*[1,2]

[1]School of Chemistry, University of Leeds, Leeds, UK, LS2 9JT.
[2]School of Physics and Astronomy, University of Leeds, Leeds, UK, LS2 9JT.

*Author for correspondence e-mail: r.mandle@leeds.ac.uk

## Abstract

Polar liquid crystals - fluid phases in which molecular dipoles organise into ferro- or antiferroelectric states - present a highly constrained molecular design space, where small changes can entirely suppress polar organisation. Here we demonstrate, contrary to expectations, that installation of a methyl group in the 5-position of a 1,3-dioxane ring substantially increases the onset temperature of polar order. Remarkably, this effect proves transferable across several liquid-crystal families, including examples where the methylated derivative exhibits a polar phase whereas its parent compound does not. Across 8 matched pairs, we find that a 5-methyl group can enhance the onset temperature of polar order in 1,3-dioxane materials by up to 120 °C. These findings establish a simple and general molecular design strategy for enhancing and enabling spontaneous dipolar ordering in liquid crystals.

## Introduction

Longitudinally polar liquid crystals (LCs) are an emerging class of ferroelectric fluids which combine the translational freedom of a liquid-like medium with the macroscopic spontaneous polarization of traditional solid-state materials [1-9]. Whilst ferroelectric and anti-ferroelectric order can be superimposed onto many traditional LC phases (e.g. SmA [10] and SmC [11-13]), it is the fluidity and field responsiveness of the polar nematic state that renders this especially attractive for practical applications [14-17], this being referred to as the ferroelectric nematic ($N_F$) phase. While the molecular origins of spontaneous polar order are still emerging, it is apparent that the factors driving the formation of this new class of polar fluids differs from their apolar counterparts [18].

Materials exhibiting polar mesophases currently exist in a constrained area of chemical space, with the development of new materials being largely antecedent to the earliest materials shown to exhibit fluid ferroelectricity such as **DIO** [1], and **RM734** [19]. Whilst design principles linking chemical structure to the emergence of polar order are in their infancy [20-24] , since the discovery of the $N_F$ phase, the structural design of polar LCs has tended to become more complex, incorporating a greater number of aromatic rings, linking groups and unusual chemical moieties. Although this has led to the observation of new and interesting polar mesophases [10-13,25-30], this has also resulted in an overall increase in melting points, transition temperatures and a decrease in chemical stability owing largely to materials possessing high clearing points. This shift away from materials with the potential to be operable at room temperature is detrimental to the use of such materials in device applications.

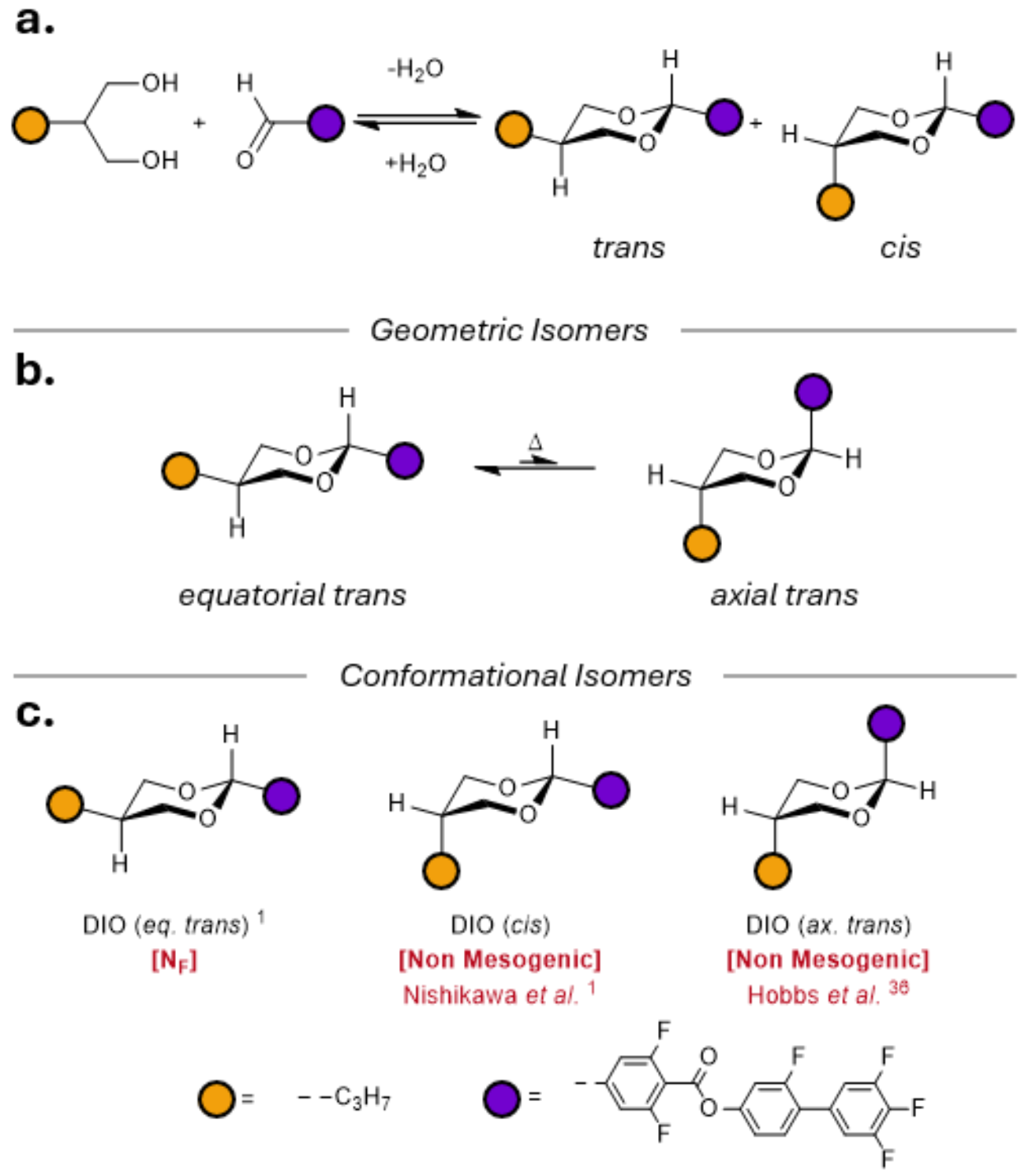

**Fig. 1.** **(a)** The two geometric isomers of 2,5-disubstituted 1,3-dioxane rings (*cis* and *trans*) obtained during synthesis; **(b)** the two conformation isomers (*equatorial* and *axial trans*) obtained on the heating of a 2,5-disubstituted 1,3-dioxane above a threshold temperature; and **(c)** the chemical structure and phase behaviour of different geometric and conformational isomers of the ferroelectric nematic material **DIO**.

The 2,5-disubstituted 1,3-dioxane motif is commonly employed in polar liquid crystals (e.g. **DIO**, **Fig. 1**). This motif is easily synthesised from readily available and cost-effective starting materials under mild conditions and has been shown to be capable of promoting a wide variety of polar LC mesophase [8,25-27,31,32]. Whilst the cis geometric isomers of 1,3-dioxane rings result in non-mesogenic materials owing to their irregular shape [33,34], the near linear shape of the trans geometric isomers are compatible with liquid crystal order (**Fig. 1a**). These geometric isomers are generally separable chromatographically, yielding isometrically pure materials with reproducible LC phase transition temperatures. For a given geometric trans isomer, there are however two possible conformational isomers, either an equatorial or axial trans conformation (*eq.* trans and *ax.* trans, respectively, **Fig. 1b**) [35,36]. The *eq.* trans conformer is the most thermodynamically stable and, in the case of **DIO**, is liquid crystalline. It exhibits the polar $N_F$ and anti-ferroelectric ($N_S$) phases [37], sometimes referred to as $N_X$ [4,5] or $SmZ_A$ [38,39], as well as a conventional apolar nematic (N) phase at high temperature (**Fig. 1c**). The *ax.* trans conformer of **DIO** is non-mesogenic and can be induced in a pure sample of *eq.* trans DIO by heating a sample to high temperatures [36]. As an equilibrium process, a mixture of both trans conformers can result, especially after thermal treatment, which has lower transition temperatures than the pure *eq.* trans material due to said isomerisation.

Given the thermal instability of 1,3-dioxane rings, we considered that introducing a methyl ($CH_3$) group may bias the conformational preference in such a way as to enhance the thermal stability of materials. Unexpectedly, we find this to confer significant advantage to the onset temperatures of polar phase types as well as a stabilisation of non-helical phases in molecules with a tendency to biaxial helical self-assembly.

## Results and Discussion

With the archetypal material **DIO** as a starting point, we began by positioning a methyl group in the 5-position of the 1,3-dioxane ring, affording **5-Me DIO** (**Fig. 2a**). Details of chemical synthesis and structural characterisation of all compounds presented here are given in the ESI. We find **5-Me DIO** exhibits the same phase sequence as the parent molecule **DIO** (i.e. $N_F$, $N_S$ and N) however, positioning a methyl group in the 5-position results in significantly increased transition temperatures to polar phases (i.e. $T_{NF}$ and $T_{NS}$) with an accompanying corresponding decrease in the apolar transition temperature (i.e. $T_{N\text{-}I}$). The melting point of **5-Me DIO** is marginally higher than the unmethylated 1,3-dioxane.

The transitional properties of **5-Me DIO** were first determined by polarised optical microscopy (POM) by the observation of characteristic optical textures (**Fig. 2b**) before accurate transition temperatures were obtained by differential scanning calorimetry (DSC) (**Fig. 2c**). The polar nature of the $N_F$ and $N_S$ phases were then confirmed by current reversal techniques where a single and double peak was observed either side of the polarization reversal (**Fig. 2d**) for the $N_F$ and $N_S$ phases respectively. X-ray scattering measurements confirm the nematic-like ordering of all three phases, where diffuse signals are seen in both the wide and small angle regions, indicating nematic-like orientational ordering of the molecules with no positional order (**Fig. 2f**). Some changes are seen in the scattering patterns where the wide-angle signal is more condensed for the **5-Me DIO** vs **DIO** indicating a slightly increased lateral corelation length between molecules while the small angle single, indicating longitudinal corelation lengths, is relatively unchanged.

Placing a lateral methyl group on the rigid core is typically detrimental to the thermal stability of liquid crystalline phases as it is thought to disrupt the interactions between molecules - nominally reflected as a simple decrease in transition temperatures [40-42]. While this is true here for $T_{N\text{-}I}$, where a decrease of some 30 °C is afforded by the addition of the methyl group, onset temperatures of the polar $N_F$ and $N_S$ phases are found to increase. This reinforces the idea that the molecular origins of polar liquid crystalline phases differ significantly from apolar counterparts [18]. With the melting point of the material remaining changed only minimally (9.2 °C decrease), the near 40 °C increase in $T_{NF}$ and $T_{NS}$ results in the polar phases being enantiotropic, i.e. they are observed at a temperature above the melting point.

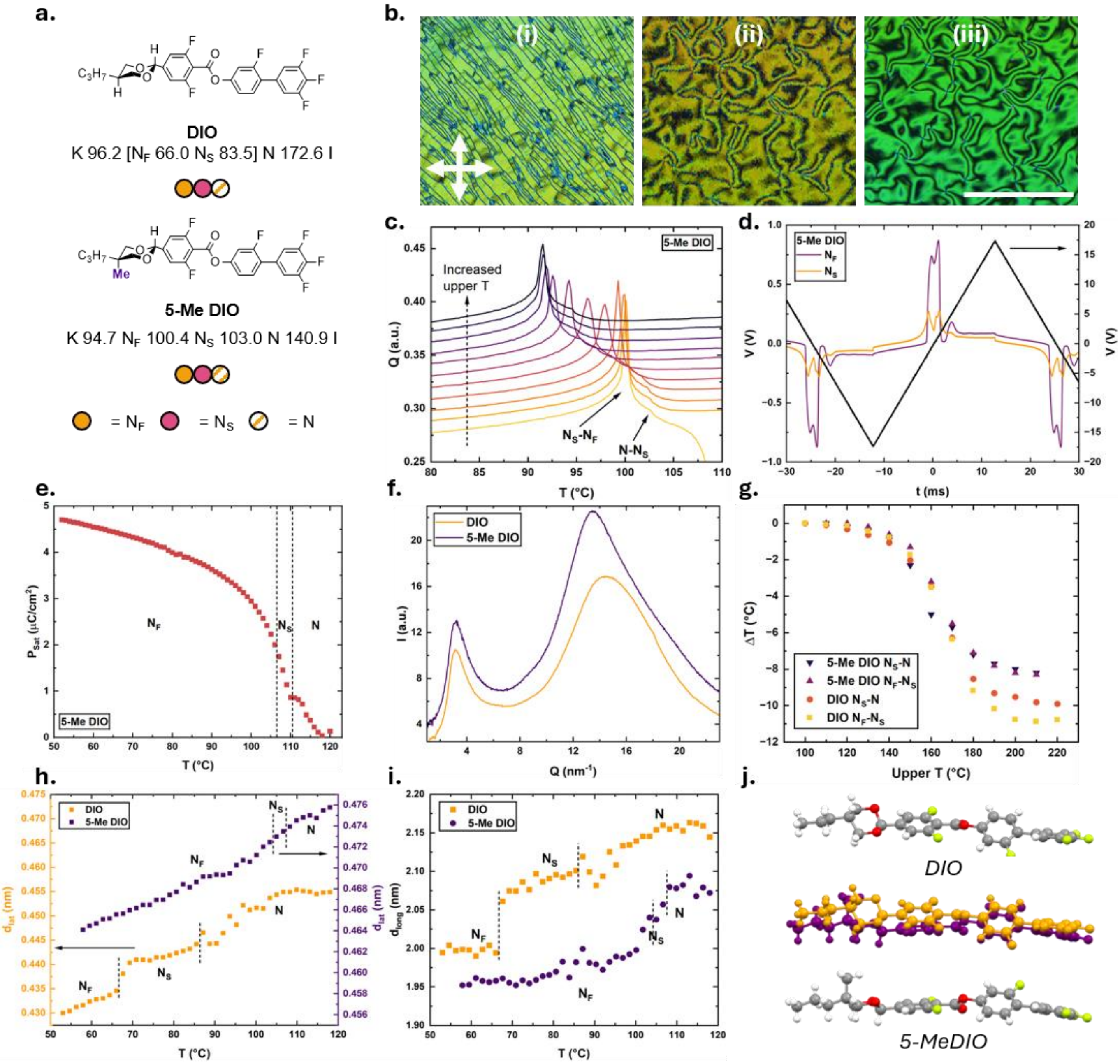

**Fig. 2** **(a)** Chemical structures and transitional behaviour of **DIO** [1] and **5-Me DIO** (top and bottom, respectively). Temperatures are reported in °C with [ ] indicating a monotropic phase transition; **(b)** POM micrographs depicting **(i)** a banded texture of the $N_F$ phase, **(ii)** a textured Schlieren texture of the $N_S$ phase, and **(iii)** a Schlieren texture of the N phase; **(c)** DSC thermogram on repeated heating and cooling cycles for **5-Me DIO**. Each cycle increases the maximum temperature of the cycle by 10 °C from 100 – 220 °C; **(d)** current response measurements measured in the $N_F$ and $N_S$ phases exhibited by **5-Me DIO**. Responses were measured at 20 Hz; **(e)** temperature dependence of the saturated polarization measured for **5-Me DIO** in the N, $N_S$, and $N_F$ phases; **(f)** 1D x-ray scattering pattern measured for **DIO** and **5-Me DIO** in the $N_F$ phase; **(g)** the percentage change in the values of $T_{NS}$ and $T_{NF}$ measured on consecutive heating cycles by DSC for **DIO** and **5-Me DIO**. The maximum temperature was increased by 10 °C between cycles from 100 – 220 °C; the temperature dependence of the **(h)** lateral and **(i)** longitudinal correlation length extracted from the wide-angle peak in the 1D SAXS pattern for **DIO** and **5-Me DIO**; and **(j)** comparison of the molecular shapes of **DIO** [top] and **5-Me DIO** [bottom] extracted from their respective crystal structures. The similarity in molecular shape is demonstrated by the overlayed structures in the centre (orange and purple, **DIO** and **5-Me DIO**; respectively).

To probe any increase in thermal stability against isomerisation we subjected **5-Me DIO** to repeated heating and cooling cycles by DSC. On each cycle the upper temperature limit was increased by 10 °C from a start of 100 °C to an end temperature of 220 °C (**Fig. 2g**). Values of $T_{NF}$ and $T_{NS}$ remain constant until the sample is heated beyond 140 °C, after which they decrease, plateauing at 180 °C. This broadly mirrors the same behaviour observed for **DIO** previously [36]. NMR spectroscopy of the thermally cycled sample reveals partial isomerisation of the *eq.* trans isomer to the *ax.* trans form; the proton resonance in the 2-position of the 1,3-dioxane motif being shifted downfield (**Fig. S1**). This assignment is supported by calculated gauge independent atomic orbital (GIAO) NMR shielding tensors (at the TPSS/pCcseg-3 level of DFT) for each conformer (**Fig. S2**). The change in chemical shift, while not as dramatic as calculated, is consistent with the assignment made based on experimental data. For **5-Me DIO**, the mixture of conformers has a similar composition to **DIO**, being < 3% *ax.* trans. Samples of **5-Me DIO** subjected to prolonged periods elevated temperatures display a comparable drop in transition temperatures to that seen for **DIO** (**Fig. 2g**) [36].

The mechanism of this methyl enhanced polar order is not immediately obvious. Wide-angle X-ray scattering demonstrates that both analogues have similar temperature dependencies of their lateral and longitudinal correlation lengths (**Fig. 2h** and **i**, respectively). The minor differences arise due to the increased free volume afforded by the methyl group, suggesting the interactions occurring between molecules within the LC phases are similar. Analysis of the crystal structures of **DIO** [43] and **5-Me DIO** (**Fig. S3a** and **S3b**, respectively) indicate that the interactions between the molecules are comparable with both materials crystallising into unit cells with primitive space groups of $\underline{P}2_1/c$ and $P\overline{1}$ respectively. Overlaying the structures extracted from their respective crystal structures also indicates that the conformations of both molecules in the crystal form are also comparable (**Fig. 2j**). Notably, there is no strong directional interaction involving the appended methyl group apparent in the crystal structure data.

In biological systems, the so-called ‘magic methyl effect’ is appreciated for its ability to modulate biological and physical properties of molecules: increasing drug potency [44]; protein folding [45]; or even solubility [46] by orders of magnitude through the selective exchange of a single proton for a methyl group[47]. The origins of this effect are debated and subjective, with the interactions between molecules mediated through a combination of steric, conformational and electronic effects resulting from the methyl group [48]. Given the similarities in correlation length and crystal structures we observe here for **DIO** and **5-Me DIO**, we suggest an analogous magic methyl effect could be responsible for the observed enhancement of polar order. We hypothesise that the 5-methyl group alters the steric or conformational preferences of the molecules. This leads to a higher onset temperature for polar order which requires breaking the inversion symmetry of the director in the conventional apolar nematic phase. We tested this computationally with relaxed scans along the potential energy surface, allowing the alkyl chain to rotate relative to the dioxane motif. We find the methyl group acts as a conformational modifier, biasing certain conformations through the steric footprint of the methyl unit (**Fig S4**). In this view, the methyl group partially restricts the conformational freedom of the terminal propyl chain, causing it to behave more like a shorter or less flexible terminal group. Generally, shorter terminal chains give higher onset temperatures for polar mesophase types [19]. Since alkyl chains contribute little to the molecular dipole but increase conformational entropy, this reduced conformational freedom may favour molecular arrangements more compatible with spontaneous polar order. In this spirit, exploration of alternate substituents (for example, alkyl chains, spirocycles, alternate heterocyclic ring systems, polar motifs, and so on) presents a logical extension of this work; and we note the paucity of such examples in the liquid-crystalline literature [41].

Next, we consider other structural modifications to the parent 1,3-dioxane. As the synthesis of **DIO** involves multiple linear synthetic steps [49], we developed a new $N_F$ core system in which the dioxane motif is constructed in the final step (**1**, **Scheme S2**); this enables the expedient synthesis of alternate ring systems. The parent 1,3-dioxane of this series (**1**), as well as its 5-methyl (**5-Me 1**), 2-methyl (**2-Me 1**), 2,5-dimethyl (**2,5-DiMe 1**) and 5-methyl-1,3,5-dioxasilinane (**5-MeDs 1**) derivatives were prepared (**Fig. 3**).

The parent 1,3-dioxane (**1**) exhibits an apolar nematic as well as a polar $N_S$ phase over a short temperature window. A virtual $N_F$ transition temperature for **1** can be obtained via binary mixtures with **DIO** at 33 °C (**Fig. S5**); its absence in the pure material is presumably due to preclusion by crystallisation. The 5-methyl analogue **5-Me 1** shows no apolar mesophases, instead the $N_F$ phase forms directly from the isotropic liquid. From this reaction, we obtained both the *eq.* trans and cis forms of **5-Me 1**. 2D $^1$H NOESY NMR confirms the assignment of ring geometry (**Fig. S6**). Crystal structures for both **1** and **5-Me 1** were obtained (**Fig. S7**) which, as was the case with **DIO** [43] and **5-Me DIO**, both adopt unit cells with primitive space groups ($\underline{P}2_1$/c and $P\ 2_1$, respectively) as well as possessing similar molecular shapes (**Fig. S8**). We again note the absence of any apparent directional interactions between the methyl groups, which is further suggestion of a steric or conformational origin for the methyl enhanced polar order effect we see here.

Positioning the methyl group in the 2-position (**2-Me 1**), or one methyl group in each of the 2- and 5- positions (**2,5 DiMe 1**) invariably gave a mixture of *cis-* and *trans-* geometric isomers, assigned by the observation of the proton and fluorine resonances of the methyl group in the 2-position (**Fig. S9a** and **b,** respectively). These isomers were inseparable by flash chromatography – either manual, or with an automated system (**Fig. S9c**) – or reverse-phase preparative HPLC (**Fig. S9d)**. Similarly, the 5-methyl-1,3-5-dioxasilinane material (**5-MeDs 1**) is non-mesogenic and exists as an inseparable mixture of isomers. All three materials present as isotropic liquids at ambient temperature, precluding isolation of isomerically pure materials by recrystallisation, and we deem these non-mesogenic. For each ring system we find the *eq. trans* to *ax. trans* interconversion proceeds via an intermediate twist-boat state and two transition states. For the parent 1,3-dioxane and its 5-methyl variant, the *eq. trans* configuration is the energy minimum, whereas for the 2-methyl and 2,5-dimethyl the *ax. trans* is the ground state (**Fig S10**). The 1,3,5-dioxasilinane is notably different, with the intermediate twist-boat conformation lower in energy than the *ax.* trans conformer. To the best of our knowledge, this ring system has not previously been reported and given both its synthetic accessibility and unusual conformational landscape, it may be of interest beyond its rather limited utility as a liquid-crystalline motif.

Having established the positional specificity of the methyl enhanced polar order effect, we next sought to test the generality of this enhancement through its inclusion in different molecular architectures. We identified a series of polar liquid crystalline materials, and one apolar example, which contain the 1,3-dioxane motif – with the exception of **5**, we re-synthesised each of these in house, as well as the analogous material bearing a methyl group in the 5-position (Compounds **2-8**; **Fig. 3**). Materials were chosen to reflect not only structural diversity, but also to represent a broad range of different ferroelectric phase types and phase sequences [8,25-27,31].

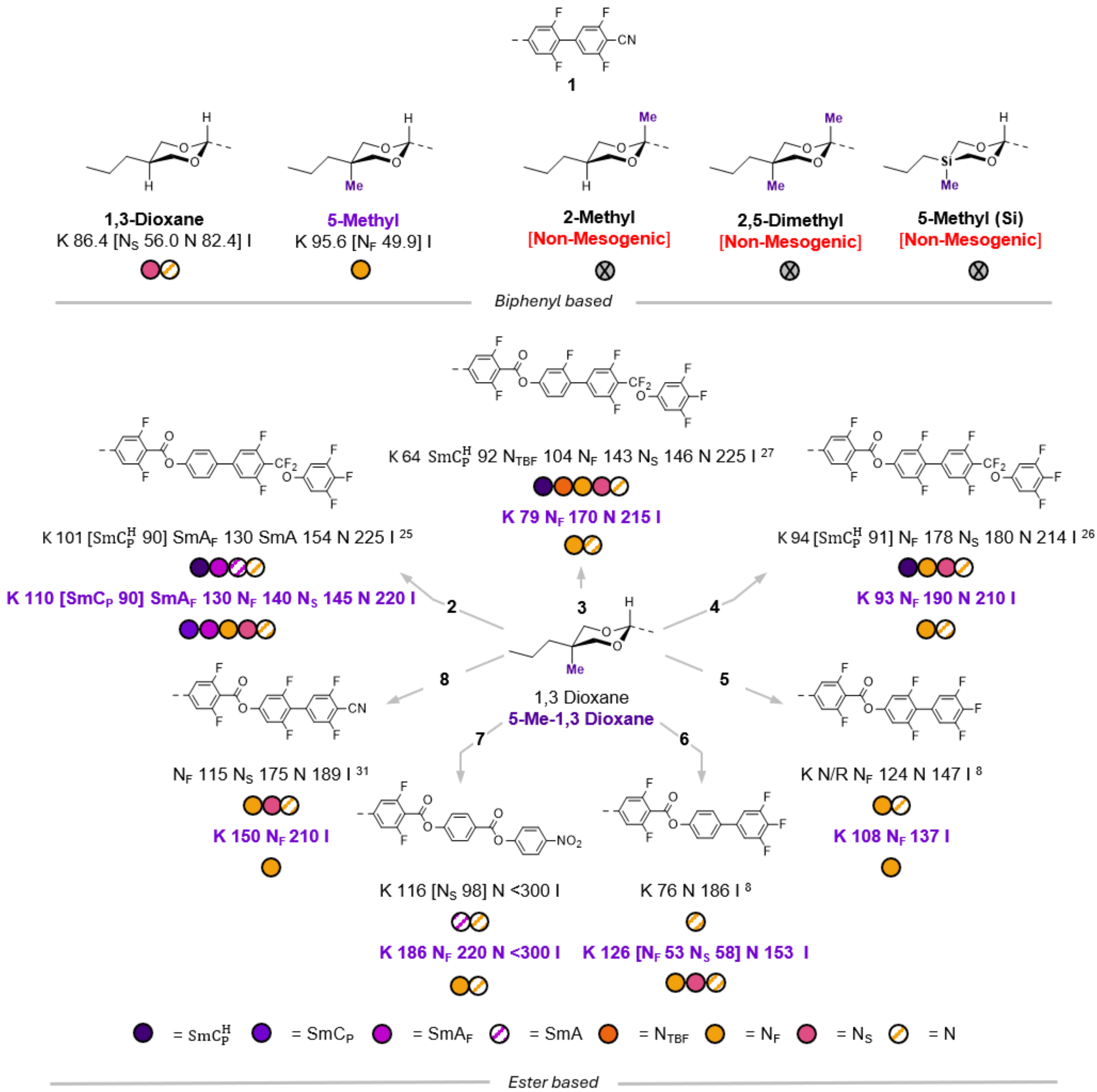

**Fig. 3** The chemical structures and associated transitional properties (temperatures given in °C) for the eight additional materials and their **5-Me**. [ ] indicate a monotropic phase transition and N/R indicates a temperature was not previously reported. For **1**, the **2-Me** and 2,**5-DiMe**, and **5-Me (Si)** analogues were also prepared however, due to the materials being an inseparable mix of the *cis-* and *trans-* geometric isomers, the resulting materials are non-mesogenic. Some of the un-substituted 1,3-dioxane containing materials have been reported previously [8,25-27,31], however, in the case of 6, we find different transition temperatures to those previously recorded.

Without exception, the addition of the extra methyl group in the 5-position results in an increase in the onset temperature of polar order and a decrease in the stability of apolar mesophases, some by over 120 °C (**Table S1**, **Fig. S13-16**). As exemplified by compound **2** and its 5-methyl analogue (**Fig. 3**), the enhancement is confined to the temperature of the $N_F$ phase, with the $SmA_F$ onset temperature remaining roughly the same. Unexpectedly, the heliconical $SmC_P^H$ phase in **2** is replaced by a non-helical $SmC_P$ phase in the methyl variant (**5-Me 2**).

Small angle X-ray scattering (SAXS) confirms the tilted nature of the $SmC_P$ phase **(Fig. 4a)** with current response measurements revealing a small peak pre-polarity reversal associated with reformation of molecular tilt [11] (**Fig. 4b-c**). Notably, this peak is also observed in materials showing $SmC_P^H$ phases, however this peak is generally larger and far broader than in the non-helical $SmC_P$ phase – an explanation for that difference has not yet been found [11,25,26]. POM also confirms the assignment of the $SmC_P$ phase as there is there is no significant drop in birefringence, nominally the result of a helical structure, when moving from the mosaic texture of the $SmA_F$ phase (**Fig. 4d**) to the blocky texture of the $SmC_P$ phase (**Fig. 4e**-**g**). Further to this, no selective reflection is observed within the $SmC_P$ phase again indicating a lack of any helical structure. For single crystals of **5-Me 2,** grown from solution with an isotropic solvent, the material was found to crystallise into a triclinic cell which was solved in the polar *P*1 space group, with two molecules in the asymmetric unit (**Fig. S17**).

Analogous behaviour to **2** is seen for **3** and **4**, with the heliconical polar nematic ($N_{TBF}$) phase of **3** being absent in the **5-Me 3** and the $SmC_P^H$ phase of **4** being absent in **5-Me 4**. Instead, both **5-Me 3** and **5-Me 4** show π-twisted domains with no fingerprint textures (**Figure 4h-i**) under POM, indicating the presence of the non-helical $N_F$ phase [7,27,50,51]. For **5-Me 2**, **3** and **4**, the additional lateral bulk afforded by the methyl group results in additional structural rigidity *via* a conformational bias (**Fig S4**), and we speculate that this strongly reduces flexoelectric contributions [4,5,52,53] with the result that both antiferroelectric and heliconical order is destabilised such that the phases are either not observed or have significantly significant reduced temperature ranges.

Compounds **1**-**8**, and their methylated analogues, demonstrate that the methyl enhanced polar order is a general and transferable effect that can increase the onset temperature of the ferroelectric nematic phase. Except for **1** and **7**, all materials display the $N_F$ phase above the melting point, i.e. they are enantiotropic, and therefore highly suited to in-depth experimental study. In every single case (**1**-**8**) the 5-methyl 1,3-dioxane outperforms the unsubstituted parent material in terms of onset temperature of polar order, and quite often by a very significant margin.

For compound **6,** the parent 1,3-dioxane exhibits solely an apolar (N) phase whereas the 5-methyl variant displays both ferroelectric ($N_F$) and antiferroelectric ($N_S$) nematic phases. In the case of compound **7**, the onset temperature of the $N_F$ phase increases by over 120 °C; and for compound **8**, the same onset temperature is almost 100 °C higher.

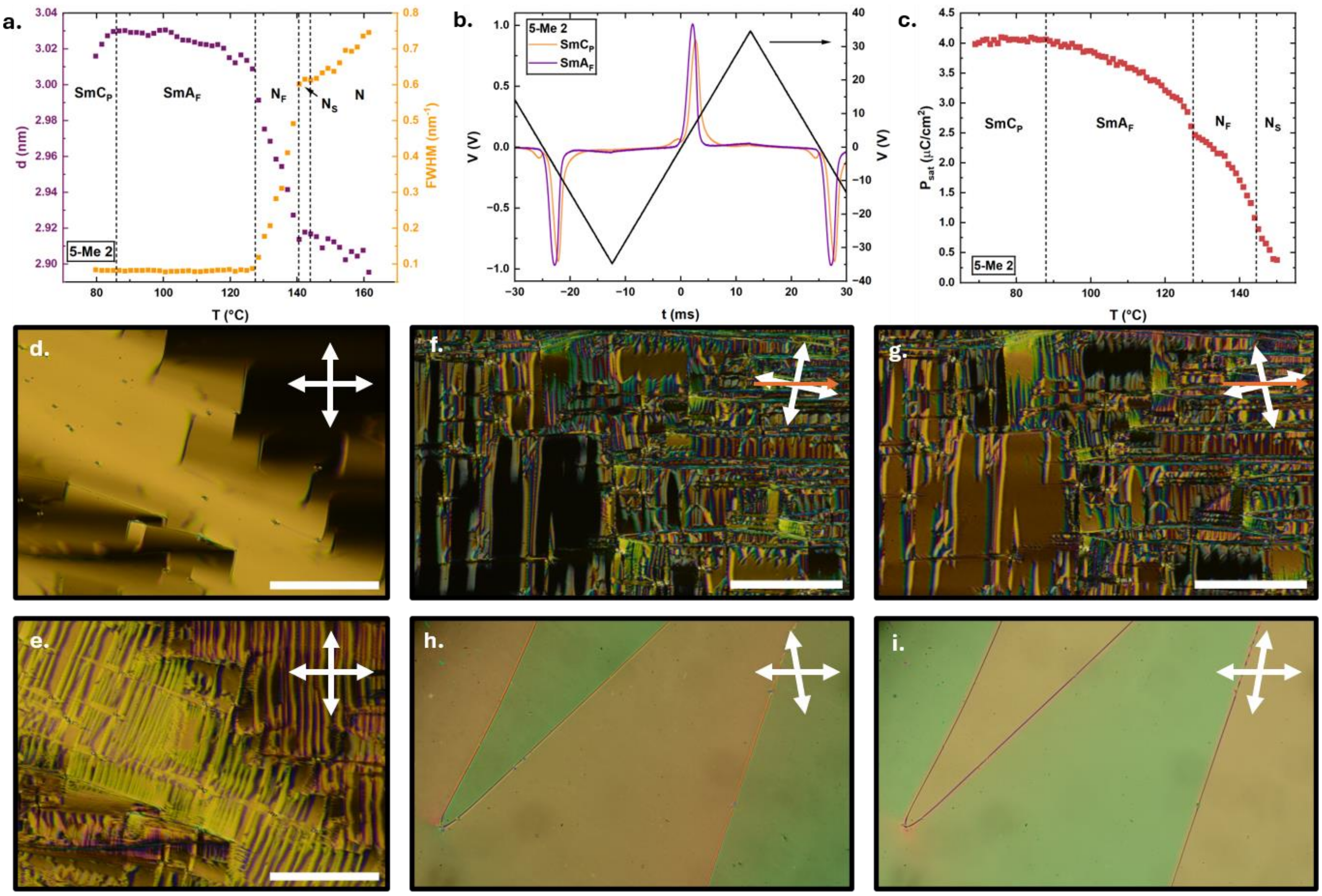

**Fig. 4** **(a)** Temperature dependence of the layer spacing and full width half maxima (FWHM) of the layer peak for compound **5-Me 2**; **(b)** Current response data for **5-Me 2** using a 20 Hz triangle wave in a 5 μm thick LC cell with out-of-plane electrodes. No alignment layer was used; **(c)** temperature dependence of the saturated polarisation of **5-Me 2**; POM images of **5-Me 2** in the **(d)** SmA$_F$ and **l** SmC$_P$ phase taken between coverslips. The images are of the same area; **(f, g)** POM images of **5-Me 2** in the SmC$_P$ phase in a 5 μm thick cell with syn-parallel rubbed planar alignment. The rubbing direction is marked with an orange arrow. The apparent tilt domains can each alternatively be brought into extinction by rotation of the cell to align the optical axis with the polariser. In a SmC phase the optical axis does not necessarily lie along the rubbing direction; **(h,i)** POM images of **5-Me 3** in the N$_F$ phase filled into a 5 μm thick cell with anti-parallel rubbed planar alignment. This results in π twisted domains of opposite handedness separated by domain walls. Upon de-crossing the polarisers the domains show opposite optical activity indicated by the different colours.

## Conclusions

Against expectations we show that the inclusion of a methyl group in the 5-position of 1,3-dioxanes leads to substantial increases in the onset temperature of polar order in molecules exhibiting fluid ferroelectricity. The enhanced polar order conferred by the 5-methyl group is shown to be general and transferable across different families of liquid crystals over a range of structure types and phase sequences. This methyl enhanced polar order effect can even generate polar organisation in materials where the parent material is strictly apolar. The effect is positionally specific to the 5-position, with alternative substitution patterns yielding non-mesogenic materials. Not only do the 5-methyl-1,3-dioxanes outperform unsubstituted dioxanes, they are synthetically accessible – the precursor 2-methyl-2-propylpropane-1,3-diol is inexpensive (circa £10 $mol^{-1}$) and the acetalization procedure is trivial, requiring no modifications. Given the extremely constrained design space around polar liquid crystals – where even minor structural changes can eliminate polar order – we believe the new and general design principle reported here is an especially valuable synthetic tool capable of being widely deployed in the design of future materials.

## Data availability

The data associated with this paper are openly available from the University of Leeds Data Repository at https://doi.org/10.5518/1885.

CCDC 2574725-2574729 contain the supplementary crystallographic data for this paper. These data can be obtained free of charge from The Cambridge Crystallographic Data Centre via www.ccdc.cam.ac.uk/structures.

## Acknowledgements

Computational work was performed on AIRE, part of the high-performance computing facilities at the University of Leeds. R.J.M. thanks UKRI for funding via a Future Leaders Fellowship, grant number MR/W006391/1, and the University of Leeds for funding via a University Academic Fellowship. R.J.M and J.H. thank the Royal Society for funding via Research Grant RGS\R2\242503. R.J.M. gratefully acknowledges support from Merck KGaA. The authors acknowledge EPSRC for funding the SAXS/WAXS system via a capital equipment grant EP/X0348011.

# Broken Inversion Symmetry via a Magic Methyl Effect

## Supplemental Information

Calum J. Gibb[1], Jordan Hobbs[2], Caitlin O'Brien[1,2], Kate Hille[2], Benji Maher[2] , Christopher M. Pask [1], Richard. J. Mandle*[1,2]

[1]School of Chemistry, University of Leeds, Leeds, UK, LS2 9JT.
[2]School of Physics and Astronomy, University of Leeds, Leeds, UK, LS2 9JT.

*Author for correspondence e-mail: r.mandle@leeds.ac.uk

**Contents**

# 1. Supplementary Methods

## 1.1. Chemical Synthesis

Chemicals were purchased from commercial suppliers (Fluorochem, Merck, ChemScene, Ambeed) and used as received. Solvents were purchased from Merck and used without further purification. Reactions were performed in standard laboratory glassware at ambient temperature and atmosphere and were monitored by TLC with an appropriate eluent and visualised with 254 nm light. Chromatographic purification was performed using a Combiflash NextGen 300+ System (Teledyne Isco) with a silica gel stationary phase and a hexanes/ethyl acetate (EtOAc) or hexanes/dichloromethane (DCM) gradient as the mobile phase, with detection made in the 200-800 nm range. Chromatographed materials were filtered through 200 nm PTFE frits and then subjected to re-crystallisation from an appropriate solvent system where appropriate.

## 1.2. Chemical Characterisation

Chemical materials were characterised by NMR spectroscopy using a Bruker Avance III HD 9.4 T NMR spectrometer operating at 400 MHz, 100.5 MHz or 376.4 MHz ($^{1}H$, $^{13}C\{^{1}H\}$ and $^{19}F$, respectively) or a Bruker AV4 NEO 11.75T NMR spectrometer operating at 500 MHz, 125.5 MHz ($^{1}H$, and $^{13}C\{^{1}H\}$, respectively). HPLC analysis was performed using an Agilent 1290 Infinity II system fitted with a poroshell 120 ec-c18 column running a water:MeCN gradient.

## 1.3. Thermal Analysis

Differential scanning calorimetry (DSC) measurements were performed using a TA Instruments Q2000 DSC instrument (TA Instruments, Wilmslow UK), equipped with a RCS90 Refrigerated cooling system (TA Instruments, Wilmslow UK). The instrument was calibrated against an Indium standard, and data were processed using TA Instruments Universal Analysis Software. Samples were analysed under a nitrogen atmosphere, in hermetically sealed aluminium TZero crucibles (TA Instruments, Wilmslow, UK) and subjected to three analysis cycles. In all cases, samples were subject to heating and cooling at a rate of 10 K $min^{-1}$. Phase transition temperatures were measured as onset values on cooling cycles for consistency between monotropic and enantiotropic phase transitions, while crystal melts were obtained as onset values on heating.

## 1.4. Polarized Optical Microscopy (POM)

Polarised light optical microscopy (POM) was performed using a Leica DM2700P polarised light microscope (Leica Microsystems (UK) Ltd., Milton Keynes, UK), equipped with 10x and 50x magnification, and a rotatable stage. A Linkam TMS 92 heating stage (Linkam Scientific Instruments Ltd., Redhill, UK) was used for temperature control, and samples were studied sandwiched between two untreated glass coverslips. Images were recorded using a Nikon D3500 Digital Camera (Nikon UK Ltd., Surbiton, UK), using DigiCamControl software.

### 1.5. X-ray Scattering

X-ray scattering measurements, both small angle (SAXS) and wide angle (WAXS) were recorded using an Anton Paar SAXSpoint 5.0. This was equipped with a primux 100 Cu X-ray source with a 2D EIGER2 R detector with a variable sample-detector distance. The X-rays had a wavelength of 0.154 nm. Samples were filled into either thin-walled quartz capillaries or held between Kapton tape. Temperature was controlled using an Anton Paar heated sampler with a range of 20 °C to 300 °C and the samples held in a chamber with an atmospheric pressure of >1 mbar.

Background scattering patterns were recorded, scaled according to the sample transmission and then subtracted from the samples' obtained 2D SAXS pattern. 1D patterns were obtained by radially integrating the 2D SAXS patterns. Peak positions and FWHM was recorded and then converted into d spacing following Bragg's law. In the tilted smectic phases, the tilt was obtained from:

$$\frac{d_c}{d_A} = \cos\theta \qquad \textbf{(1)}$$

where $d_c$ is the layer spacing in the tilted smectic phase, $d_A$ is the extrapolated spacing from the non-tilted preceding smectic phase, extrapolated to account for the weak temperature dependence of the preceding phases due to shifts in conformation and order, and $\theta$ the structural tilt angle.

### 1.6 Measurement of Spontaneous Polarization ($P_S$)

Spontaneous polarisation measurements are undertaken using the current reversal technique. Triangular waveform AC voltages are applied to the sample cells with an Agilent 33220A signal generator (Keysight Technologies), and the resulting current outflow is passed through a current-to-voltage amplifier and recorded on a RIGOL DHO4204 high-resolution oscilloscope (Telonic Instruments Ltd, UK). Heating and cooling of the samples during these measurements is achieved with an Instec HCS402 hot stage controlled to 10 mK stability by an Instec mK1000 temperature controller. The LC samples are held in 4μm thick cells with no alignment layer, supplied by Instec. The measurements consist of cooling the sample at a rate of 1 $Kmin^{-1}$ and applying a set voltage at a frequency of 20 Hz to the sample every 1 K. The voltage was set such that it would saturate the measured $P_S$ and was determined before final data collection.

There are three contributions to the measured current trace: accumulation of charge in the cell ($I_c$), ion flow ($I_i$), and the current flow due to polarisation reversal ($I_p$). To obtain a $P_S$ value, we extract the latter, which manifests as one or multiple peaks in the current flow, and integrate as:

$$P_S = \int \frac{I_p}{2A} dt \qquad \textbf{(2)}$$

where A is the active electrode area of the sample cell.

## 1.7 Electronic Structure Calculations

Electronic structure calculations were performed using the Orca 6.1 software package on the AIRE high-performance computing facility University of Leeds.[1-3] Geometries were first optimised using the GFN2-xTB tight binding method [4,5]. Conformer searching was performed using the GOAT algorithm [6] with GFN2-xTB and their energies obtained calculated using the wB97x-3c composite DFT method [7]. Selected low energy conformers (equatorial trans, axial trans) were reoptimized at this same level. Transition states for the equatorial-to-axial interconversion were identified using the Nudged Elastic Band (NEB-TS) method [8], both saddle and stationary points were then reoptimized both at the wB97x-3c level. Harmonic vibrational frequency calculations were carried out for all stationary points on the potential energy surface (axial, equatorial, intermediate, and transition states) at the wB97x-3c level to verify their character (zero or one imaginary frequency), to obtain zero-point and finite-temperature thermochemical corrections and so obtain the Gibbs free energy.

Calculated $^{1}$H and $^{19}$F NMR shieldings were obtained using ORCA for the *equatorial* and *axial* trans conformational isomers of compound **1**. For geometry preoptimized at the wB97x-3c level we perform a single-point gauge-independent atomic orbital (GAIO) NMR calculations using the TPSS functional with the pcSseg-3 basis set [9] and the CPCM continuum solvation model for chloroform. Auxiliary basis functions were generated automatically using *AutoAux* in Orca [10]. For the meta-GCA NMR calculations we employ the Dobson treatment of the kinetic-energy-density. NMR shifts were scaled using tetramethylsilane (TMS) as a reference ($\delta = 0.00\,ppm$) calculated at the same level, yielding the shift parameters $\delta_{ref}$= *31.6* and $\alpha$= *-1.0*.

## 2. Supplementary Results

### 2.1 Tabulated transitional properties of 5-Me DIO; 1; 5-Me 1; 6 and 7; and 5-Me 2-8.

**Table S1.** Tabulated transitional properties, including transition temperatures (°C) and associated enthalpies of transition (KJ mol$^{-1}$) for **5-Me DIO**, compounds **1** and **5-Me 1**, compounds **6** & **7**, and the **5-Me** compounds **2-8** materials. Data compounds **2-5** and **8** have been published previously [11-16]. [ ] indicate a monotropic phase transition and $^{m}$ indicates a transition temperature as observed by POM.

| Material | Transitional properties | |
|---|---|---|
| | Transition Temperatures (°C) | Enthalpies (KJ mol$^{-1}$) |
| **5-Me DIO** | K 94.69 $N_F$ 100.42 $N_S$ 102.99 N 140.87 I | K 23.6 $N_F$ 0.14 $N_S$ 0.002 N 0.35 I |
| **1** | K 86.4 [$N_S$ 56.0] N 82.4 I | K 25.93 [$N_S$ 0.004] N 0.928 I |
| **5-Me 1** | K 95.6 [$N_F$ 49.9] I | K 18.9 [$N_F$ 2.557] I |
| **6** | K 79.2 N 185.0 I | K 21.5 N 0.47 I |
| **7** | K 116.2 [$N_S$ 97.6] N <300 I (decomp.) | K 27.4 [$N_S$ 0.010] N [N/A] I (decomp.) |
| **5-Me 2** | K 101.9 [$SmC_P$ 90$^{m}$] $SmA_F$ 124.5 $N_F$ 140.5 $N_S$ 142$^{m}$ N 208.1 I | K 28.2 $SmC_P$ [N/A] $SmA_F$ 0.082 $N_F$ 0.0.67 $N_S$ [N/A] N 1.01 I |
| **5-Me 3** | K 79.2 $N_F$ 153.1 N 205.9 I | K 24.3 $N_F$ 0.95 N 1.59 I |
| **5-Me 4** | K 92.5 $N_F$ 183.7 N 203.5 I | K 27.0 $N_F$ 2.38 N 2.12 I |
| **5-Me 5** | K 107.7 $N_F$ 131.5 I | K 31.4 $N_F$ 3.84 I |
| **5-Me 6** | K 125.7 $N_F$ 53.2 $N_S$ 58.1 N 153.1 I | K 37.5 [$N_F$ 0.09 $N_S$ 0.012] N 0.388 I |
| **5-Me 7** | K 186.1 $N_F$ 219.9 N <300 I (decomp) | K 21.5 $N_F$ 0.29 N [N/A] I (decomp) |
| **5-Me 8** | K 150.0 $N_F$ 194.3 $N_S$ 198.0 N 201.0 I | K 155 $N_F$ 0.81 $N_S$ 0.02 N 1.310 I |

## 2.2 Degradation of *eq* trans to *ax* trans conformational isomers in DIO and 5-Me DIO

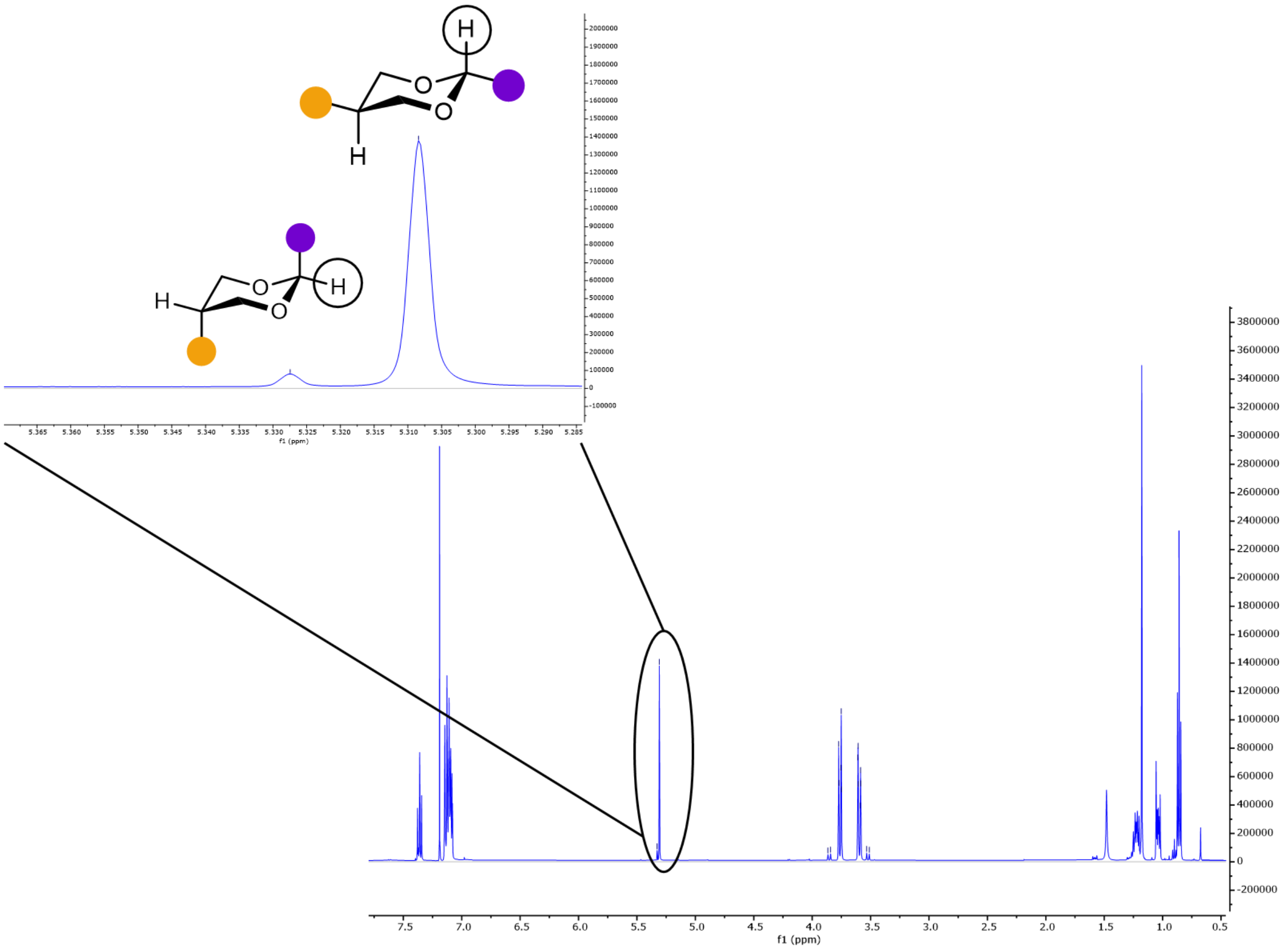

**Fig. S1** $^{1}H$ NMR of **5-Me DIO** after thermal degradation induced by heating the sample to 200 °C for 5 minutes. The growth of the second resonance at 5.335 ppm is indicative of the conversion of the *eq* trans isomer to the *ax* trans isomer.

### 2.3 Simulated NMR spectra

As discussed in the manuscript (**Fig. 2g**), repeated cycling of the pure *eq* trans conformational isomer of **5-Me DIO** leads to partial isomerisation. The identity of the *eq. trans* form in $^1$H NMR is entirely unambiguous, as its structure is known from single crystal X-ray diffraction. The assignment as *ax trans* from NMR alone is reasonable, with the proton in the 2-position of the 1,3-dioxane being shifted downfield in line with expectations.

To further support the assignment of the *ax trans,* we computed the GIAO NMR shielding tensors for at the TPSS/pcSseg-3 level with an auxiliary basis set and CPCM solvation (chloroform) for both the *ax* and *eq trans* geometries of compound **1** (selected as it has fewer environments in $^1$H NMR than DIO) as well as tetramethylsilane (TMS). Using TMS as a reference ($\delta = 0.00\, ppm$) we obtain the calculated NMR spectra (**Fig. S2**). The calculated spectra support our assigned experimental spectra, namely, the resonance corresponding to the hydrogen atom in the 2- position is shifted downfield in the *axial trans* isomer when compared with the *equatorial* trans isomer.

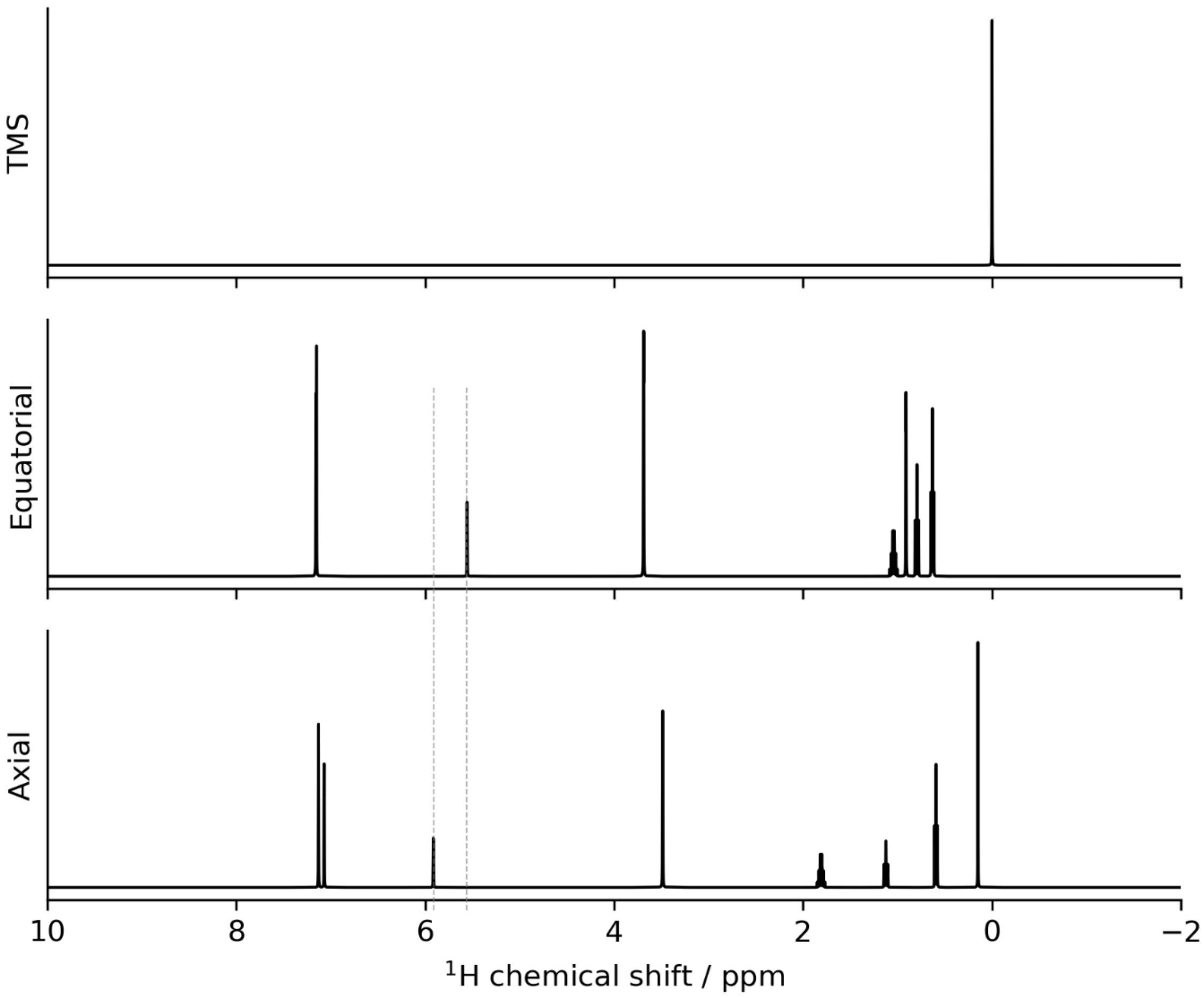

**Fig. S2** Computed GAIO NMR $^1$H chemical shifts at the TPSS/pcSseg-3 level of DFT with an auxiliary basis set and the CPCM($CHCl_3$) solvation model. (Top) TMS; (centre) *eq. trans*; and (bottom) *ax. trans*.

## 2.4 Crystal Structures of DIO and 5-Me DIO

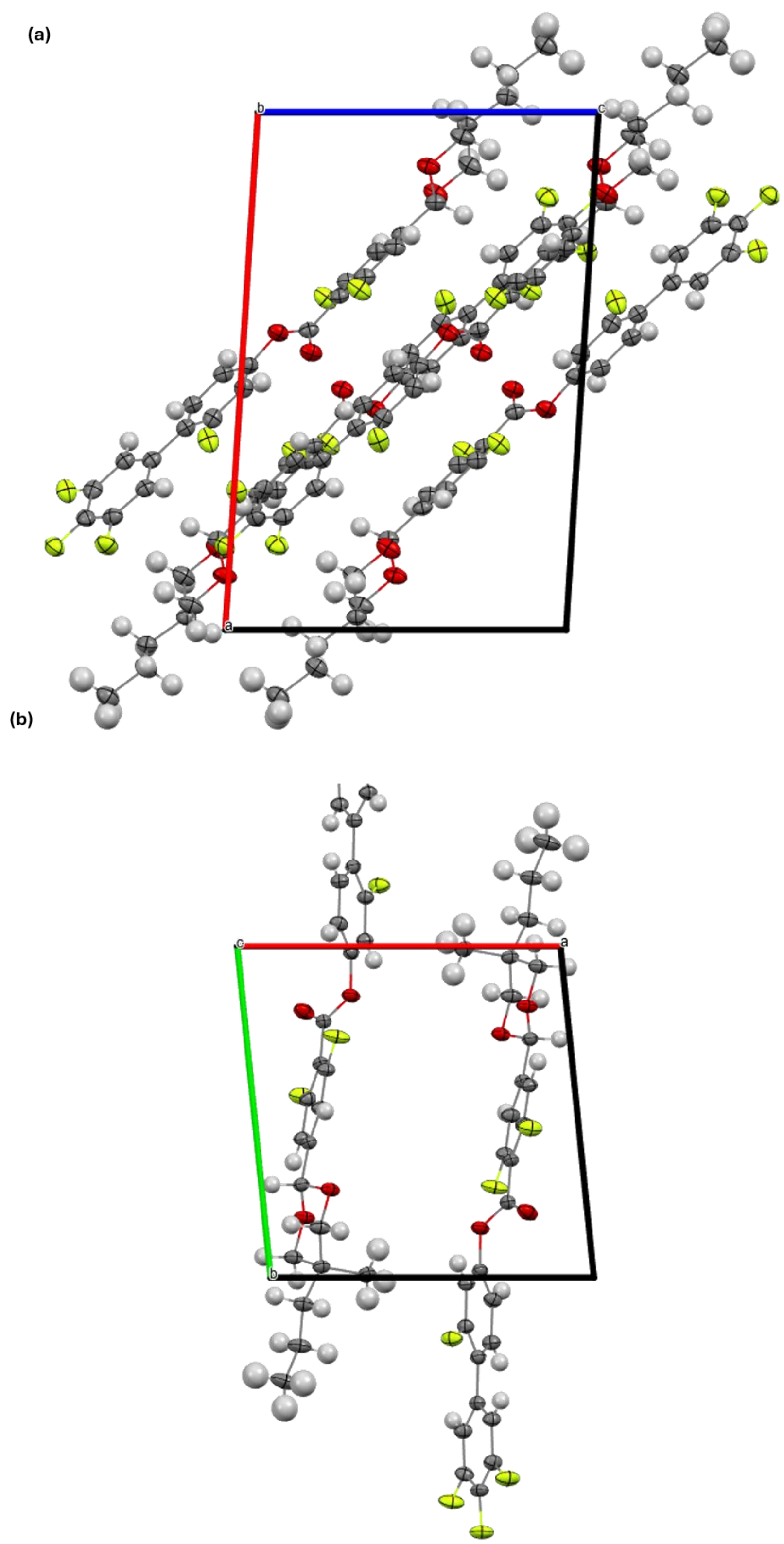

**Fig. S3** Crystal structures of **(a) DIO**; and **(b) 5-Me DIO**. Unit cells viewed along cryptographic b and c axes, respectively. Thermal ellipsoids are displayed at the 50% level.

### 2.5 Potential Energy Scans Demonstrating Conformational Bias via a Magic Methyl

Relaxed scans at the wB97x-3c level were used to assess how the inclusion and position of the methyl group alter the accessible conformations. Scanning over the 1D potential energy surface (PES) of the torsion centred on the bond between the 1,3-dioxane and the adjacent phenyl ring for compound **1** we find, unsurprisingly perhaps, that the unsubstituted and 5-methyl dioxanes behave comparably. For the 2-methyl, the steric footprint of the methyl leads to significantly larger energy barriers. For all three geometries the one-dimensional PES is symmetric with minima located at approximately -30° and 150° (**Fig. S4**).

The PES for rotation about the bond linking the dioxane ring to the terminal chain is broadly similar for **1** and **5-Me 1**, with comparable low-energy conformers. However, methyl substitution at the 5-position strongly perturbs one region of the torsional surface: the barrier near ca**.** +120° is increased from approximately 7 kJ mol$^{-1}$ in **1** to 76 kJ mol$^{-1}$ in **5-Me 1**. This indicates that the 5-methyl group selectively penalises conformations in which the terminal chain is directed towards the substituted face of the dioxane ring, consistent with a steric interaction between the methyl group and the central methylene region of the chain. The 5-methyl group effectively biases the accessible conformational landscape by suppressing this region of torsional space.

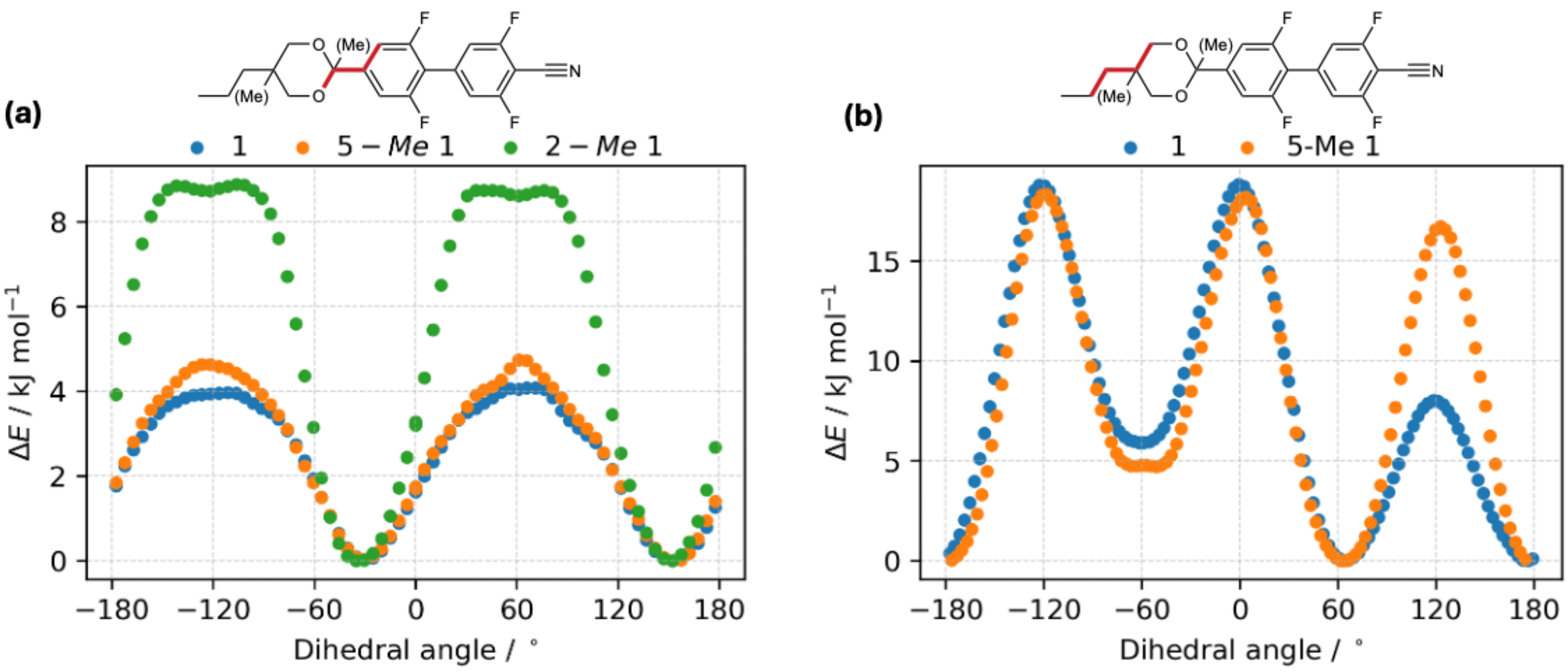

**Fig. S4:** Plot of relative energy ($\Delta E, kJ\ mol^{-1}$) as a function of angle (°) for relaxed scans about the one-dimensional potential energy surface for the indicated dihedral at the wB97x-3c level.

## 2.6 Binary Mixture Study between 1 and DIO

A virtual $N_F$ transition temperature for **1** can be obtained via binary mixture studies using **DIO** (**Fig. S5**); its physical observation in the pure material is presumably precluded by crystallisation. The phase diagram shows complete miscibility between **1** and **DIO** across the phase diagram allowing for a virtual value of $T_{NF}$ to be predicted at 33 °C.

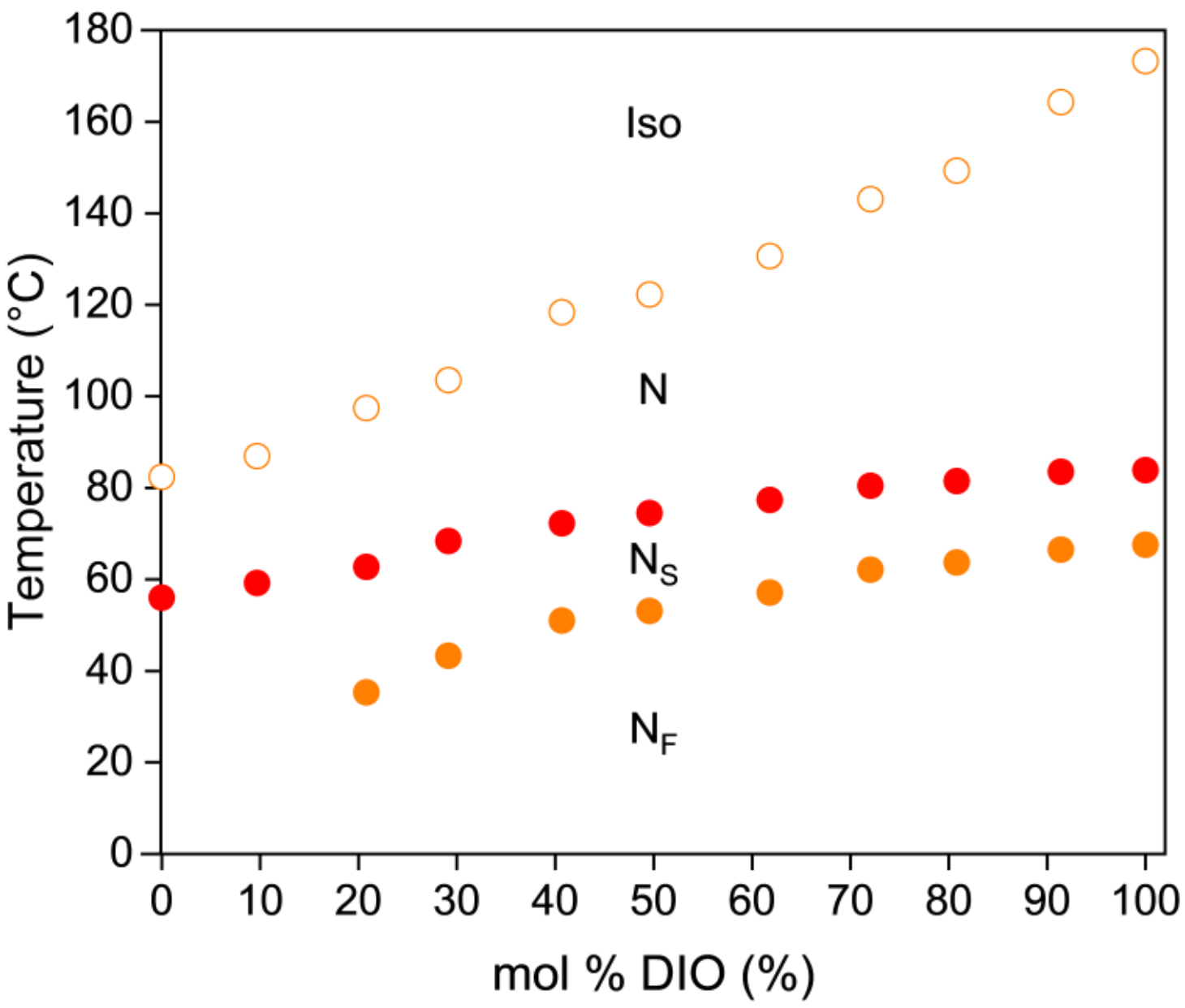

**Fig. S5** Binary mixture diagram of **1** and **DIO**.

## 2.7 [1]H NOSEY Data for the *eq* trans and cis geometric isomers of 1

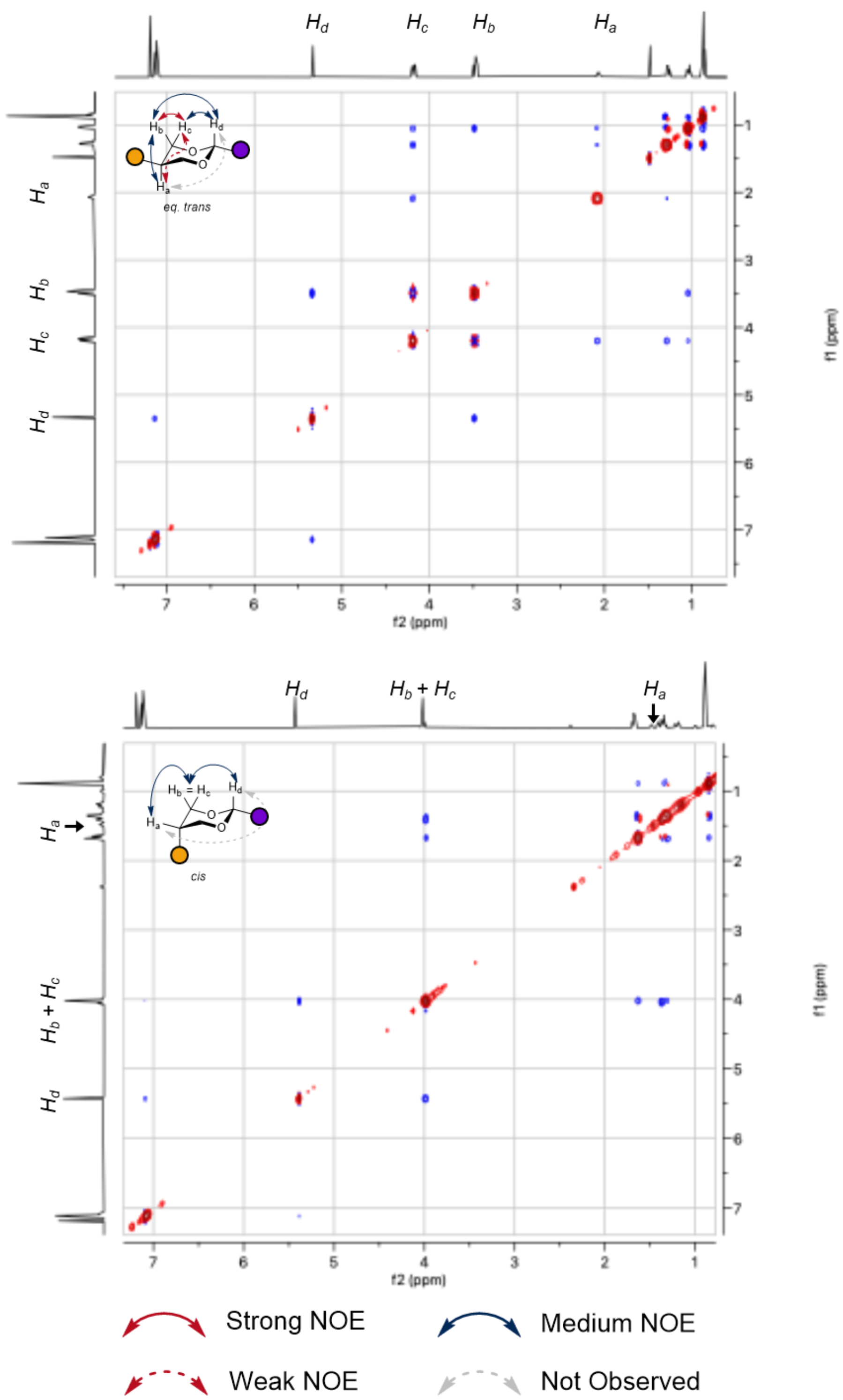

**Fig. S6** [1]H NOESY NMR for the *eq* trans [top]; and *cis* [bottom] geometric isomers of **1**. Observed NOE enhancements are assigned as shown.

## 2.8 Crystal Structures of 1 and 5-Me 1

(a)

(b)

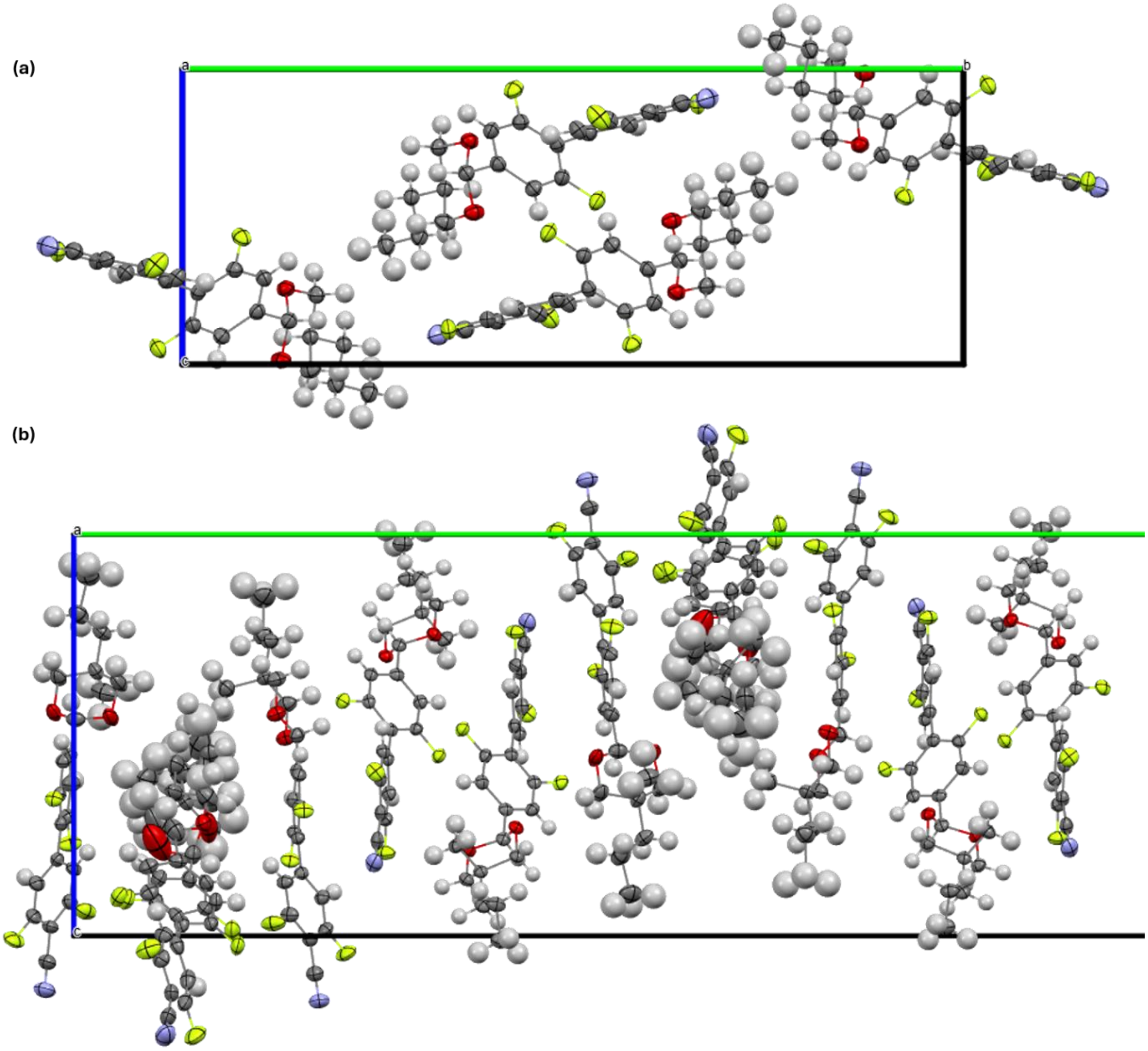

**Fig. S7** Crystal structures of **(a) 1**; and **(b) 5-Me 1**. Unit cells viewed along cryptographic a axis. Thermal ellipsoids are displayed at the 50% level.

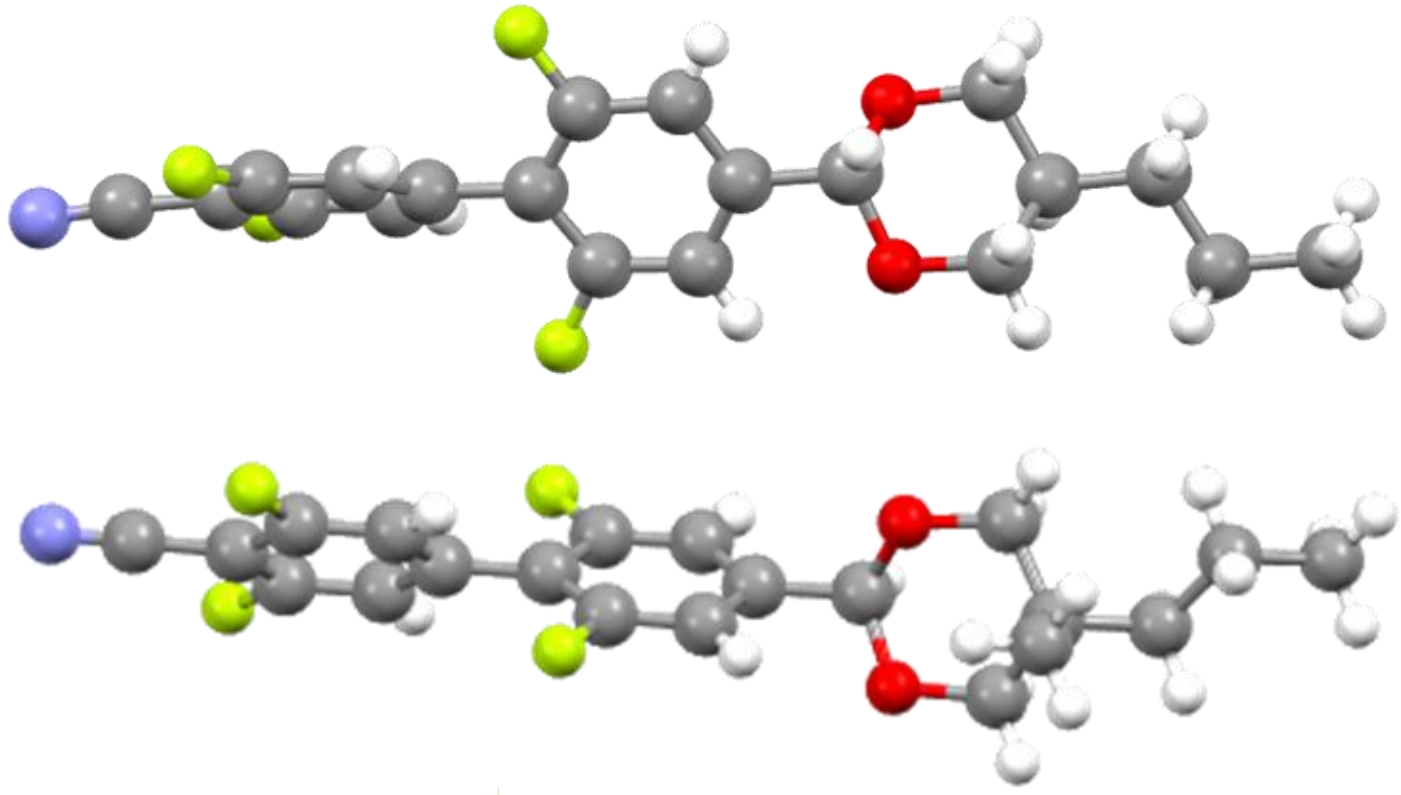

**Fig. S8** Single molecular geometries obtained from their respective crystal structures of **1** [top]; and **5-Me 1** [bottom]. Both molecules show very similar geometries about the dioxane ring.

## 2.9 Inseparability of 2-Me, 2,5-DiMe and 5-Me (Si) 1

During the preparation of the materials **2-Me 1**, **2,5-DiMe 1** and **5-Me 1 (Si)** both the cis and *eq* trans geometric isomers are obtained. For these particular homologues, we found these isomers to be inseparable by transitional separation techniques as well as HPLC. We demonstrate this inseparability in **Fig. S9** using **2-Me 1** as a representative example.

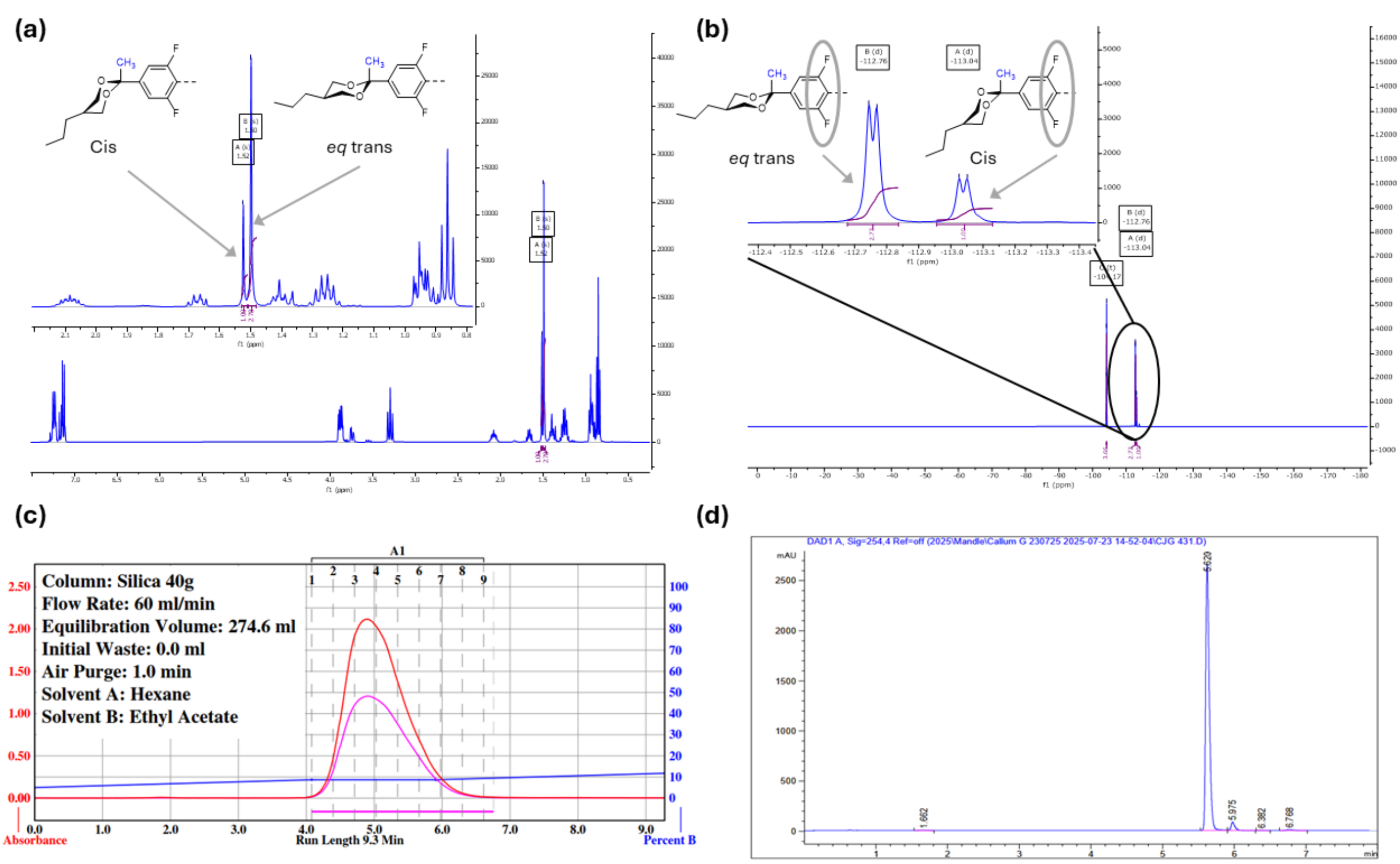

**Fig. S9** **(a)** $^{1}$H and **(b)** $^{19}$F NMR spectra measured **2-Me 1** measured on a Bruker Avance III HD 9.4T NMR spectrometer operating at 400 MHz and 376.4 MHz, respectively, showing the mix of cis and trans geometric isomers obtained during chemical synthesis; (c) chromatogram obtained for **2-Me 1** using a Combiflash NextGen 300+ flash chromatography system showing the single peak eluting at 5 minutes containing the inseparable mixture of both cis and trans geometric isomers. The run conditions are listed within the figure; and (d) HPLC assay obtained for **2-Me 1** using a Agilent 1290 Infinity II system fitted with a poroshell 120 ec-c18 column running a water:MeCN gradient. Both the cis and trans geometric isomers are found within the peak eluting at 5.620 mins.

## 2.10 Conformational Energy Landscapes of 5-methy, 2-methyl, and 2,5-dimethyl 1,3-dioxanes

We computed the Gibbs free energy of each stationary point on the potential energy surface for the *eq.* trans to *ax.* trans interconversion for **DIO**, **5-Me DIO**, and **2-Me DIO**. For each, geometries for the two *trans* forms were identified from a conformer search. We obtained initial structures for the intermediate twist-boat state and two transition states (TS1, TS2) via the nudged elastic band method using the wB97x-3c composite DFT method. The obtained geometries were reoptimized at the same level. Harmonic vibrational frequency calculation for these stationary points yields the Gibbs free energy, plotted below (**Fig. S10**).

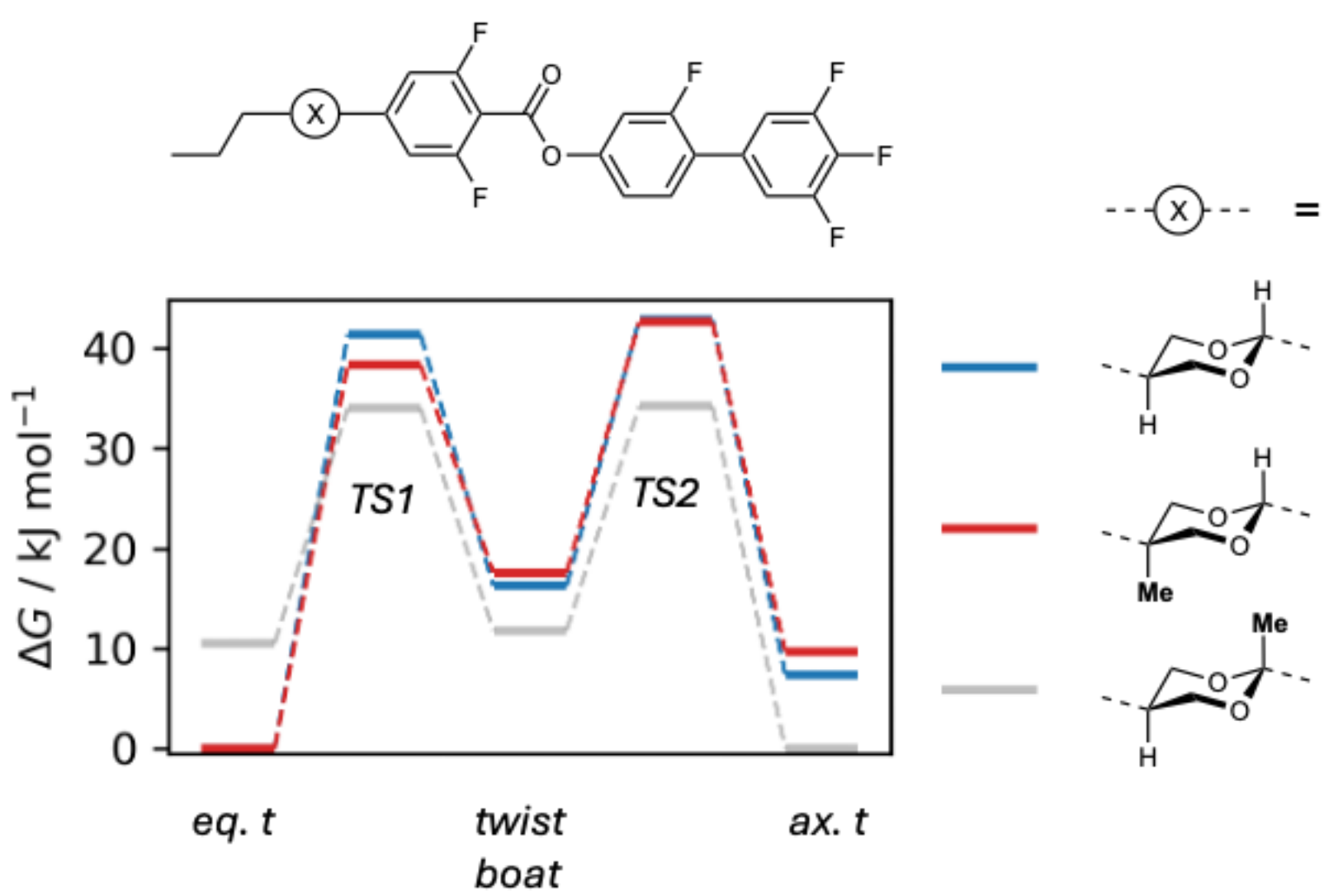

**Fig. S10** Gibbs free energy ($\Delta G$) for the *equatorial trans* to *axial trans* interconversion for **DIO** (blue), **5-Me DIO** (red), **2-Me DIO** (grey) as calculated at the wB97x-3c composite DFT level. Stationary points are labelled on the plot.

We first study **DIO**-like materials - as shown in **Figure S11** the parent unsubstituted 1,3-dioxane and the 5-methy-1,3-dioxane show the same overall trend, the *eq.* trans form being the energy minimum with the *ax. trans* form being somewhat higher in free energy ($\Delta G \sim 8\ kJ\ mol^{-1}$ for the parent 1,3-dioxane versus $\Delta G \sim 10\ kJ\ mol^{-1}$ for the 5-methyl analogue) (**Fig. S11**). The 2-methyl variant behaves entirely differently; here, the *ax.* trans isomer is significantly lower in free energy than the *eq. trans* ($\Delta G \sim 11\ kJ\ mol^{-1}$) due to the steric clash between the methyl unit and the adjacent aryl ring. This leads to the hypothesis that the materials bearing a 2-methyl group are non-mesogenic due to the adoption of an *axial trans* configuration which, as per DSC cycling experiments (**Fig. 2g**), is not conducive to supporting liquid crystalline order.

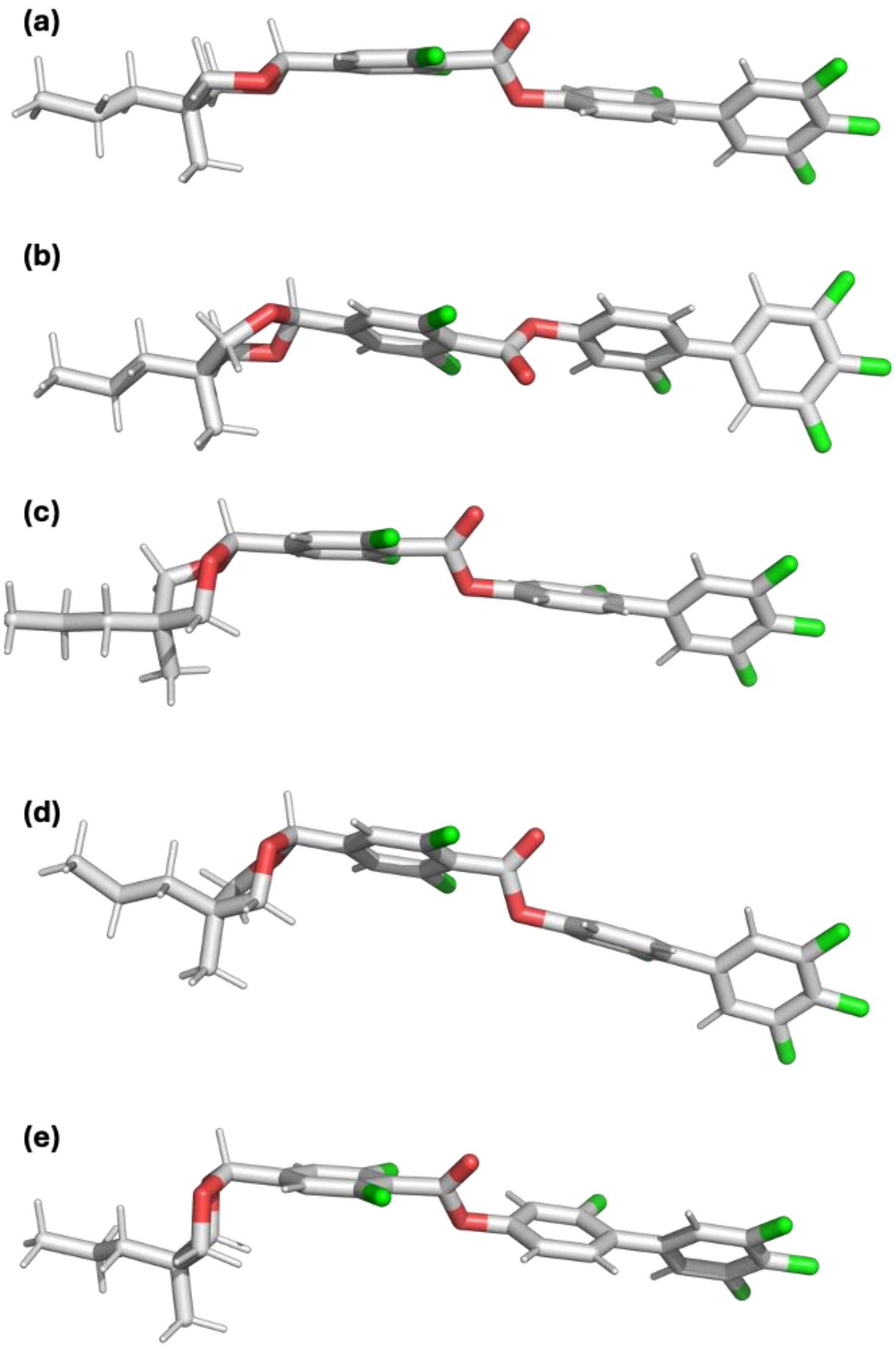

**Fig. S11:** Optimised geometries the stationary points of **5-Me DIO** at the wB97x-3c level: (a) *equatorial* trans (*ΔG=0.0 kJ mol$^{-1}$*); (b) first transition state (Δ*G=38.3 kJ mol$^{-1}$*); (c) intermediate twist-boat (Δ*G=17.6 kJ mol$^{-1}$*); (d) second transition state (Δ*G=42.6 kJ mol$^{-1}$*); (e) *axial* trans (Δ*G=9.7 kJ mol$^{-1}$*).

We now move to the *eq. trans* to *ax. trans* interconversion in compound **1** and analogues thereof (**Fig. S12**). These largely mirror the behaviour of the DIO-derived materials. For the parent and 5-methyl analogue is lowest in energy and for the 2-methyl and 5-methyl the axial trans is lowest in energy. For the 5-methyl-1,3,5-dioxasilinane the behaviour is somewhat different; the length of the Si-C bond distorts the ring, and the intermediate twist-boat state is lower in energy to the *ax. trans* form. Experimentally, the 2-methyl, 2,5-dimethyl and 5-methyl-1,3,5-dioxasilinane materials present as inseparable isomers. From this, and results from DIO-like materials, we establish that the enhanced thermal stability conferred by the methyl group is not due to any dramatic change in the geometry of the dioxane ring.

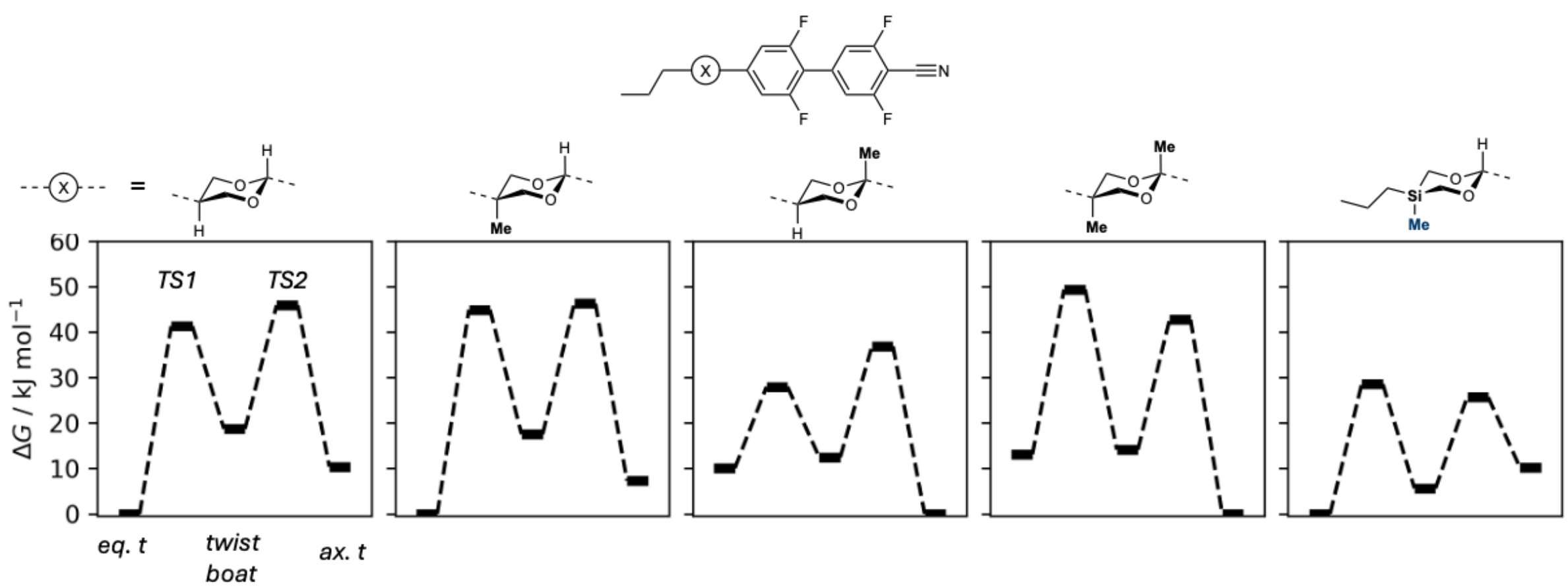

**Fig. S12:** Gibbs free energy ($\Delta G$) for the *equatorial trans* to *axial trans* interconversion for **1**, **5-Me 1**, **2-Me DIO**, **2,5-DiMe 1**, **5-MeDs** as calculated at the wB97x-3c level. Stationary points are labelled on the leftmost plot.

## 2.11 Characterisation data for 5-Me 2 – 8

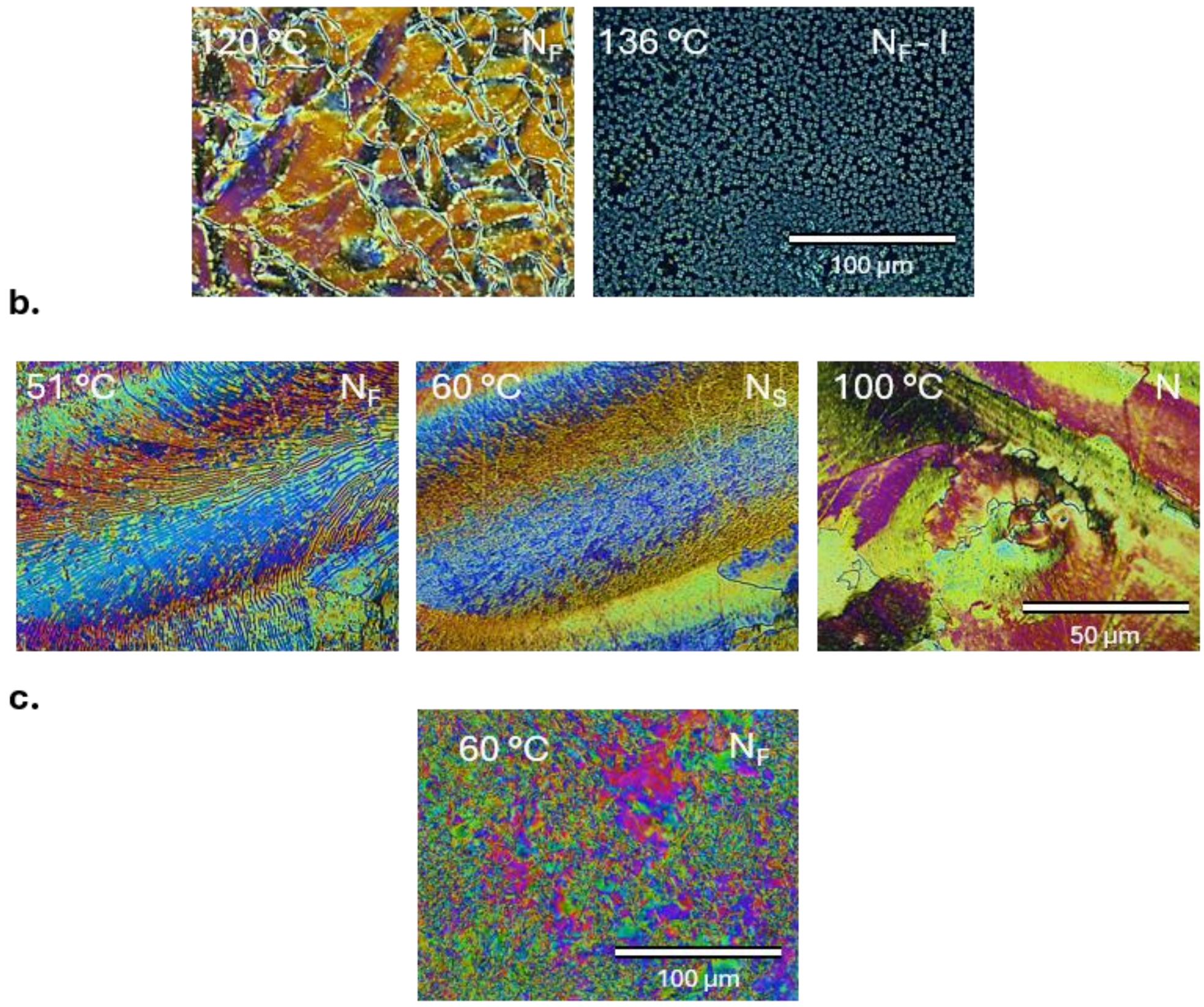

**Fig. S13** Representative POM micrographs depicting the phase sequences exhibited by **(a) 5-Me 5**; **(b) 5-Me 6** ; and **(c) 5-Me 8** Images are taken for thin samples sandwiched between glass coverslips.

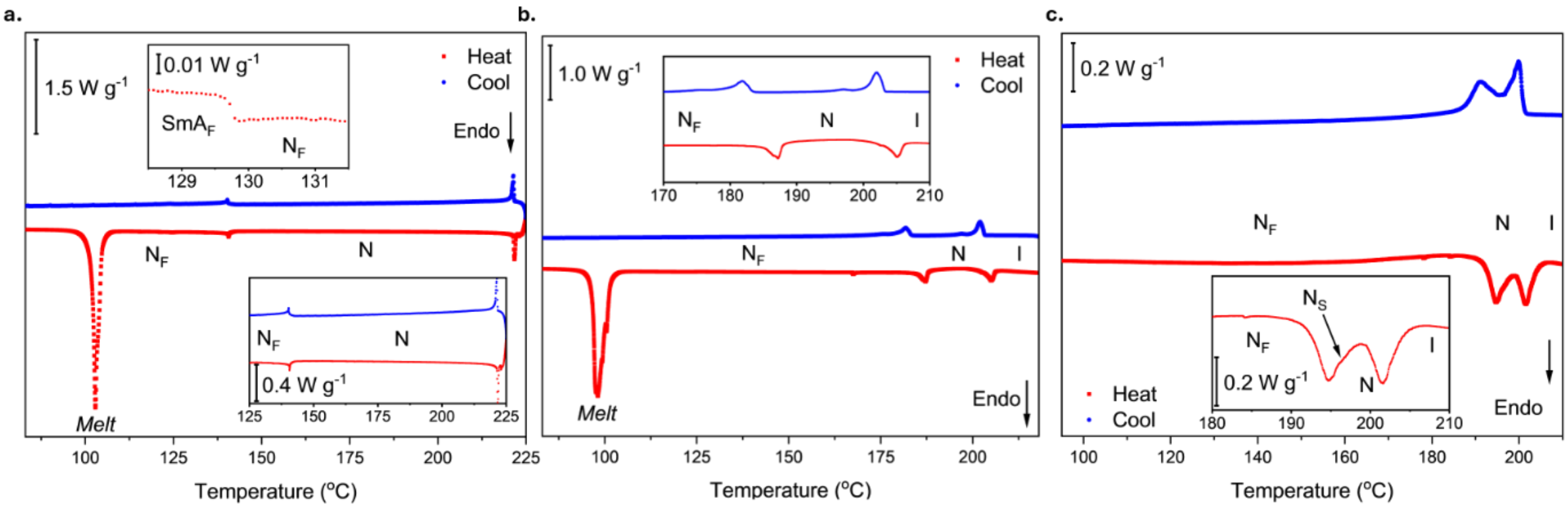

**Fig. S14** Representative DSC thermograms for **(a) 5-Me 2**; **(b) 5-Me 4**; and **(c) 5-Me 8**. Traces are given on the second heat and cooling cycle.

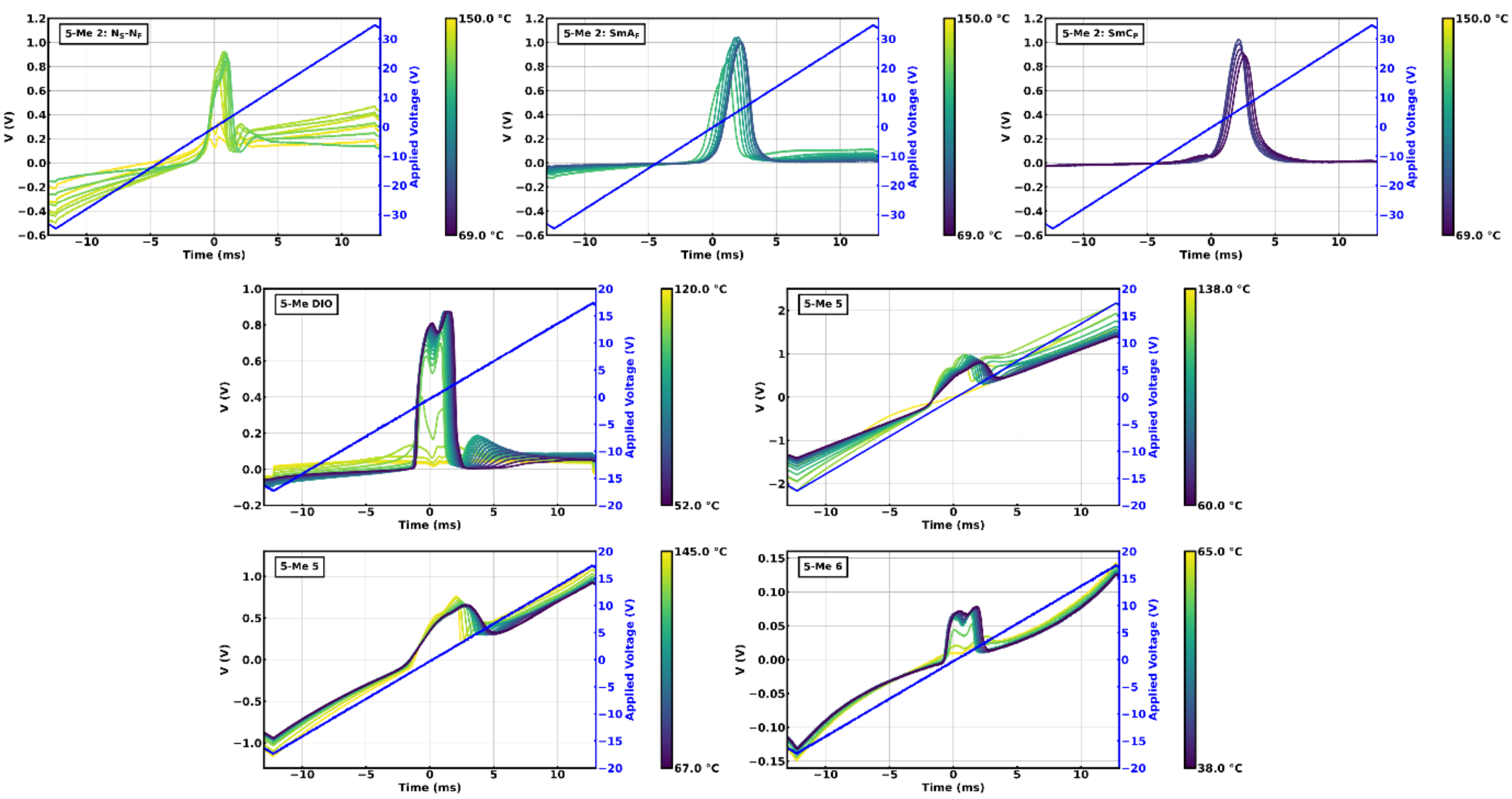

**Fig. S15** Representative current response measurements for the listed materials. Data was recorded in 5 μm thick cells with no alignment layers. Triangle waves were used with 20 Hz frequency.

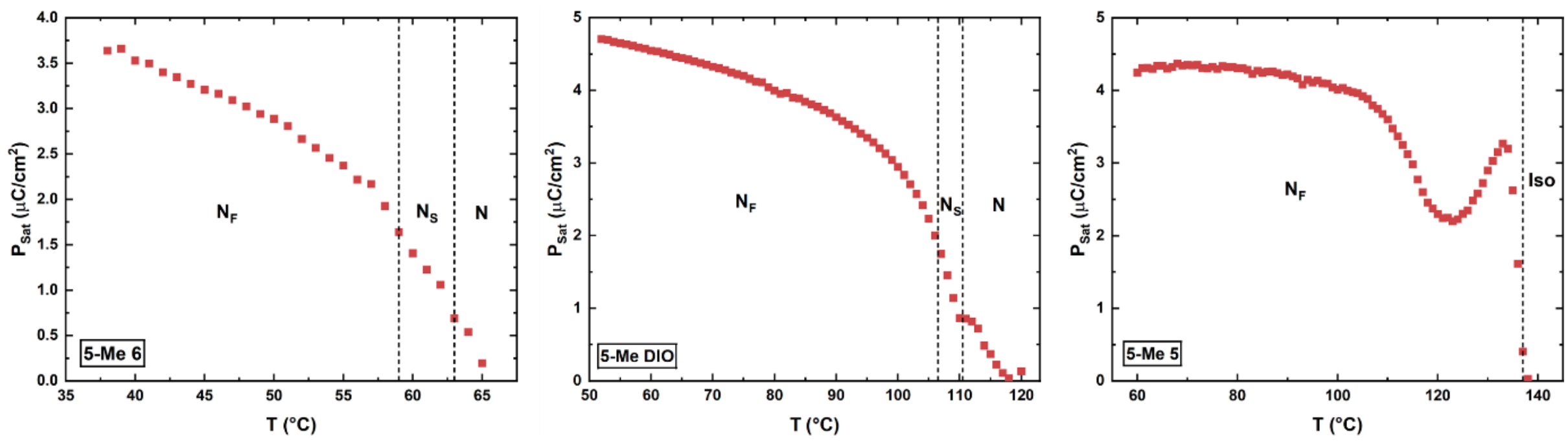

**Fig. S16** Temperature dependence for the saturated polarisation measured from the current response measurements for the materials marked.

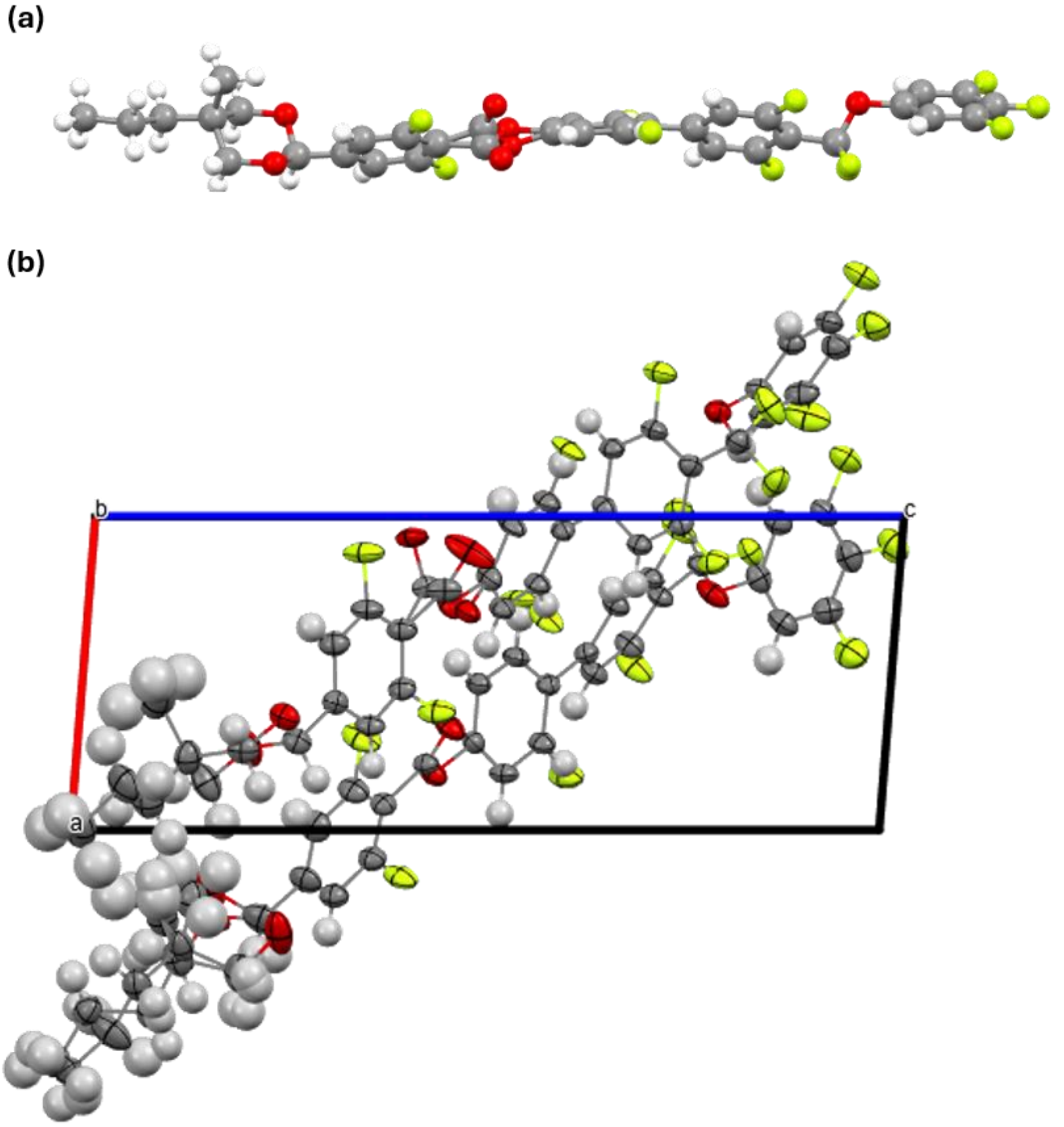

**Fig. S17** **(a)** Molecular shape extracted from the crystal structure; and **(b)** crystal structures of **5-Me 2**. Unit cells viewed along crystallographic b axis. Thermal ellipsoids are displayed at the 50% level.

## 3. Organic Synthesis

The total synthesis of **5-Me DIO**, compounds **6 & 7**, and the **5-Me** compounds **2-8** is outlined in **Scheme S1**. The synthesis and transitional properties of compounds **2-5** and **8** have been reported previously [11-16]. The synthesis of the of the phenolic materials and the synthesis of 2,6-difluoro-4-(5-propyl-1,3-dioxan-2-yl)benzoic acid have been reported elsewhere [12,13,17-21].

X = H or $CH_3$

(i) *p*-TsOH, Toluene
(ii) (1) n-BuLi, -78 °C
(2) $CO_{2(g)}$
(3) HCl $_{(aq)}$
(iii) EDC.HCl, DMAP, DCM

**Scheme S1:** Preparation of **5-Me DIO**, compounds **6 & 7**, and the **5-Me** compounds **2-8**.

### 3.1 Preparation of 2,6-difluoro-4-(5-methyl-5-propyl-1,3-dioxan-2-yl)benzoic acid

#### 3.1.1 Preparation of 2-(3,5-difluorophenyl)-5-propyl-1,3-dioxane

A flame dried flask fitted with a dean-stark apparatus was charged with 3,5-difluorobenzaldehyde (7.10g, 50 mmol), 2-methyl-2-propylpropane-1,3-diol (6.73g, 51 mmol), and a catalytic amount of *p*-toluenesulphonic acid under an atmosphere of dry nitrogen. The reagents were dissolved in 150 mL of dry toluene and the flask heated to reflux for 18h where completion of the reaction was judged by TLC by the absence of 3,5-difluorobenzaldehyde.

The reaction was cooled to room temperature, diluted with 100 mL of EtOAc and washed with 1x 50 mL of 2M $NaOH_{(aq)}$ and 2x 50 mL of brine. The organics were then dried over $MgSO_4$ before the reaction solution was concentrated and purified by flash chromatography over silica gel using a using a Combiflash NextGen300+ system running a gradient of hexanes and EtOAc (0-10% EtOAc). The chromatographed material was then recrystallised from MeOH to afford the title compound as white crystals.

*2-(3,5-difluorophenyl)-5-propyl-1,3-dioxane*

Yield: (White crystalline solid) 7.00 g, 54%.

$R_F$ (10% EtOAc): 0.78

$^1$H NMR (501 MHz): 7.09 – 7.03 (m, 2H, Ar-**H**), 6.80 (tt, J = 8.9, 2.4 Hz, 1H, Ar-**H**), 5.36 (s, 1H, Ar-C**H**-$O_2$), 3.83 (dd, J = 11.2, 1.4 Hz, 2H, 2x O-C$\mathbf{H_{eq}}$$H_{ax}$-C), 3.67 (dd, J = 11.4, 0.8 Hz, 2H, 2x O-C$\mathbf{H_{ax}}$$H_{eq}$-C), 1.36 – 1.28 (m, 2H, Me-C-C$\mathbf{H_2}$-$CH_2$),

1.28 (s, 3H, (O-$CH_2$)$_2$-C-C$\mathbf{H_3}$), 1.15 – 1.08 (m, 2H, $CH_2$-C$\mathbf{H_2}$-$CH_3$), 0.94 (t, J = 7.3 Hz, 3H, $CH_2$-C$\mathbf{H_3}$).

$^{13}C\{^1H\}$ NMR (126 MHz): 163.04 (dd, $J$ = 248.4, 12.4 Hz), 142.35 (t, $J$ = 9.1 Hz), 109.44 (dd, $J$ = 19.8, 5.9 Hz), 104.18 (t, $J$ = 25.3 Hz), 100.21 (t, $J$ = 2.6 Hz), 38.85, 32.95, 20.25, 15.99, 15.10.

$^{19}F$ NMR (376 MHz) -109.73 (t, $J_{F\text{-}H}$ = 8.0 Hz, 2F, Ar-**F**).

### 3.1.2 Preparation of 2,6-difluoro-4-(5-methyl-5-propyl-1,3-dioxan-2-yl)benzoic acid

An oven dried flask was cooled under an atmosphere of dry nitrogen, charged with 2-(3,5-difluorophenyl)-5-propyl-1,3-dioxane (4.32g, 17 mmol), anhydrous THF (100 mL), and cooled to -78 °C with an external $CO_2$/acetone bath. To this, a solution of *n*-butyl lithium (1.6 M in hexane, 12.6 mL, 20 mmol) was added dropwise. Once the addition was complete, the solution was allowed to stir for 0.5 h before $CO_{2(g)}$, pre-dried via bubbling through conc. $H_2SO_4$, was bubbled through the reaction for 20 mins. The cooling bath was removed, and the reaction mixture allowed to warm to ambient temperature over the course of 2h. The solution was acidified with 2M HCl (~ 50 mL). The organic layer was separated and retained, and the aqueous extracted with 2x 20 mL EtOAc. The combined organics were dried over $MgSO_4$ and concentrated *in vacuo* and re-crystallised from hexanes: toluene (10:1) to afford the title compound as a crystalline white solid.

*2,6-difluoro-4-(5-methyl-5-propyl-1,3-dioxan-2-yl)benzoic acid*

Yield: (White solid) 4.49 g, 88 %

$R_F$ (EtOAc): 0.75

$^1H$ NMR (400 MHz): 7.14 (d, $J$ = 9.0 Hz, 2H, Ar-**H**), 5.35 (s, 1H, Ar-C**H**-$O_2$), 3.81 (dd, $J$ = 10.2, 1.3 Hz, 2H, 2x O-C$\mathbf{H_{eq}}H_{ax}$-C), 3.65 (d, $J$ = 11.0 Hz, 2H, 2x O-C$\mathbf{H_{ax}}H_{eq}$-C), 1.35 – 1.24 (m, 2H, Me-C-C$\mathbf{H_2}$-$CH_2$), 1.23 (s, 3H, (O-$CH_2$)$_2$-C-C$\mathbf{H_3}$), 1.14 – 1.04 (m, 2H, $CH_2$-C$\mathbf{H_2}$-$CH_3$), 0.92 (t, $J$ = 7.2 Hz, 3H, $CH_2$-C$\mathbf{H_3}$).

$^{13}C\{^1H\}$ NMR (126 MHz): 166.06, 161.38 (dd, $J$ = 259.3, 5.7 Hz), 145.57 (t, $J$ = 10.0 Hz), 110.42 (dd, $J$ = 23.9, 3.4 Hz), 109.43 (t, $J$ = 16.3 Hz), 99.36, 38.80, 32.99, 20.20, 15.99, 15.09.

$^{19}F$ NMR (376 MHz): -107.95 (d, $J_{F\text{-}H}$ = 10.0 Hz, 2F, Ar-**H**).

### 3.2 Preparation of 5-Me DIO, compounds 6 & 7, and compounds 5-Me 2-8.

**5-Me DIO**, compounds **6** & **7**, and the **5-Me** compounds **2-8** were prepared according to the following general esterification procedure. A small vial was charged with the appropriate phenol (0.5 mmol, 1.0 eq), benzoic acid (0.6 mmol, 1.1 eq.), EDC.HCl (0.75 mmol, 1.5 eqv.) and DMAP (~ 2 mol%). Dichloromethane was added (conc. ~ 0.1 M) and the suspension stirred until complete consumption of the phenol as judged by TLC. Once complete, the reaction solution was concentrated and purified by flash chromatography over silica gel with a gradient of hexane/DCM using a Combiflash NextGen300+ system. To remove ionic impurities, a pre-column containing a small amount of neutral alumina was used before samples entered the silica gel column. The chromatographed material was dissolved into the minimum quantity of DCM, filtered through a 0.2 micron PTFE filter, concentrated to dryness and finally recrystalised from MeOH or EtOH as indicated to afford the title materials as white solids.

**5-Me DIO**

*2,3',4',5'-tetrafluoro-[1,1'-biphenyl]-4-yl 2,6-difluoro-4-(5-methyl-5-propyl-1,3-dioxan-2-yl)benzoate*

| | |
|---|---|
| Yield: | (White crystalline solid) 215 mg, 82 %. |
| $R_F$ (DCM: hexanes [1:1]): | 0.36 |
| Re-crystallisation Solvent: | MeOH |
| $^1$H NMR (501 MHz): | 7.43 (t, *J* = 8.6 Hz, 1H, Ar-**H**), 7.23 – 7.14 (m, 6H, Ar-**H**)*, 5.38 (s, 1H, Ar-C**H**-$O_2$), 3.83 (dd, *J* = 9.9, 1.3 Hz, 2H, 2x O-C$\mathbf{H_{eq}}$$H_{ax}$-C), 3.67 (dd, *J* = 9.9, 2.1 Hz, 2H, 2x O-C$\mathbf{H_{ax}}$$H_{eq}$-C), 1.34 – 1.26 (m, 2H, Me-C-C$\mathbf{H_2}$-$CH_2$), 1.25 (s, 3H, $(CH_2)_2C(CH_2)$-**Me**), 1.14 – 1.08 (m, 2H, $CH_2$-C$\mathbf{H_2}$-$CH_3$), 0.93 (t, *J* = 7.2 Hz, 3H, $CH_2$-C$\mathbf{H_3}$). * Overlapping Signals. |
| $^{13}$C{$^1$H} NMR (126 MHz): | 162.11 (dd, *J* = 258.1, 5.7 Hz), 159.62 (d, *J* = 250.9 Hz), 158.50, 151.32 (ddd, *J* = 243.9, 9.5, 4.7 Hz), 151.10 (d, *J* = 10.8 Hz), 145.80 (t, *J* = 9.9 Hz), 131.05 – 130.72 (m), 124.40 (d, *J* = 11.8 Hz), 118.28 (d, *J* = 3.6 Hz), 113.37 (m), 110.79 (d, *J* = 26.0 Hz), 110.47 (dd, *J* = 26.4, 3.0 Hz), 109.61 (t, *J* = 17.1 Hz), 99.32 (t, *J* = 2.3 Hz), 38.81, 33.01, 20.22, 16.00, 15.10. |
| $^{19}$F NMR (376 MHz): | -108.50 (d, $J_{F-H}$ = 9.6 Hz, 2F, Ar-**F**), -114.34 (t, $J_{F-H}$ = 9.7 Hz, 1F, Ar-**F**), -134.17 (dd, $J_{F-F}$ = 20.6 Hz, $J_{F-H}$ = 8.6 Hz, 2F, Ar-**F**), -161.15 (tt, $J_{F-F}$ = 13.8 Hz, $J_{F-H}$ = 6.5 Hz, 1F, Ar-**F**). |

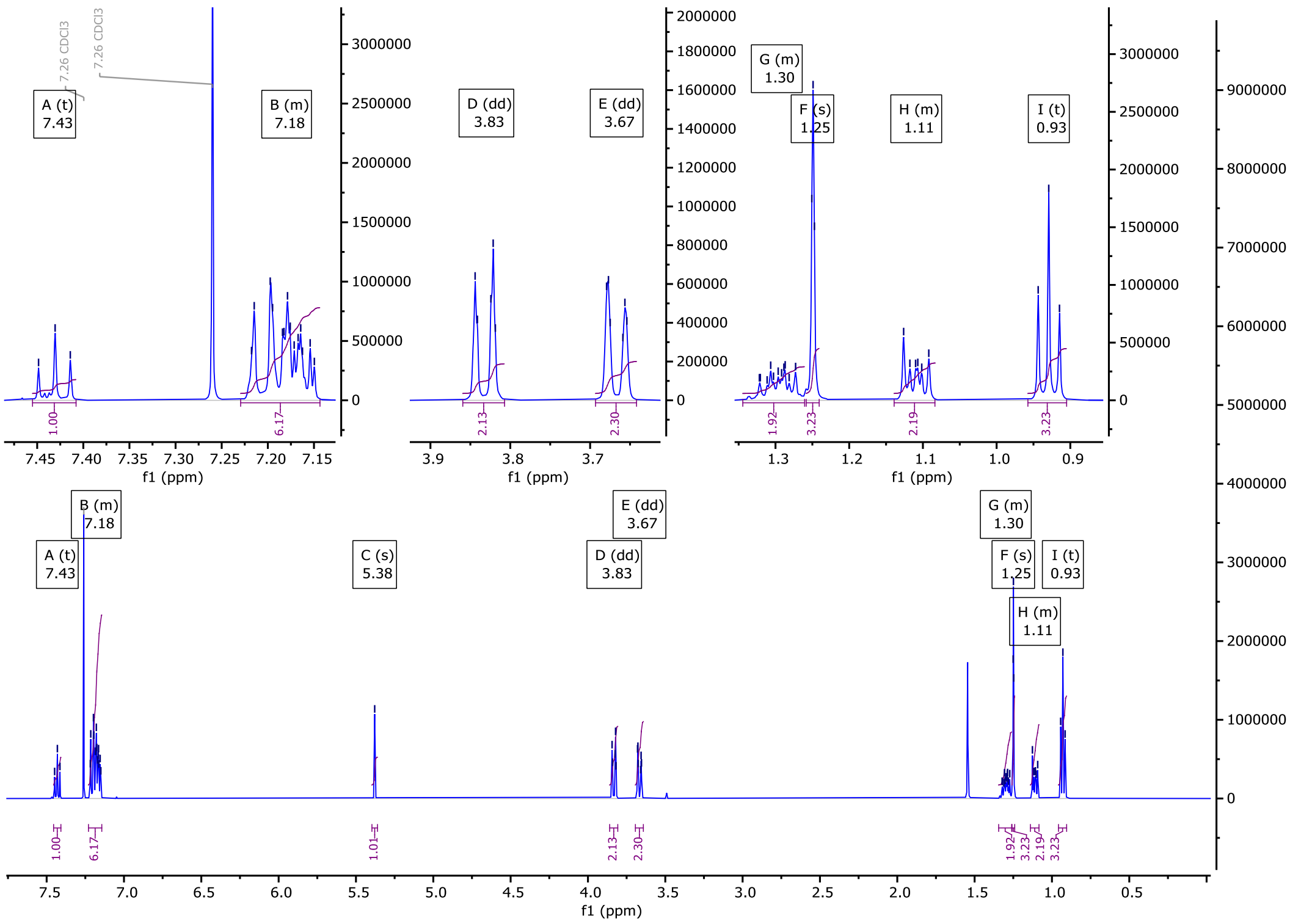

**Fig. S18.** $^1H$ NMR of **5-Me DIO** in $CDCl_3$.

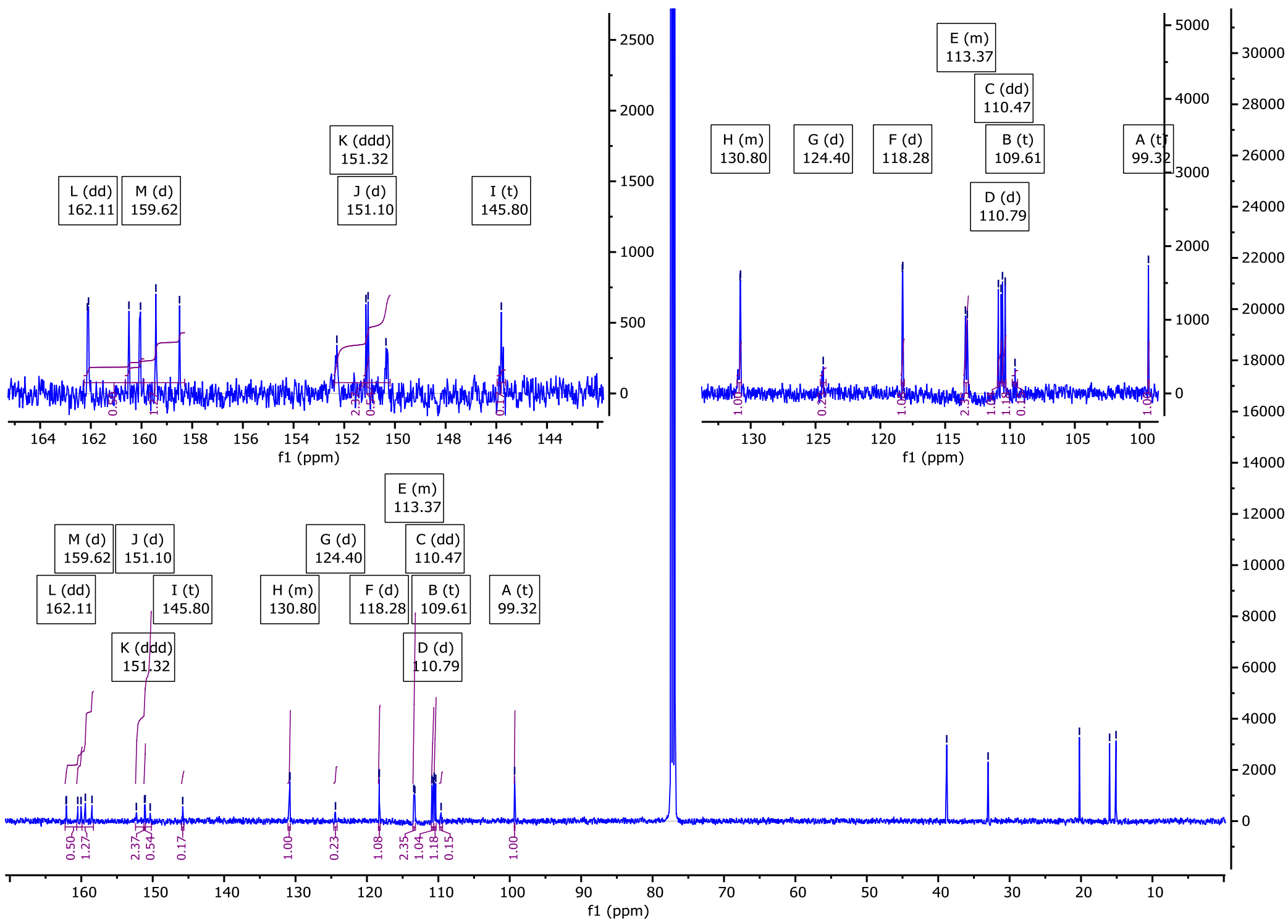

**Fig. S19.** $^{13}C\{^1H\}$ NMR of **5-Me DIO** in $CDCl_3$.

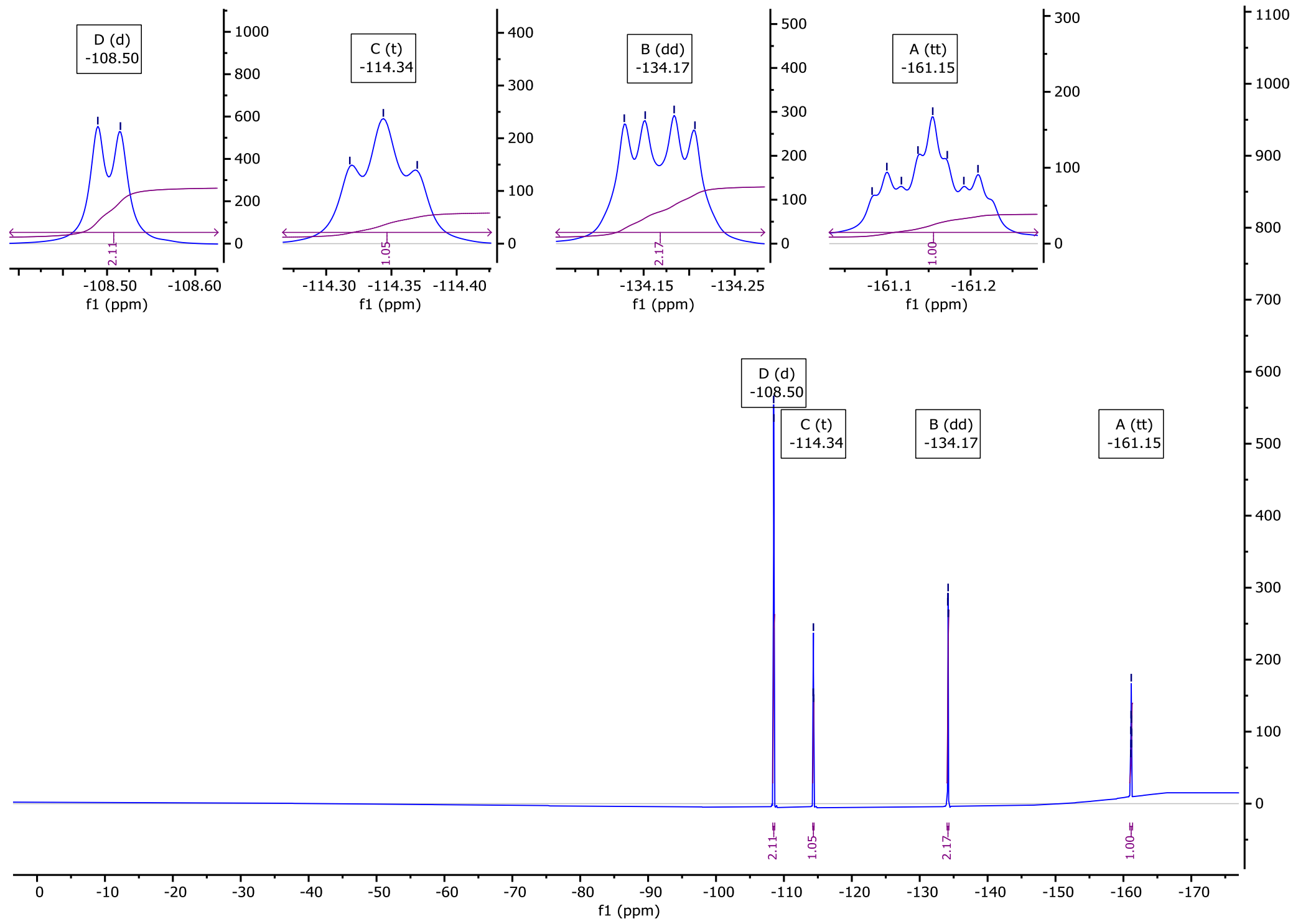

**Fig. S20.** $^{19}F$ of **5-Me DIO** in $CDCl_3$.

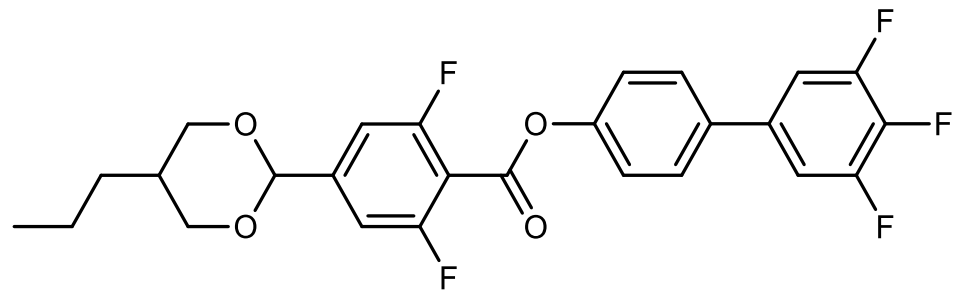

**6**

*3',4',5'-trifluoro-[1,1'-biphenyl]-4-yl 2,6-difluoro-4-(5-propyl-1,3-dioxan-2-yl)benzoate*

Yield: (White crystalline solid) 191 mg, 77 %.

$R_F$ (DCM: hexanes [1:1]): 0.40

Re-crystallisation Solvent: EtOH

$^1$H NMR (501 MHz): 7.55 (ddd, *J* = 8.6, 2.9, 2.0 Hz, 2H, Ar-**H**), 7.34 (ddd, *J* = 8.6, 2.8, 2.1 Hz, 2H, Ar-**H**), 7.22 – 7.14 (m, 4H, Ar-**H**)*, 5.40 (s, 1H, Ar-C**H**-$O_2$), 4.26 (dd, *J* = 11.8, 4.6 Hz, 2H, 2x O-C$\mathbf{H_{eq}}$$H_{ax}$-C), 3.54 (t, *J* = 11.5 Hz, 2H, 2x O-C$\mathbf{H_{ax}}$$H_{eq}$-C), 2.20 – 2.09 (m, 1H, $CH_2$-C**H**-$CH_2$), 1.35 (h, *J* = 7.3 Hz, 2H, $CH_2$-C$\mathbf{H_2}$-$CH_3$)), 1.10 (q, *J* = 7.1 Hz, 2H, CH-C$\mathbf{H_2}$-$CH_2$), 0.94 (t, *J* = 7.3 Hz, 3H, $CH_2$-C$\mathbf{H_3}$). *Overlapping Signals.

$^{13}$C{$^1$H} NMR (126 MHz): 160.99 (dd, *J* = 257.9, 5.8 Hz), 159.84, 151.71 (ddd, *J* = 249.8, 9.7, 4.0 Hz), 150.61, 145.45 (t, *J* = 9.9 Hz139.50 (dt, *J* = 252.2, 15.5 Hz), 136.52 (dd, *J* = 12.1, 8.5 Hz), 128.21, 122.43, 111.27 (dd, *J* = 16.6, 5.2 Hz), 110.36 (dd, *J* = 23.6, 3.4 Hz), 109.99 (t, *J* = 17.3 Hz), 98.99, 72.71, 34.02, 30.36, 19.66, 14.32.

$^{19}$F NMR (376 MHz): -108.76 (d, $J_{F-H}$ = 9.8 Hz, 2F, Ar-**F**), -133.88 (dd, $J_{F-F}$ = 20.5 Hz, $J_{F-H}$ = 8.7 Hz, 2F, Ar-**F**), -162.33 (tt, $J_{F-F}$ = 20.3 Hz, $J_{F-H}$ = 6.7 Hz, 1F, Ar-**F**).

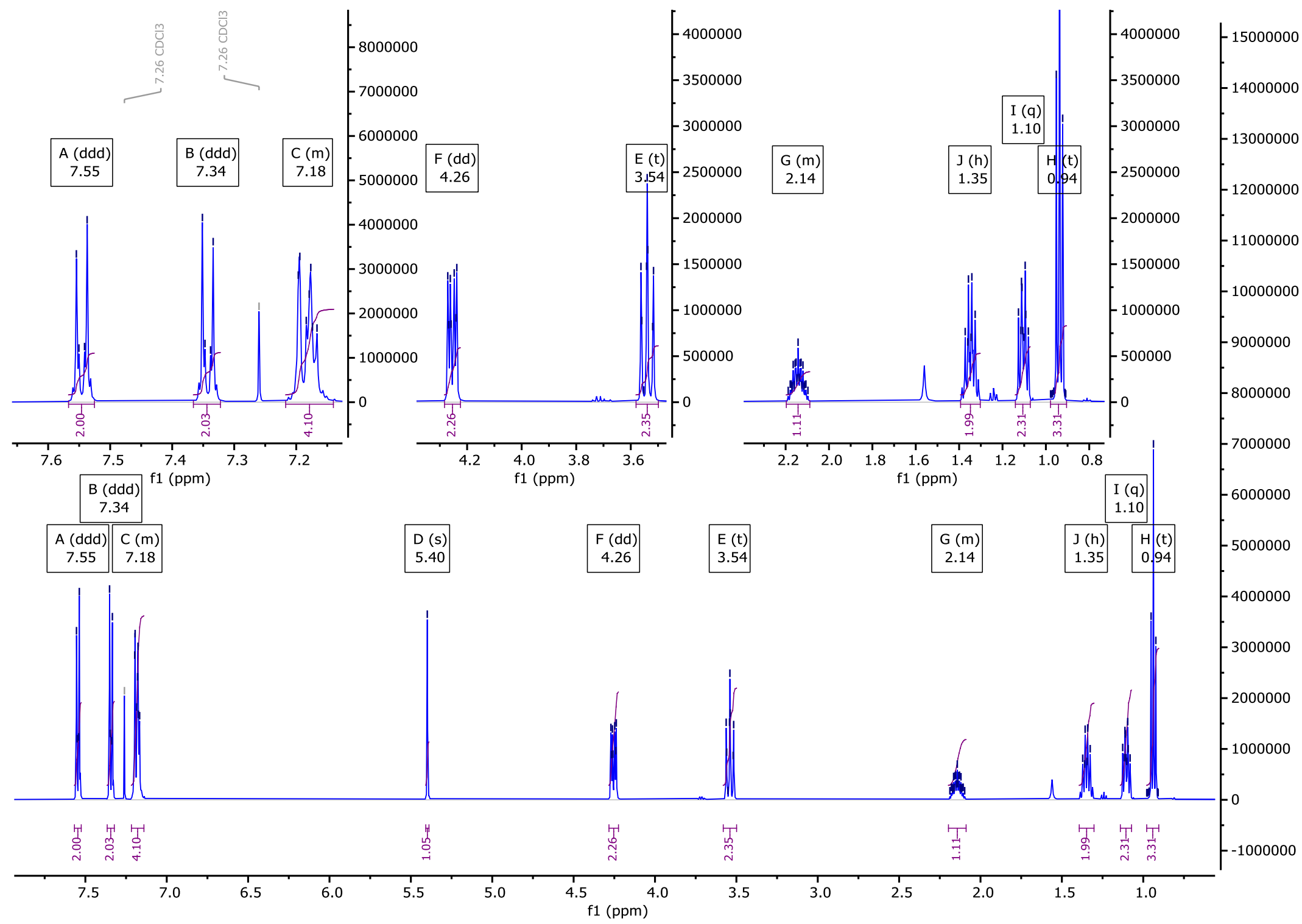

**Fig. S21.** $^{1}H$ NMR of **6** in $CDCl_3$.

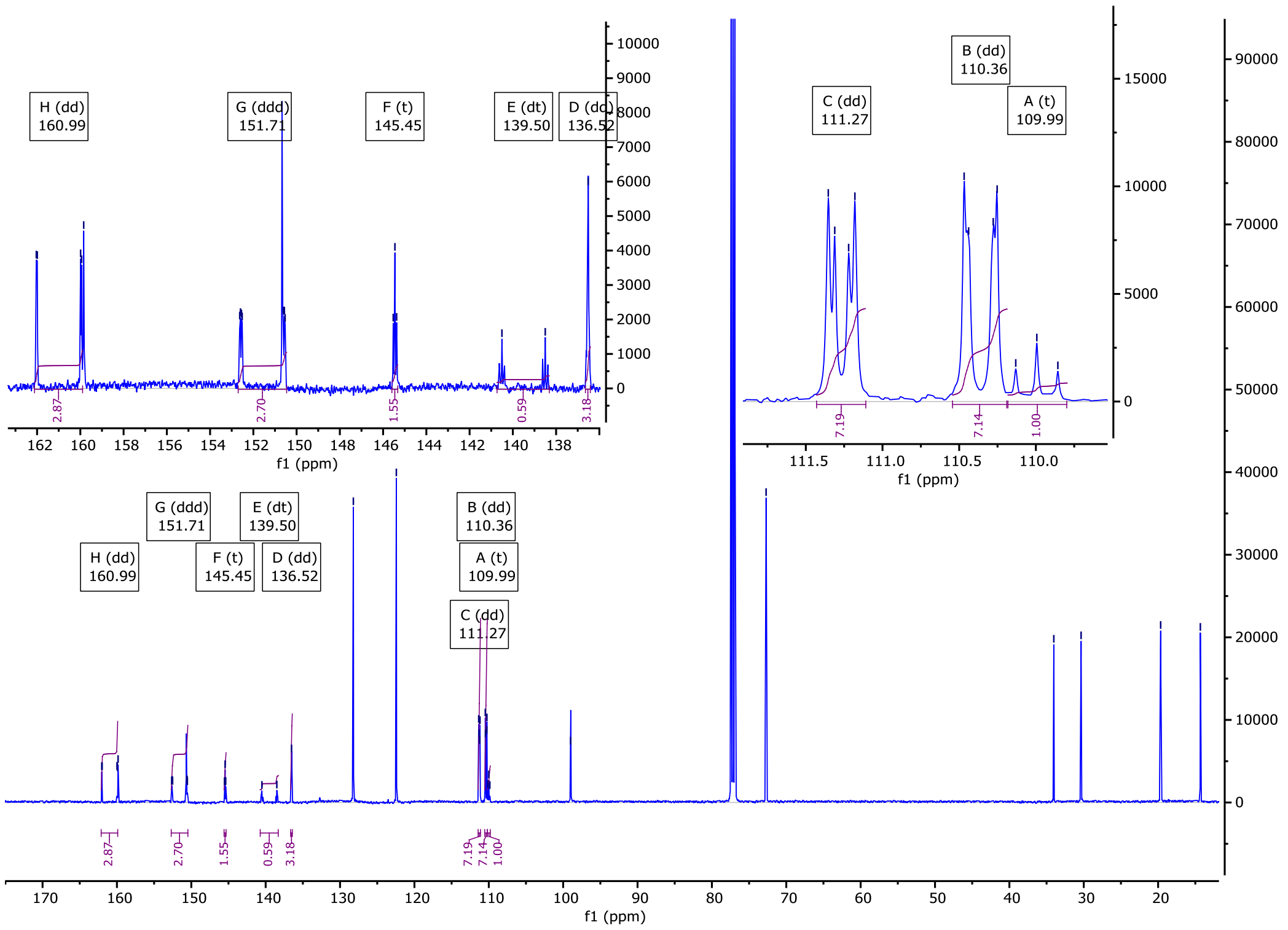

**Fig. S22.** $^{13}C\{^{1}H\}$ NMR of **6** in $CDCl_3$.

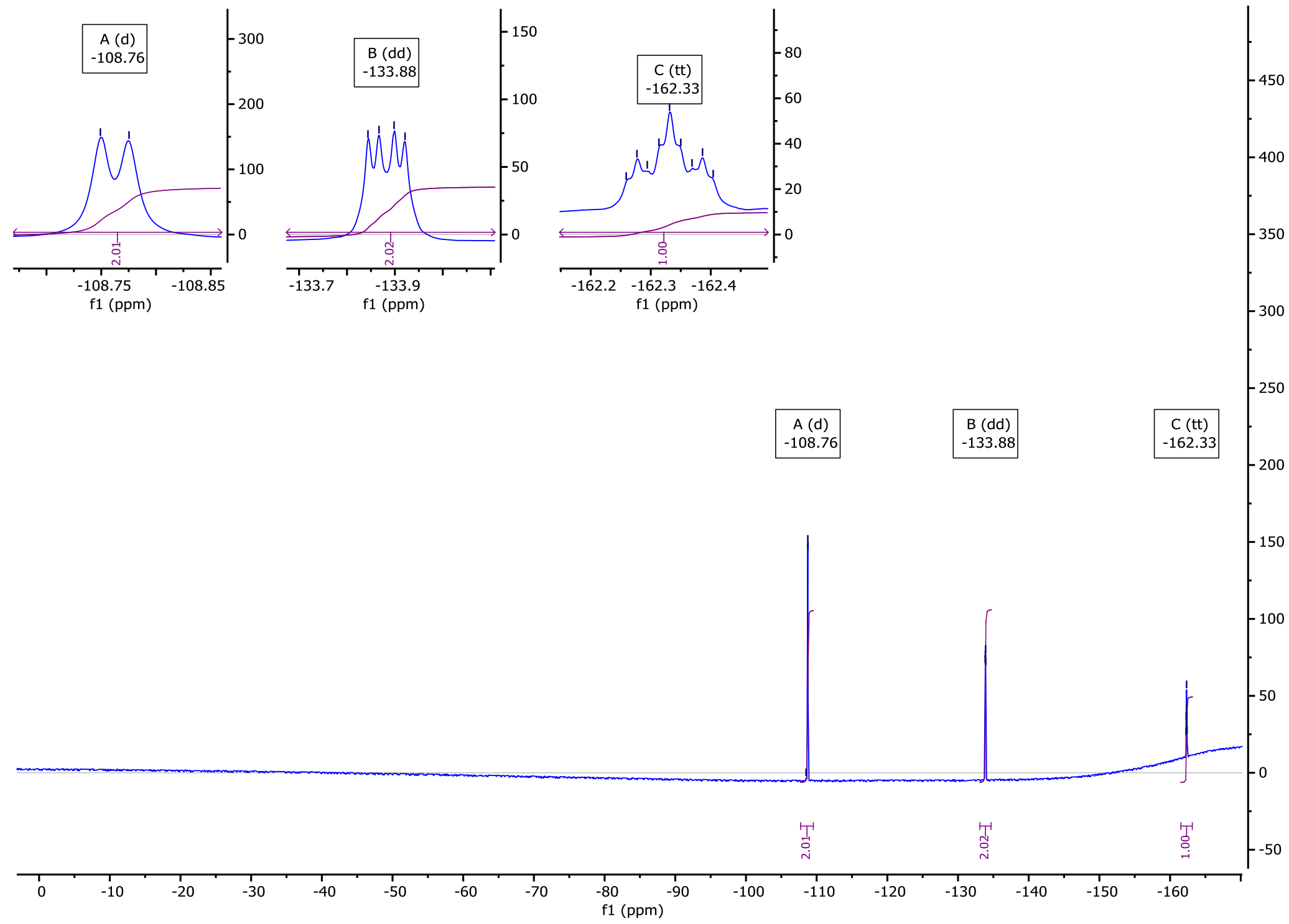

**Fig. S23.** $^{19}F$ NMR of **6** in $CDCl_3$.

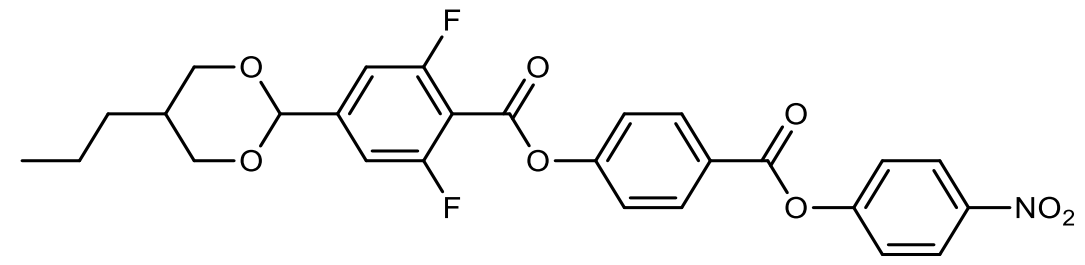

**7**

*4-((4-nitrophenoxy)carbonyl)phenyl 2,6-difluoro-4-(5-propyl-1,3-dioxan-2-yl)benzoate*

| | |
|---|---|
| Yield: | (White crystalline solid) 187 mg, 71 %. |
| $R_F$ (DCM: hexanes [1:1]): | 0.31 |
| Re-crystallisation Solvent: | MeOH |
| $^1$H NMR (501 MHz): | 8.34 (ddd, $J$ = 9.1, 3.2, 2.1 Hz, 2H, Ar-**H**), 8.29 (ddd, $J$ = 8.7, 2.5, 1.9 Hz, 2H, Ar-**H**), 7.47 – 7.41 (m, 4H, Ar-**H**)*, 7.20 (d, $J$ = 9.0 Hz, 2H, Ar-**H**), 5.41 (s, 1H, Ar-C**H**-$O_2$), 4.26 (dd, $J$ = 11.8, 4.7 Hz, 2H, 2x O-C$\mathbf{H_{eq}}H_{ax}$-C), 3.54 (t, $J$ = 11.4 Hz, 2H, 2x O-C$\mathbf{H_{ax}}H_{eq}$-C), 2.15 (ddd, $J$ = 11.3, 6.8, 4.5 Hz, 1H, $CH_2$-C**H**-$CH_2$), 1.35 (h, $J$ = 7.2 Hz, 2H, $CH_2$-C$\mathbf{H_2}$-$CH_3$), 1.11 (q, $J$ = 6.9 Hz, 2H, CH-C$\mathbf{H_2}$-$CH_2$), 0.94 (t, $J$ = 7.3 Hz, 3H, $CH_2$-C$\mathbf{H_3}$). *Overlapping Signals. |
| $^{13}$C{$^1$H} NMR (126 MHz): | 163.54, 161.10 (dd, $J$ = 258.4, 5.6 Hz), 159.26, 155.73, 155.03, 145.86 (t, $J$ = 9.8 Hz), 145.64, 132.24, 126.64, 125.47, 122.79, 122.33, 110.47 (dd, $J$ = 23.5, 3.0 Hz), 109.49 (t, $J$ = 16.8 Hz), 98.92, 72.73, 34.03, 30.36, 19.67, 14.33. |
| $^{19}$F NMR (376 MHz): | -108.41 (d, $J_{F-H}$ = 9.9 Hz, 2F, Ar-F). |

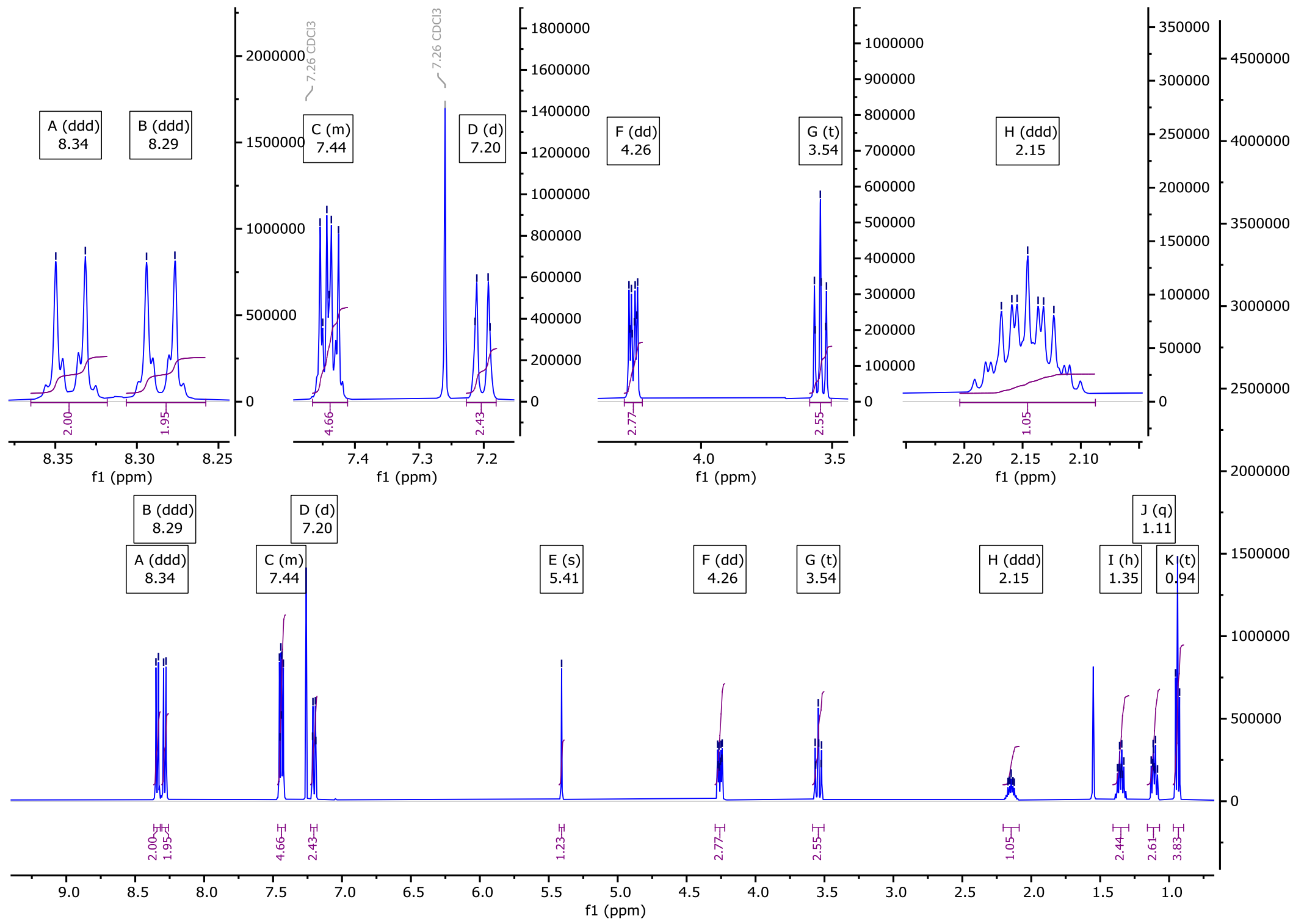

**Fig. S24.** $^{1}H$ NMR of **7** in $CDCl_3$.

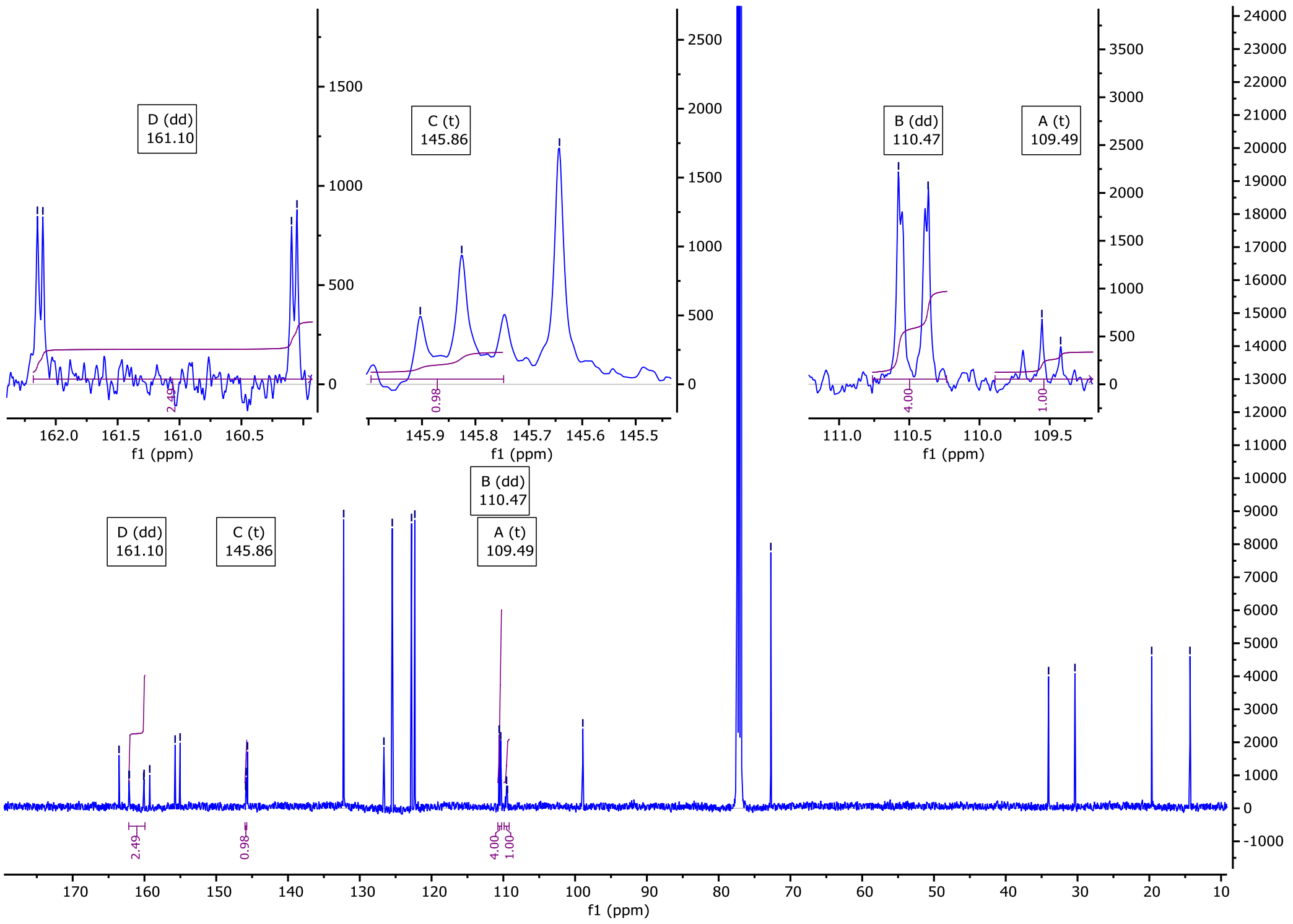

**Fig. S25.** $^{13}C\{^{1}H\}$ NMR of **7** in $CDCl_3$.

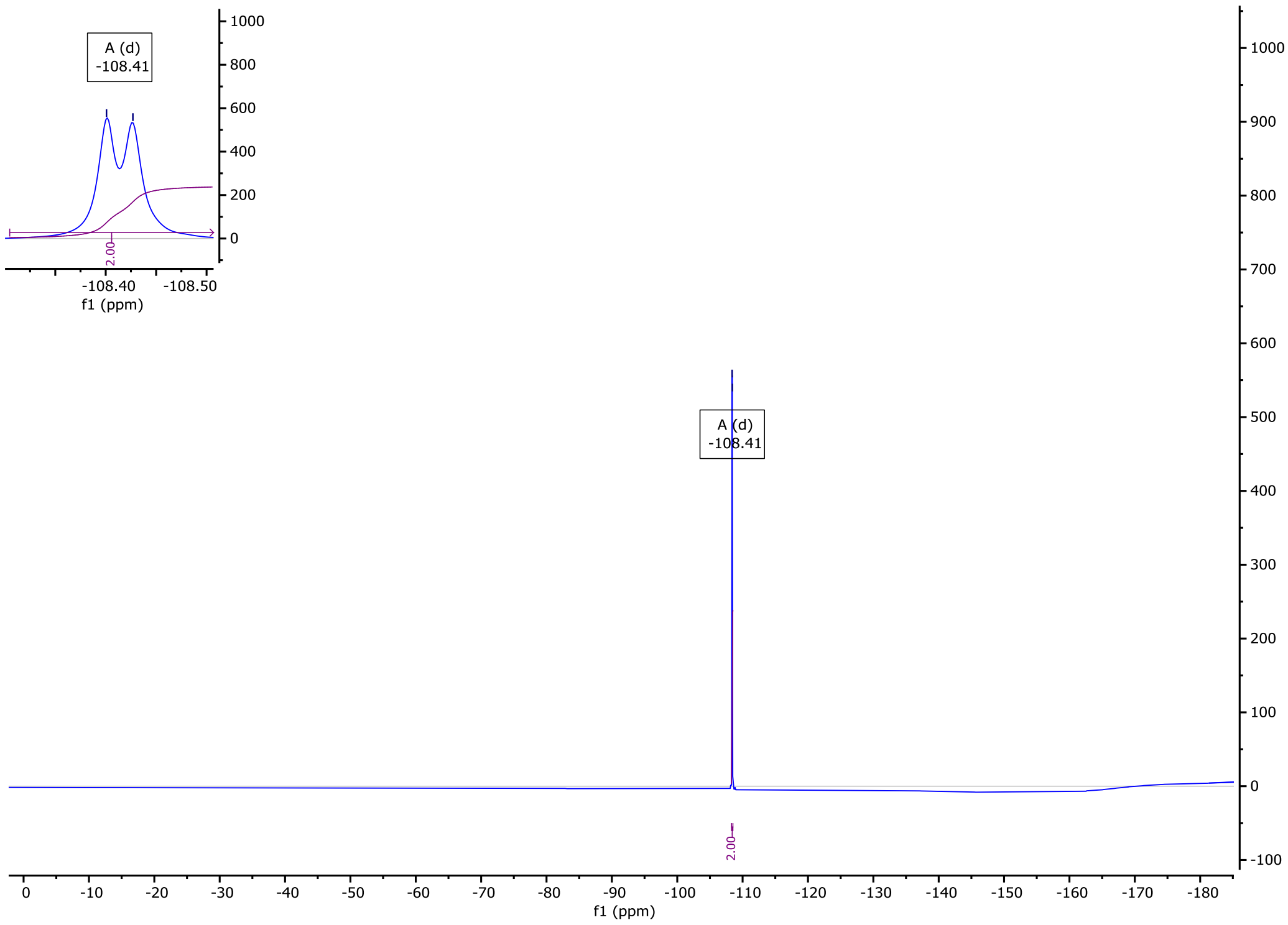

**Fig. S26.** [19]F NMR of **7** in $CDCl_3$.

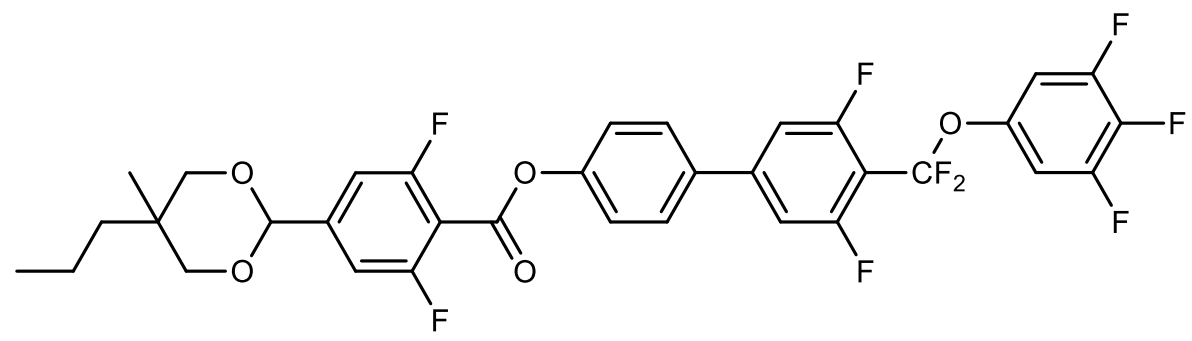

**5-Me 2**

*4'-(difluoro(3,4,5-trifluorophenoxy)methyl)-3',5'-difluoro-[1,1'-biphenyl]-4-yl 2,6-difluoro-4-(5-methyl-5-propyl-1,3-dioxan-2-yl)benzoate*

| | |
|---|---|
| Yield: | (White crystalline solid) 205 mg, 60 %. |
| $R_F$ (DCM: hexanes [1:1]): | 0.32 |
| Re-crystallisation solvent: | EtOH |
| $^1$H NMR (501 MHz): | 7.63 (ddd, *J* = 8.7, 3.0, 2.1 Hz, 2H, Ar-**H**), 7.39 (ddd, *J* = 8.7, 2.7, 2.0 Hz, 2H, Ar-**H**), 7.22 (d, *J* = 6.3 Hz, 2H, Ar-**H**), 7.20 (d, *J* = 5.3 Hz, 2H, Ar-**H**), 6.99 (dd, *J* = 7.9, 5.8 Hz, 2H, Ar-**H,** Ar-**H**), 5.38 (s, 1H, Ar-C**H**-$O_2$), 3.83 (dd, *J* = 11.3, 1.4 Hz, 2H, 2x O-C$\mathbf{H_{eq}}$$H_{ax}$-C), 3.67 (dd, *J* = 11.6, 1.3 Hz, 2H, 2x O-C$\mathbf{H_{ax}}$$H_{eq}$-C), 1.35 – 1.26 (m, 2H, Me-C-C$\mathbf{H_2}$-$CH_2$), 1.25 (s, 3H, $(CH_2)_2C(CH_2)$-**Me**), 1.14 – 1.07 (m, 2H, $CH_2$-C$\mathbf{H_2}$-$CH_3$), 0.93 (t, *J* = 7.2 Hz, 3H, $CH_2$-C$\mathbf{H_3}$). |
| $^{13}$C{$^1$H} NMR (126 MHz): | 162.08 (dd, *J* = 258.6, 5.7 Hz), 160.22 (dd, *J* = 258.2, 5.8 Hz), 159.76, 151.39, 151.18 (ddd, *J* = 250.9, 10.7, 5.2 Hz), 146.13 (t, *J* = 10.5 Hz), 145.59 (t, *J* = 9.8 Hz), 144.98 – 144.60 (m), 138.60 (dt, *J* = 250.6, 15.3 Hz), 135.72, 128.39, 123.33, 120.38, 118.27, 111.22 (dd, *J* = 23.7, 3.2 Hz), 110.43 (dd, *J* = 23.4, 3.4 Hz), 109.94 (t, *J* = 17.2 Hz), 108.57 (m), 107.61 (dd, *J* = 18.8, 5.7 Hz), 99.36 (t, *J* = 2.2 Hz), 38.81, 33.01, 20.23, 16.00, 15.10. |
| $^{19}$F NMR (376 MHz): | -61.66 (t, $J_{F\text{-}F}$ = 26.2 Hz, 2F, O-C$\mathbf{F_2}$-Ar), -108.69 (d, $J_{F\text{-}H}$ = 9.7 Hz, 2F, Ar-**F**), -110.05 (td, $J_{F\text{-}F}$ = 26.3 Hz, $J_{F\text{-}H}$ = 11.0 Hz, 2F, Ar-**F**), -132.45 (dd, $J_{F\text{-}F}$ = 20.8 Hz, $J_{F\text{-}H}$ = 8.5 Hz, 2F, Ar-**F**), -163.13 (tt, $J_{F\text{-}F}$ = 15.0 Hz, $J_{F\text{-}H}$ = 6.2 Hz, 1F, Ar-**F**). |

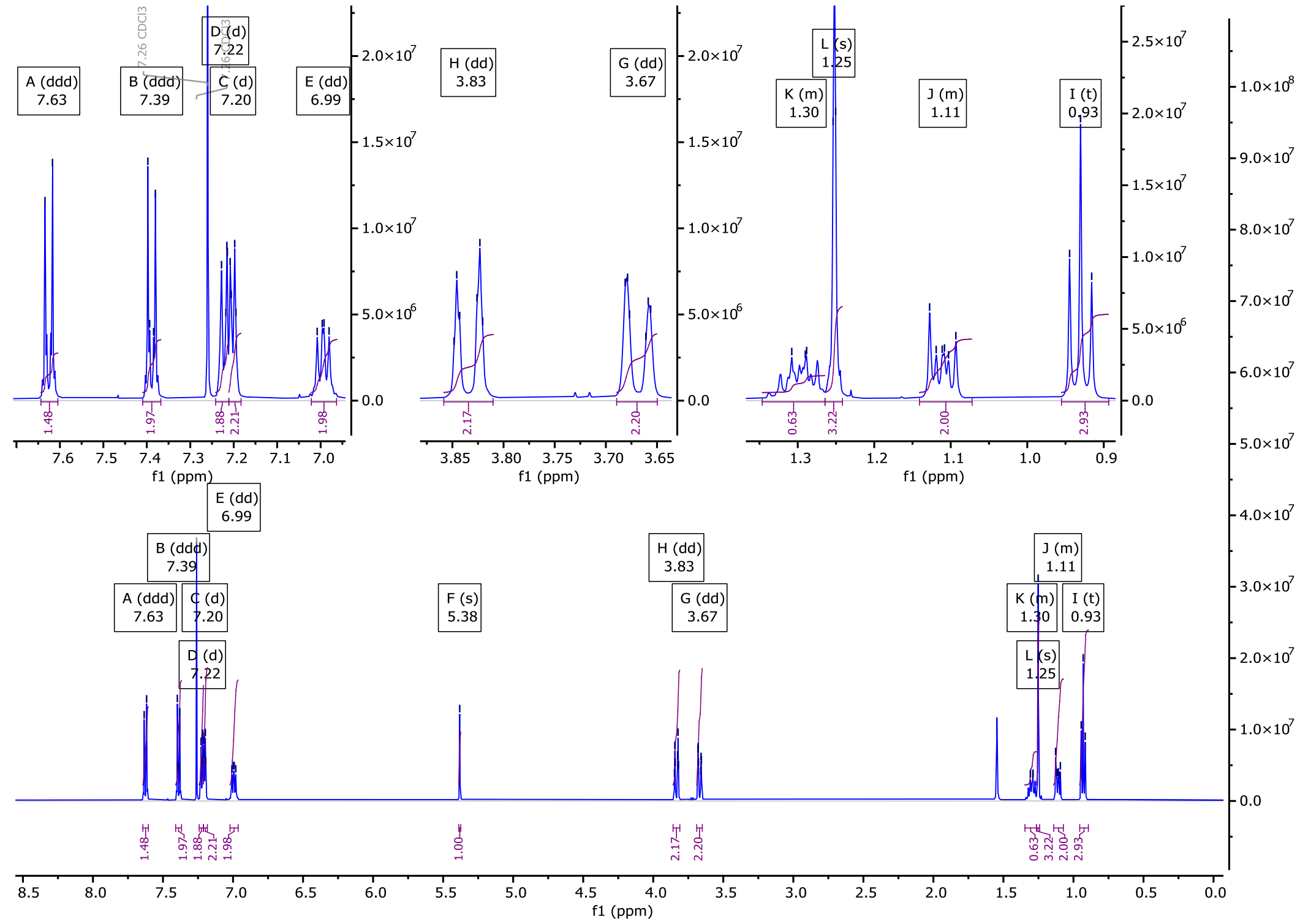

**Fig. S27.** $^{1}H$ NMR of **5-Me 2** in $CDCl_3$.

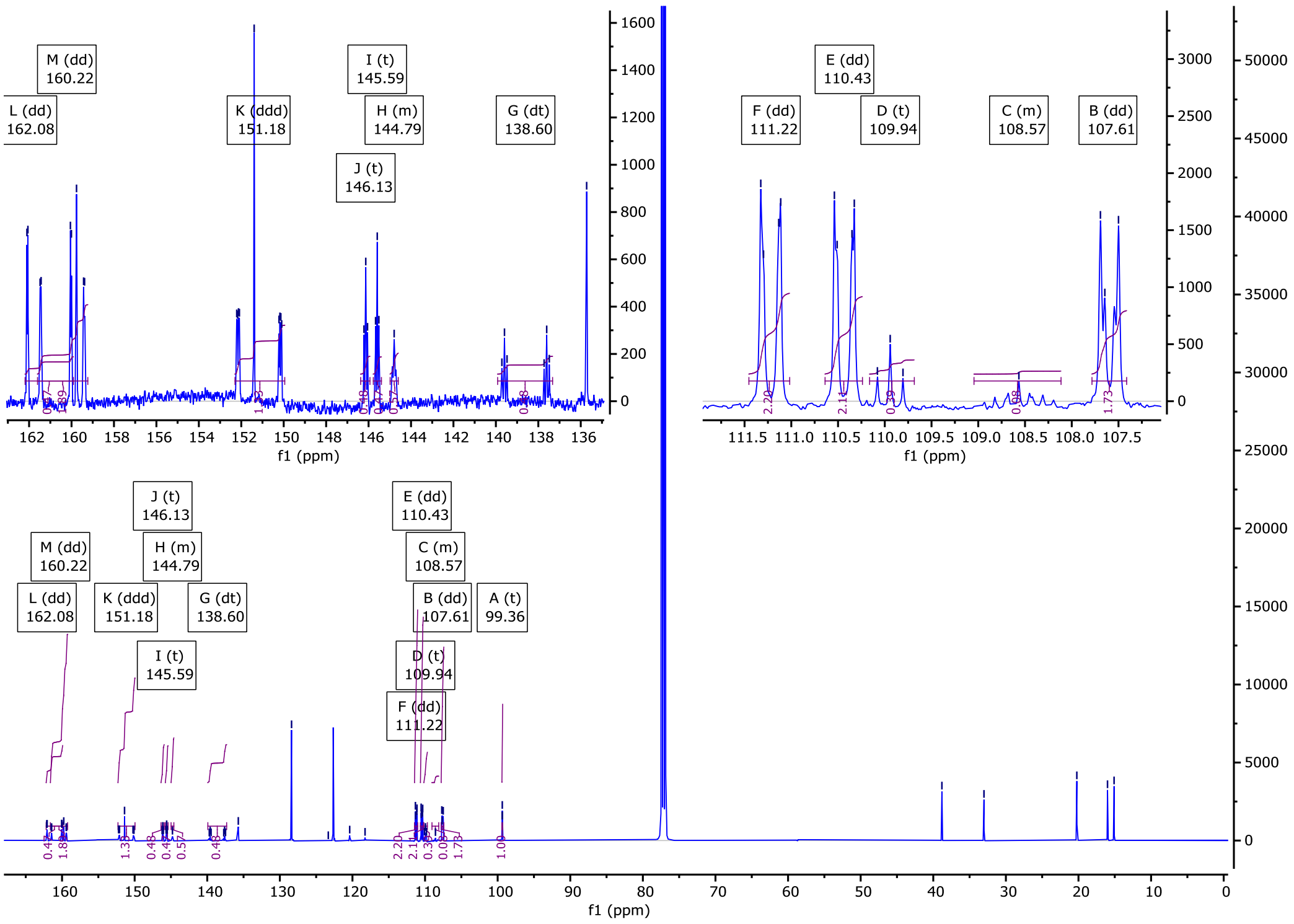

**Fig. S28.** $^{13}C\{^{1}H\}$ NMR of **5-Me 2** in $CDCl_3$.

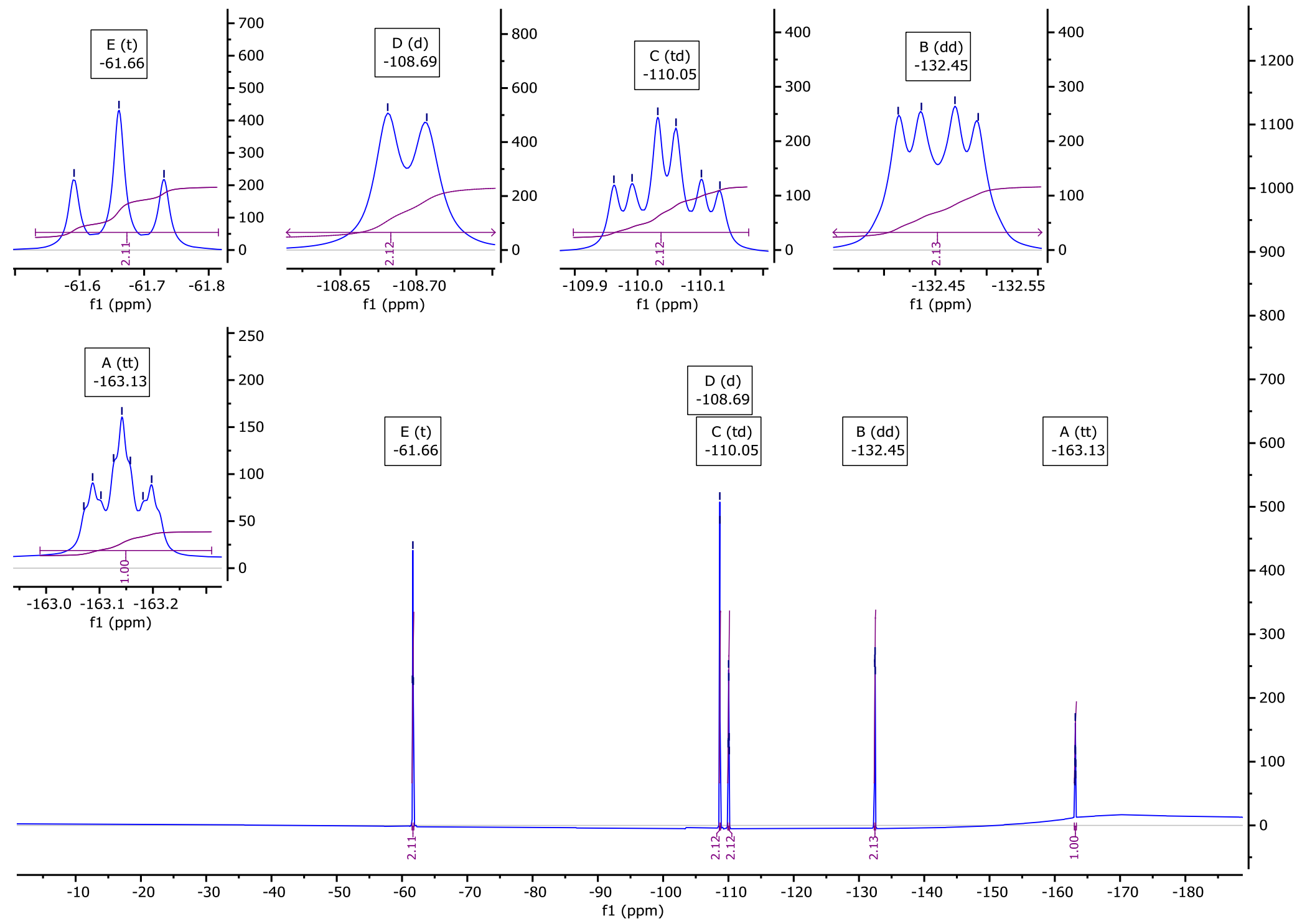

**Fig. S29.** $^{19}$F NMR of **5-Me 2** in $CDCl_3$.

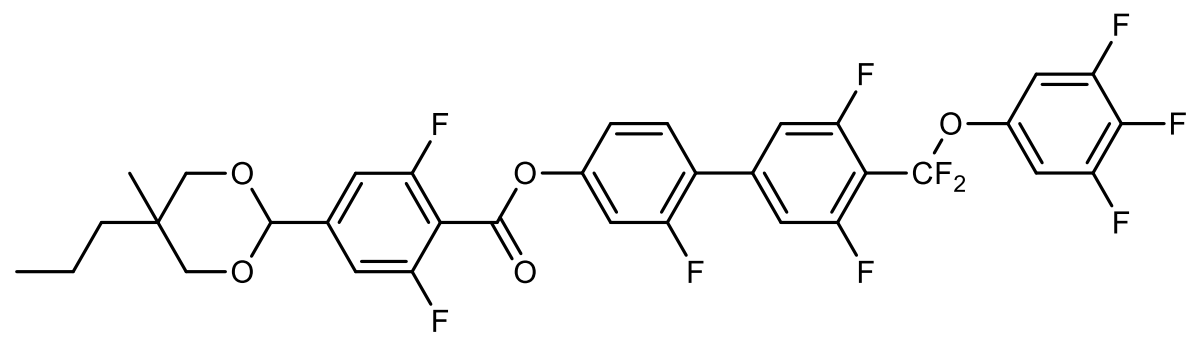

**5-Me 3**

*4'-(difluoro(3,4,5-trifluorophenoxy)methyl)-2,3',5'-trifluoro-[1,1'-biphenyl]-4-yl 2,6-difluoro-4-(5-methyl-5-propyl-1,3-dioxan-2-yl)benzoate*

| | |
|---|---|
| Yield: | (White crystalline solid) 238 mg, 68 %. |
| $R_F$ (DCM: hexanes [1:1]): | 0.41 |
| Re-crystallisation solvent: | EtOH |
| $^1$H NMR (501 MHz): | 7.49 (t, *J* = 8.5 Hz, 1H, Ar-**H**), 7.24 – 7.18 (m, 7H, Ar-**H**)*, 7.00 (dd, *J* = 7.9, 5.8 Hz, 2H, Ar-**H**), 5.38 (s, 1H, Ar-C**H**-$O_2$), 3.83 (dd, *J* = 11.2, 1.4 Hz, 2H, 2x O-C$\mathbf{H_{eq}}$$H_{ax}$-C), 3.67 (dd, *J* = 11.0, 1.3 Hz, 2H, 2x O-C$\mathbf{H_{ax}}$$H_{eq}$-C), 1.35 – 1.26 (m, 2H, Me-C-C$\mathbf{H_2}$-$CH_2$), 1.25 (s, 3H, $(CH_2)_2C(CH_2)$-**Me**), 1.14 – 1.09 (m, 2H, $CH_2$-C$\mathbf{H_2}$-$CH_3$), 0.93 (t, *J* = 7.3 Hz, 3H, $CH_2$-C$\mathbf{H_3}$).* Overlapping Signals. |
| $^{13}$C{$^1$H} NMR (126 MHz): | 161.03 (dd, *J* = 258.5, 5.8 Hz), 159.21 (dd, *J* = 257.9, 5.8 Hz), 159.09, 158.65 (d, *J* = 253.2 Hz), 151.77 (d, *J* = 11.0 Hz), 151.16 (ddd, *J* = 250.7, 10.6, 5.3 Hz), 145.91 (t, *J* = 9.9 Hz), 145.13 – 144.47 (m), 140.80 (t, *J* = 10.9 Hz), 138.62 (dt, *J* = 250.6, 15.7 Hz), 130.76 (d, *J* = 3.8 Hz), 123.78 (d, *J* = 12.7 Hz), 118.49 (d, *J* = 3.8 Hz), 113.28 (dd, *J* = 24.0, 3.5 Hz), 110.98 (d, *J* = 26.0 Hz), 110.53 (dd, *J* = 23.5, 3.4 Hz), 109.50 (t, *J* = 17.6 Hz), 109.26 (t, *J* = 14.6 Hz), 107.62 (dd, *J* = 23.0, 5.2 Hz), 99.30 (d, *J* = 1.6 Hz), 38.80, 33.01, 20.22, 16.00, 15.09. |
| $^{19}$F NMR (376 MHz): | -61.78 (t, $J_{F-F}$ = 26.3 Hz, 2F, O-C$\mathbf{F_2}$-Ar), -108.42 (d, $J_{F-H}$ = 9.8 Hz, 2F, Ar-**F**), -110.33 (td, $J_{F-F}$ = 26.3 Hz, $J_{F-H}$ = 10.9 Hz, 2F, Ar-**F**), -113.60 (t, $J_{F-H}$ = 9.6 Hz, 1F, Ar-**F**), -132.43 (dd, $J_{F-F}$ = 20.7 Hz, $J_{F-H}$ = 8.6 Hz, 2F, Ar-**F**), -163.11 (tt, $J_{F-F}$ = 20.9 Hz, $J_{F-H}$ = 5.8 Hz, 1F, Ar-**F**). |

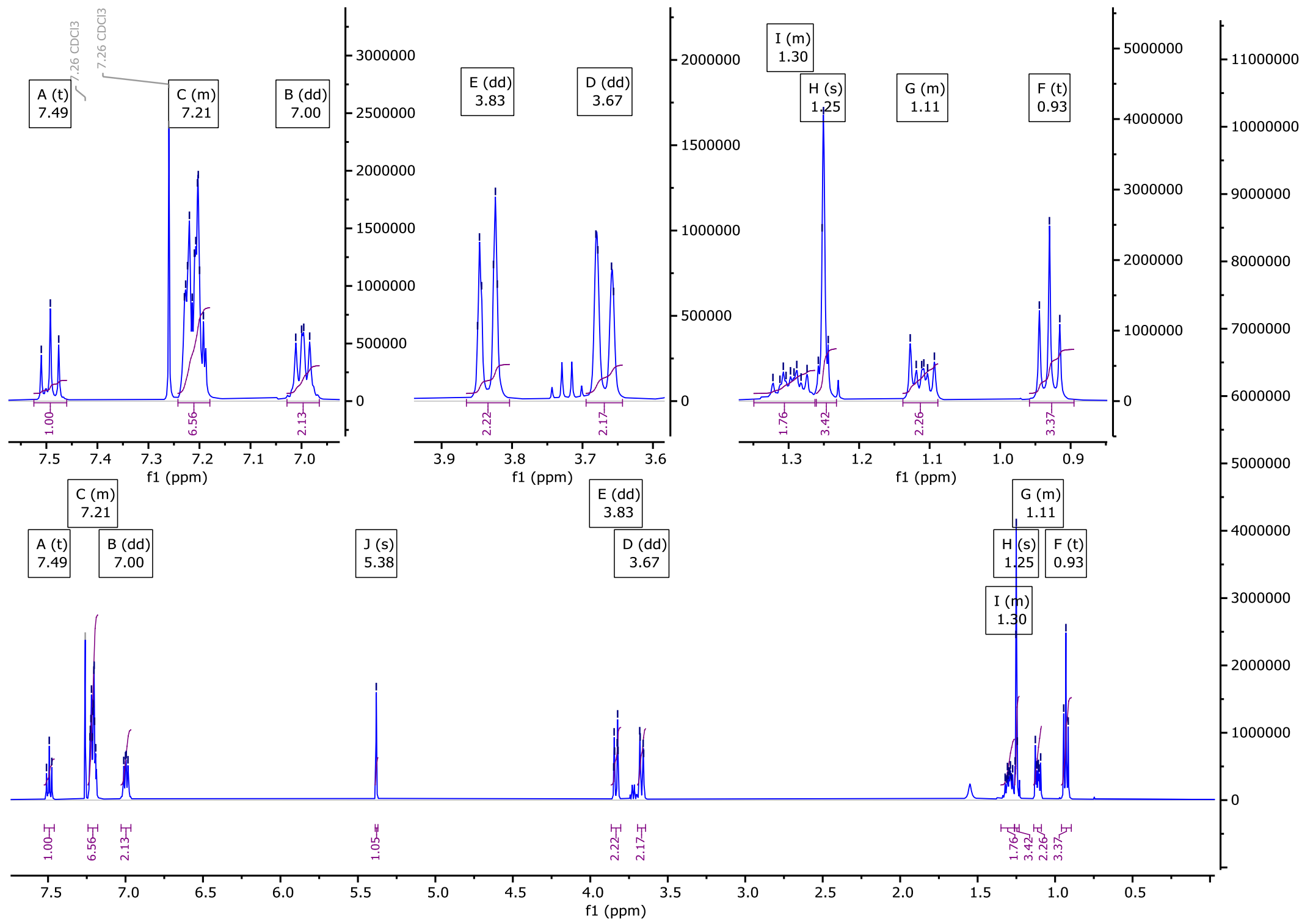

**Fig. S30.** $^{1}H$ NMR of **5-Me 3** in $CDCl_3$.

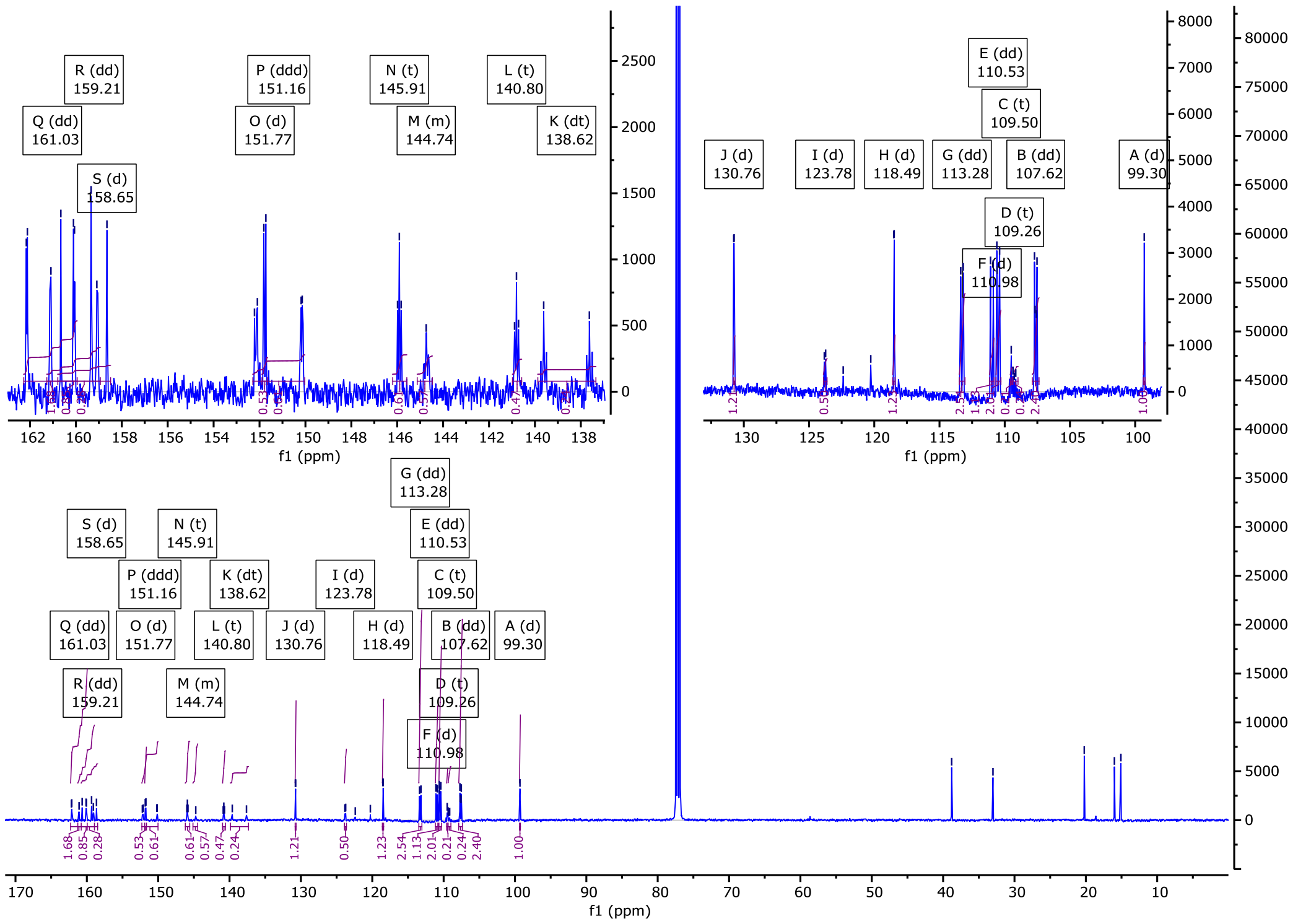

**Fig. S31.** $^{13}C\{^{1}H\}$ NMR of **5-Me 3** in $CDCl_3$.

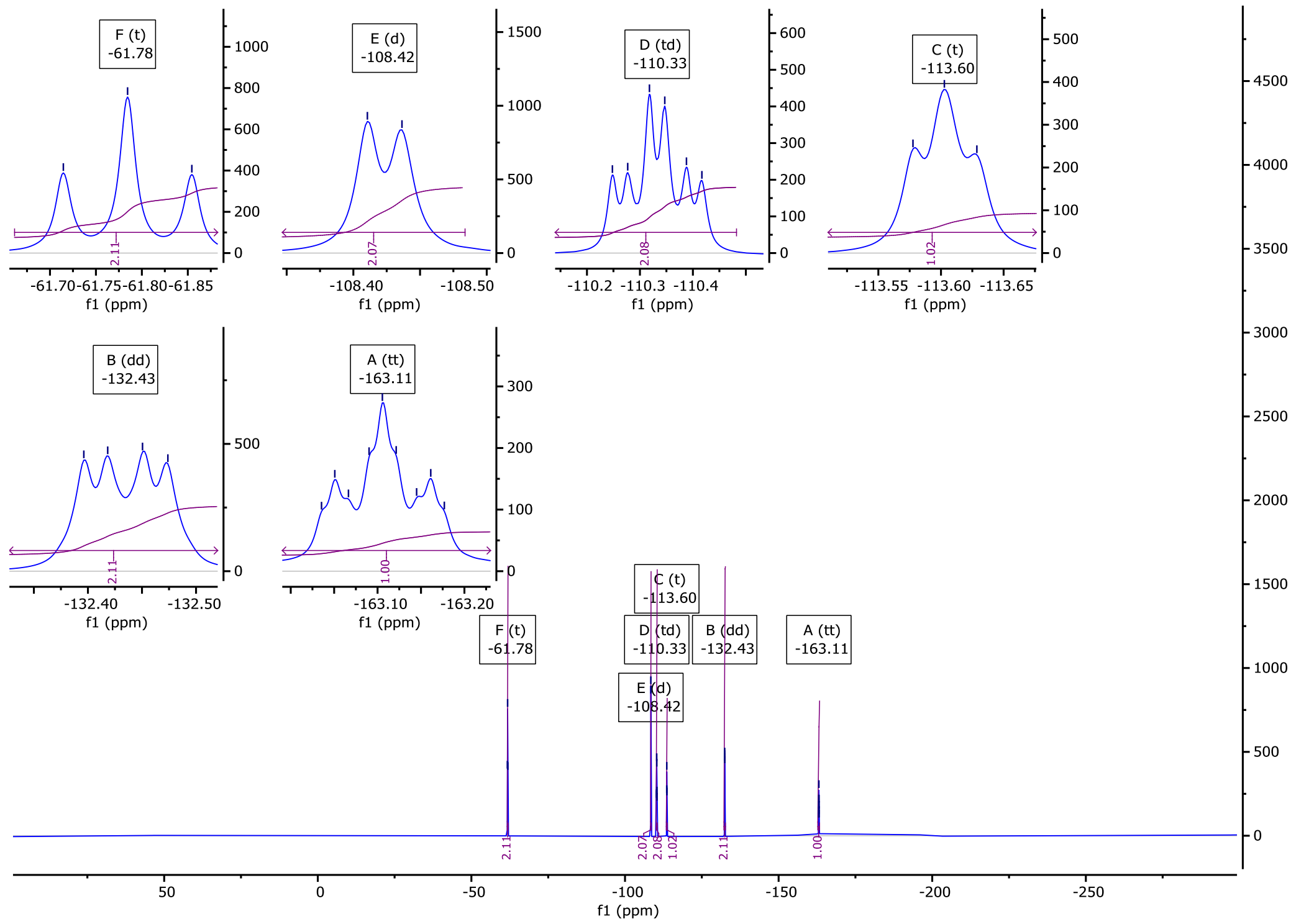

**Fig. S32.** $^{19}F$ NMR of **5-Me 3** in $CDCl_3$.

**5-Me 4**

*4'-(difluoro(3,4,5-trifluorophenoxy)methyl)-2,3',5',6-tetrafluoro-[1,1'-biphenyl]-4-yl 2,6-difluoro-4-(5-methyl-5-propyl-1,3-dioxan-2-yl)benzoate*

Yield: (White crystalline solid) 232 mg, 65 %.

$R_F$ (DCM: hexanes [1:1]): 0.53

Re-crystallisation solvent: EtOH

$^1$H NMR (501 MHz): 7.22 (d, *J* = 9.1 Hz, 2H, Ar-**H**), 7.16 (d, *J* = 10.4 Hz, 2H, Ar-**H**), 7.05 (ddd, *J* = 8.1, 3.1, 3.1 Hz, 2H, Ar-**H**), 7.00 (dd, *J* = 7.8, 5.8 Hz, 2H, Ar-**H**), 5.38 (s, 1H, Ar-C**H**-$O_2$), 3.83 (dd, *J* = 10.1, 1.2 Hz, 2H, 2x O-C$\mathbf{H_{eq}}H_{ax}$-C), 3.67 (dd, *J* = 12.3, 1.3 Hz, 2H, 2x O-C$\mathbf{H_{ax}}H_{eq}$-C), 1.36 – 1.26 (m, 2H, Me-C-C$\mathbf{H_2}$-$CH_2$), 1.25 (s, 3H, $(CH_2)_2C(CH_2)$-**Me**), 1.14 – 1.08 (m, 2H, $CH_2$-C$\mathbf{H_2}$-$CH_3$), 0.93 (t, *J* = 7.2 Hz, 3H, $CH_2$-C$\mathbf{H_3}$).

$^{13}C\{^1H\}$ NMR (126 MHz): 162.19 (dd, *J* = 259.1, 5.4 Hz), 160.88 (dd, *J* = 252.2, 8.2 Hz), 159.50 (d, *J* = 251.5 Hz), 152.16 (ddd, *J* = 250.7, 10.7, 5.2 Hz), 151.44 (t, *J* = 14.3 Hz), 146.22 (t, *J* = 9.9 Hz), 144.86 – 144.50 (m), 138.65 (dt, *J* = 250.6, 15.4 Hz), 134.41 (t, *J* = 11.5 Hz), 122.28, 120.16, 118.04, 114.90 (dd, *J* = 25.1, 2.1 Hz), 113.49 (t, *J* = 17.9 Hz), 110.56 (dd, *J* = 23.4, 3.5 Hz), 109.87 (m), 109.08 (t, *J* = 16.7 Hz), 107.68 (dd, *J* = 18.4, 6.2 Hz), 106.85 (dd, *J* = 22.8, 7.0 Hz), 99.25 (t, *J* = 2.2 Hz), 38.80, 33.01, 20.21, 16.00, 15.08.

$^{19}$F NMR (376 MHz): -61.95 (t, $J_{F-F}$ = 26.5 Hz, 2F, O-C$\mathbf{F_2}$-Ar), -108.17 (d, $J_{F-H}$ = 10.0 Hz, 2F, Ar-**F**), -110.52 (td, $J_{F-F}$ = 26.4 Hz, $J_{F-H}$ = 10.7 Hz, 2F, Ar-**F**), -111.71 (d, $J_{F-H}$ = 8.9 Hz, 2F, Ar-**F**), -132.43 (dd, $J_{F-F}$ = 20.9 Hz, $J_{F-H}$ = 8.6 Hz, 2F, Ar-**F**), -163.07 (tt, $J_{F-F}$ = 20.9 Hz, $J_{F-H}$ = 6.0 Hz, 1F, Ar-**F**).

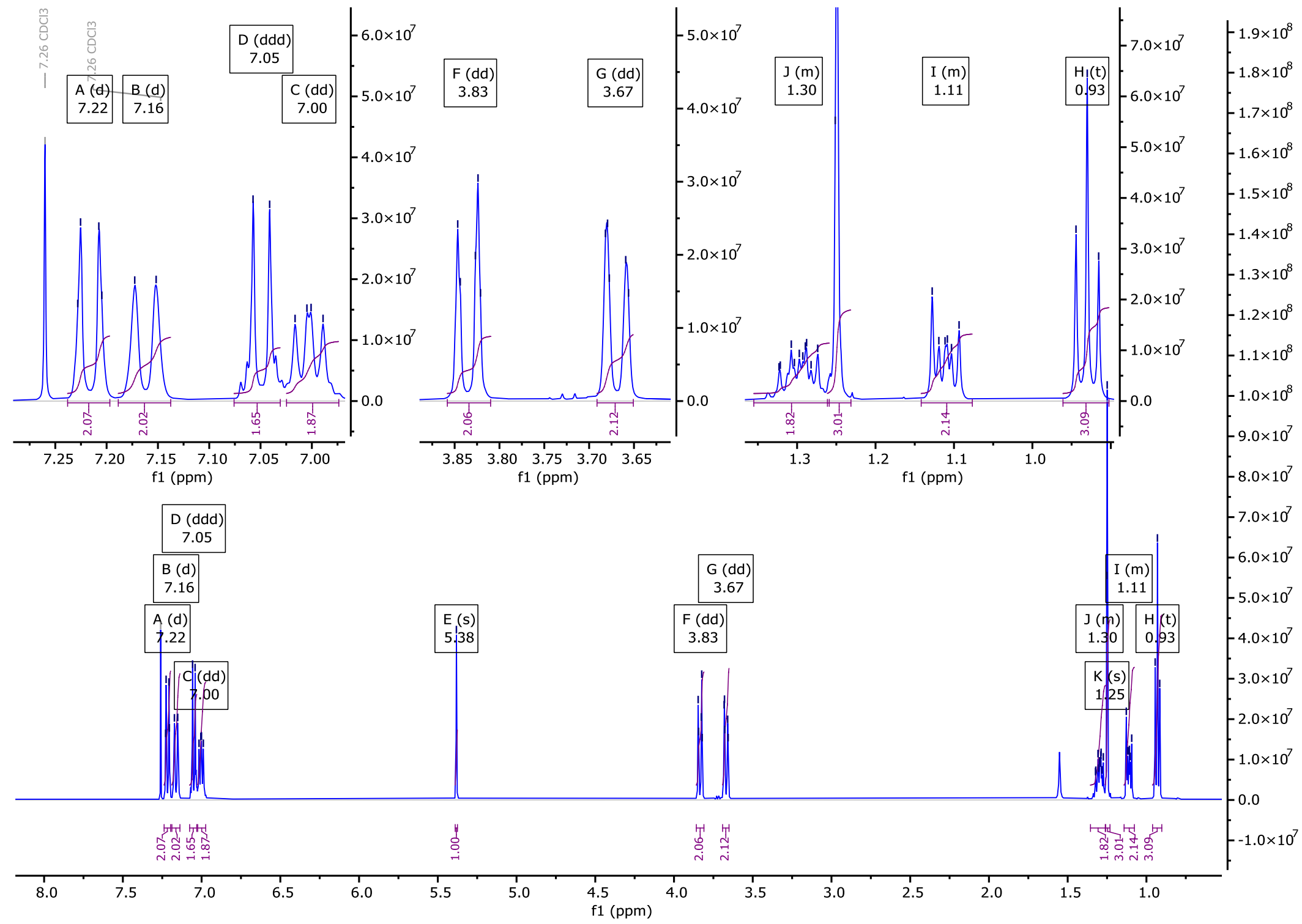

**Fig. S33.** $^{1}H$ NMR of **5-Me 4** in $CDCl_3$.

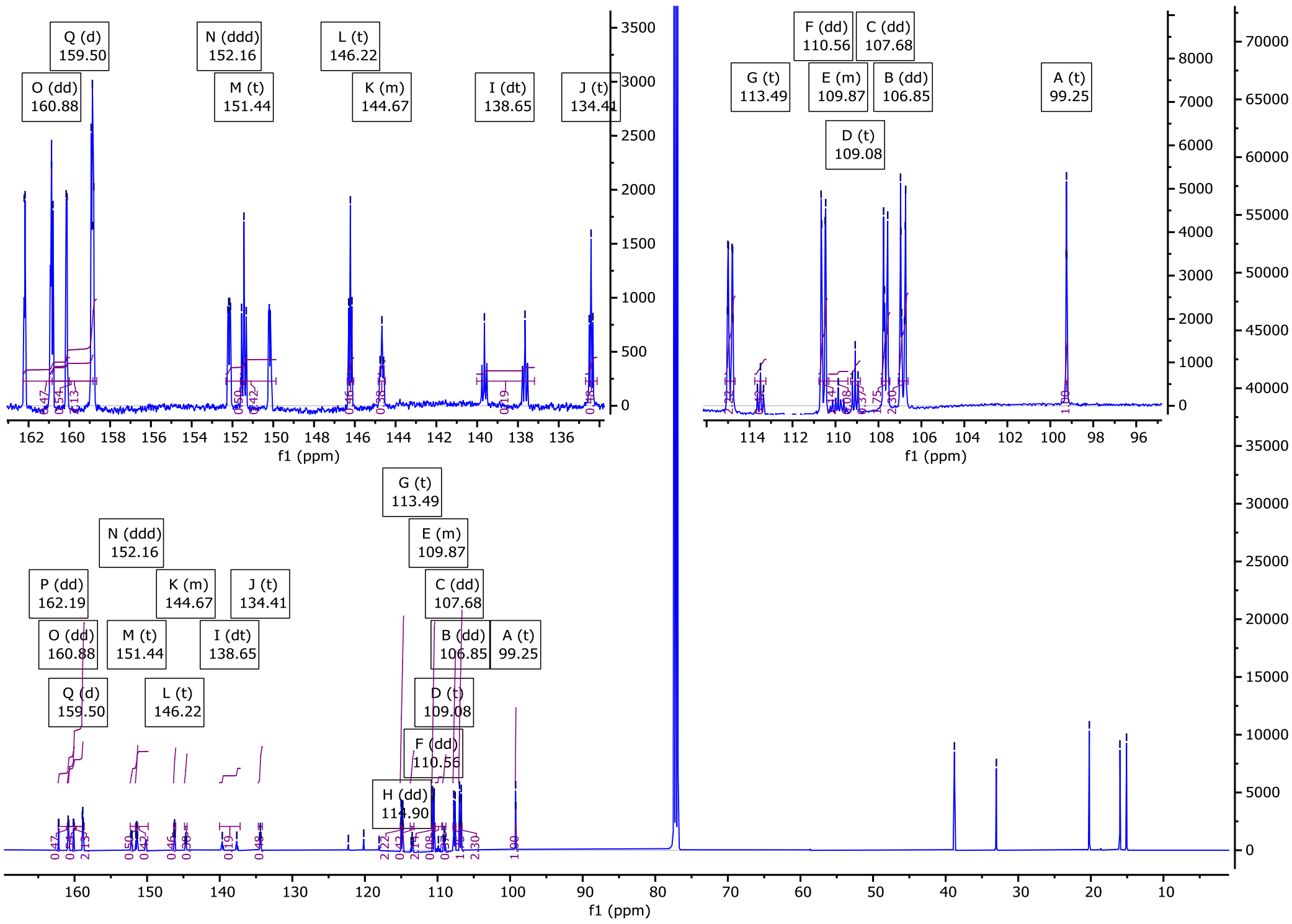

**Fig. S34.** $^{13}C\{^{1}H\}$ NMR of **5-Me 4** in $CDCl_3$.

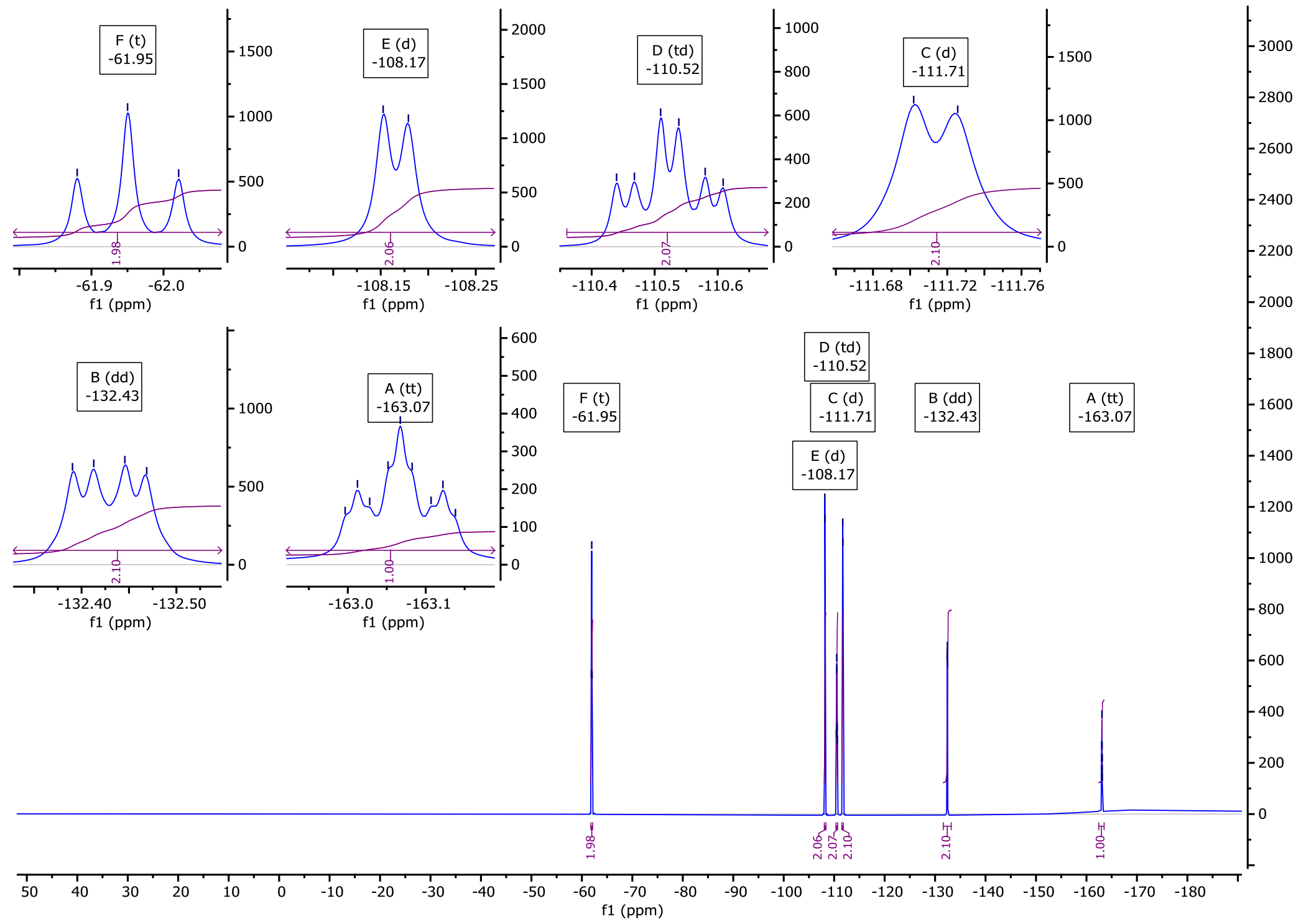

**Fig. S35.** $^{19}$F NMR of **5-Me 4** in $CDCl_3$.

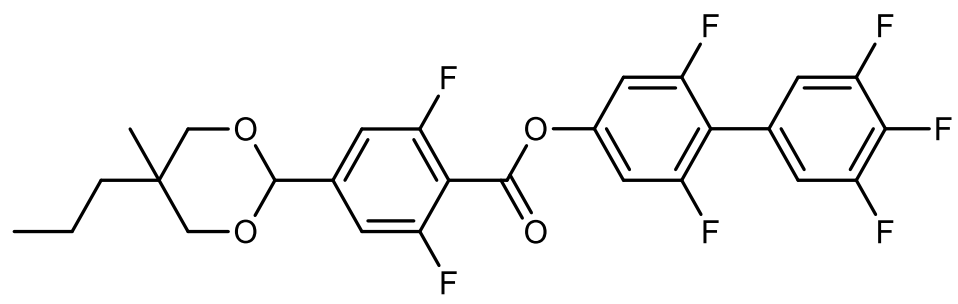

**5-Me 5**

*2,3',4',5',6-pentafluoro-[1,1'-biphenyl]-4-yl 2,6-difluoro-4-(5-methyl-5-propyl-1,3-dioxan-2-yl)benzoate*

| | |
|---|---|
| Yield: | (White crystalline solid) 209 mg, 77 %. |
| $R_F$ (DCM: hexanes [1:1]): | 0.40 |
| Re-crystallisation Solvent: | MeOH |
| $^1$H NMR (501 MHz): | 7.21 (ddd, *J* = 10.0, 1.2, 1.2 Hz, 2H, Ar-**H**), 7.16 – 7.09 (m, 2H, Ar-**H**), 7.01 (ddd, *J* = 7.9, 3.6, 3.5 Hz, 2H, Ar-**H**), 5.38 (s, 1H, Ar-C**H**-$O_2$), 3.83 (dd, *J* = 10.1, 1.4 Hz, 2H, 2x O-C$\mathbf{H_{eq}}$$H_{ax}$-C), 3.67 (dd, *J* = 11.4, 1.8 Hz, 2H, 2x O-C$\mathbf{H_{ax}}$$H_{eq}$-C), 1.35 – 1.25 (m, 2H, Me-C-C$\mathbf{H_2}$-$CH_2$), 1.25 (s, 3H, $(CH_2)_2C(CH_2)$-**Me**), 1.14 – 1.07 (m, 2H, $CH_2$-C$\mathbf{H_2}$-$CH_3$), 0.93 (t, *J* = 7.2 Hz, 3H, $CH_2$-C$\mathbf{H_3}$). |
| $^{13}$C{$^1$H} NMR (126 MHz): | 162.17 (dd, *J* = 258.4, 5.6 Hz), 159.81 (dd, *J* = 250.5, 8.8 Hz), 158.83, 150.89 (ddd, *J* = 249.8, 9.8, 4.5 Hz), 150.84 (t, *J* = 14.2 Hz), 146.12 (t, *J* = 9.9 Hz), 139.95 (dt, *J* = 254.4, 14.4 Hz), 124.50 – 124.16 (m), 114.96 (dd, *J* = 18.0, 5.2 Hz), 114.05 (t, *J* = 18.3 Hz), 110.54 (dd, *J* = 23.5, 3.6 Hz), 109.18 (t, *J* = 16.6 Hz), 106.70 (dd, *J* = 23.1, 7.2 Hz), 99.26 (t, *J* = 1.9 Hz) 38.80, 33.01, 20.22, 16.00, 15.09. |
| $^{19}$F NMR (376 MHz): | -108.24 (d, $J_{F-H}$ = 9.9 Hz, 2F, Ar-**F**), -112.16 (d, $J_{F-H}$ = 8.6 Hz, 2F, Ar-**F**), -134.30 (dd, $J_{F-F}$ = 20.6 Hz, $J_{F-H}$ = 8.3 Hz, 2F, Ar-**F**), -159.91 (tt, $J_{F-F}$ = 20.3 Hz, $J_{F-H}$ = 6.8 Hz, 1F, Ar-**F**). |

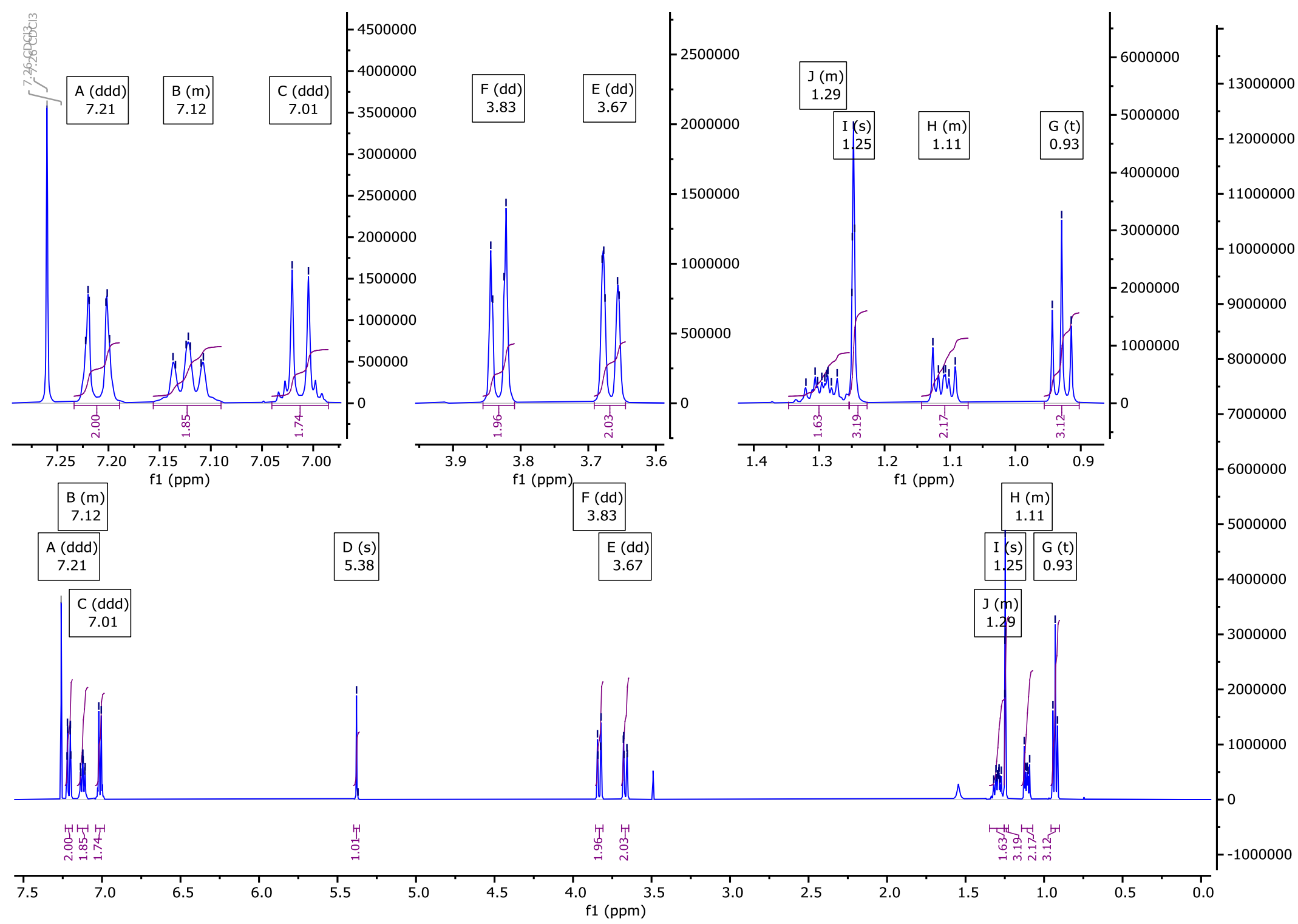

**Fig. S36.** $^{1}H$ NMR of **5-Me 5** in $CDCl_3$.

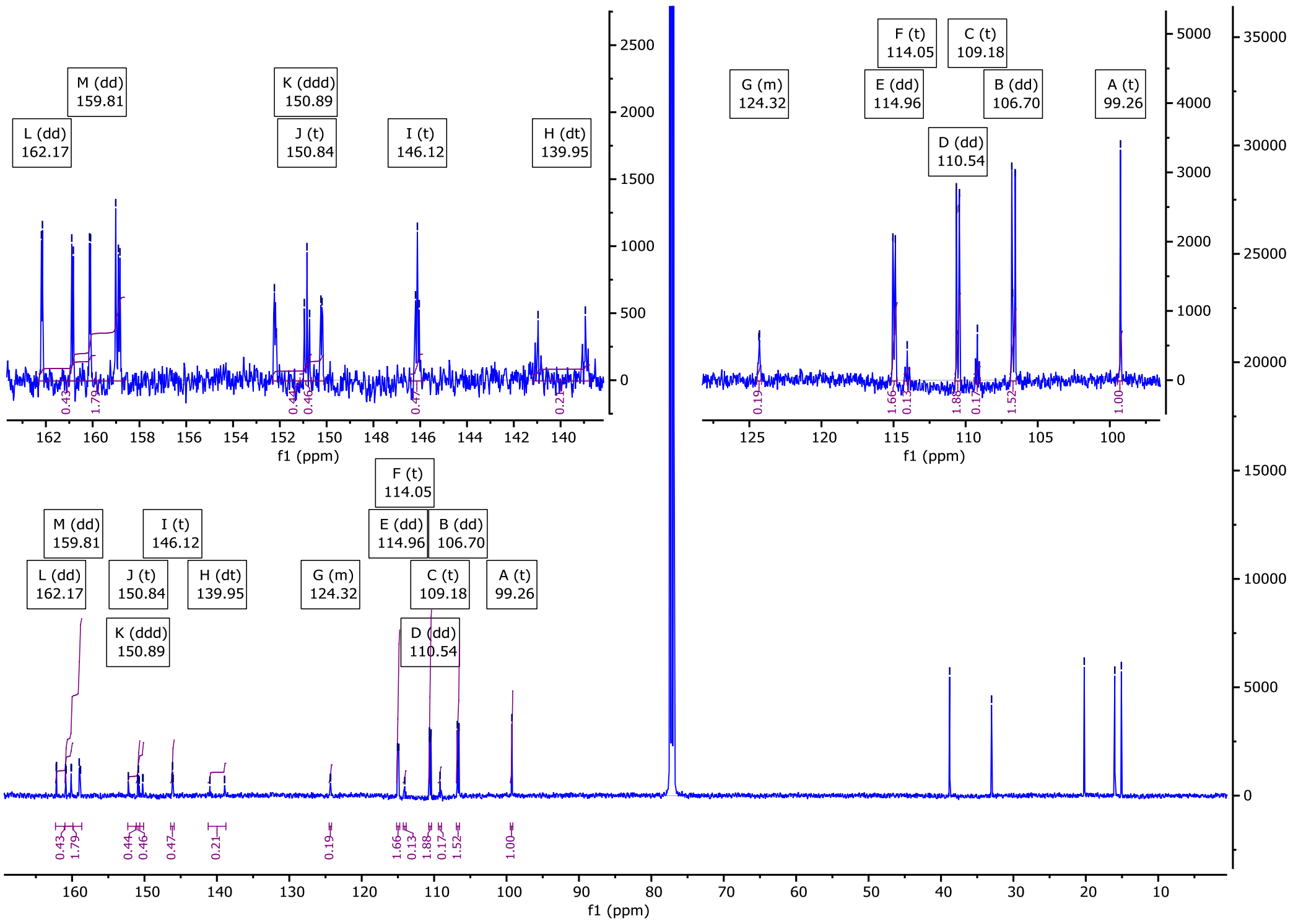

**Fig. S37.** $^{13}C\{^{1}H\}$ NMR of **5-Me 5** in $CDCl_3$.

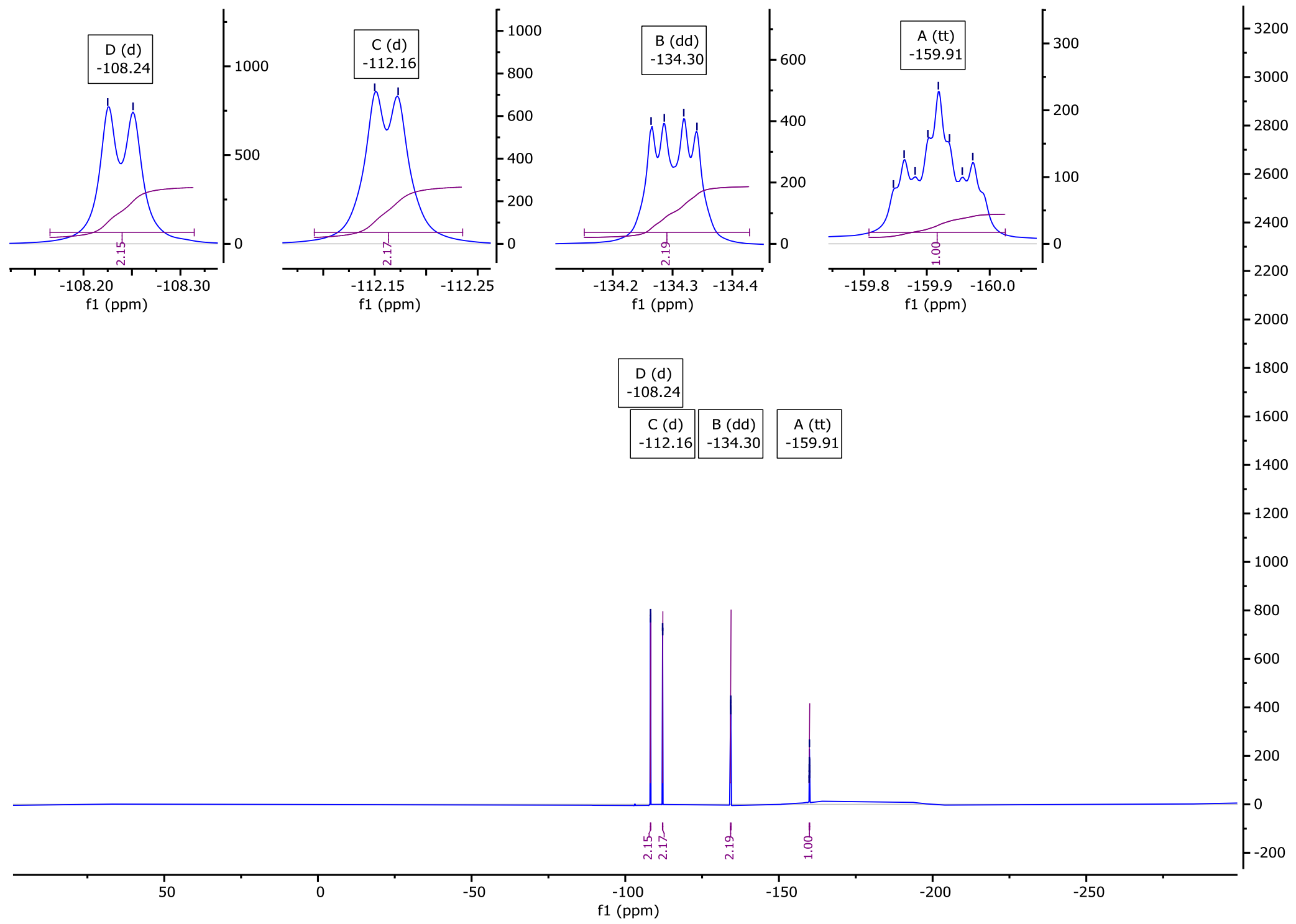

**Fig. S38.** $^{19}$F NMR of **5-Me 5** in $CDCl_3$.

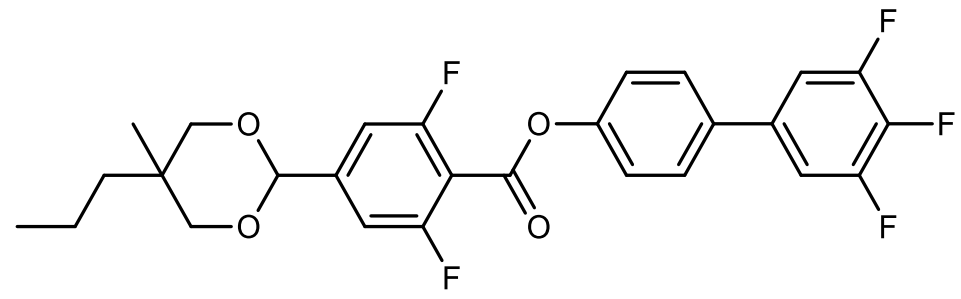

**5-Me 6**

*3',4',5'-trifluoro-[1,1'-biphenyl]-4-yl 2,6-difluoro-4-(5-methyl-5-propyl-1,3-dioxan-2-yl)benzoate*

| | |
|---|---|
| Yield: | (White crystalline solid) 187 mg, 74 %. |
| $R_F$ (DCM: hexanes [1:1]): | 0.30 |
| Re-crystallisation Solvent: | MeOH |
| $^1$H NMR (501 MHz): | 7.55 (ddd, *J* = 8.7, 2.9, 2.1 Hz, 2H, Ar-**H**), 7.35 (ddd, *J* = 8.7, 2.7, 2.1 Hz, 2H, Ar-**H**), 7.22 – 7.16 (m, 4H, Ar-**H**)*, 5.38 (s, 1H, Ar-C**H**-$O_2$), 3.83 (dd, *J* = 10.0, 1.3 Hz, 2H, 2x O-C$\mathbf{H_{eq}}$$H_{ax}$-C), 3.67 (dd, *J* = 11.3, 1.2 Hz, 2H, 2x O-C$\mathbf{H_{ax}}$$H_{eq}$-C), 1.35 – 1.26 (m, 2H, Me-C-C$\mathbf{H_2}$-$CH_2$), 1.25 (s, 3H, $(CH_2)_2C(CH_2)$-**Me**), 1.14 – 1.08 (m, 2H, $CH_2$-C$\mathbf{H_2}$-$CH_3$), 0.93 (t, *J* = 7.2 Hz, 3H, $CH_2$-C$\mathbf{H_3}$). *Overlapping Signals. |
| $^{13}$C{$^1$H} NMR (126 MHz): | 161.02 (dd, *J* = 257.9, 5.8 Hz), 159.85, 150.61 (ddd, *J* = 249.6, 9.9, 4.2 Hz), 150.54, 145.49 (t, *J* = 9.8 Hz), 139.51 (dt, *J* = 252.2, 15.3 Hz), 136.52, 128.22, 122.44, 111.28 (dd, *J* = 16.7, 4.9 Hz), 110.40 (dd, *J* = 23.5, 3.4 Hz), 110.05 (t, *J* = 17.8 Hz), 99.36 (t, *J* = 2.5 Hz), 38.81, 33.00, 20.22, 15.99, 15.09. |
| $^{19}$F NMR (376 MHz) | -108.78 (d, $J_{F\text{-}H}$ = 9.9 Hz, 2F, Ar-**F**), -133.88 (dd, $J_{F\text{-}F}$ = 20.5 Hz, $J_{F\text{-}H}$ =8.7 Hz, 2F, Ar-**F**), -162.34 (tt, $J_{F\text{-}F}$ = 20.5 Hz, $J_{F\text{-}H}$ = 6.5 Hz, 1F, Ar-**F**). |

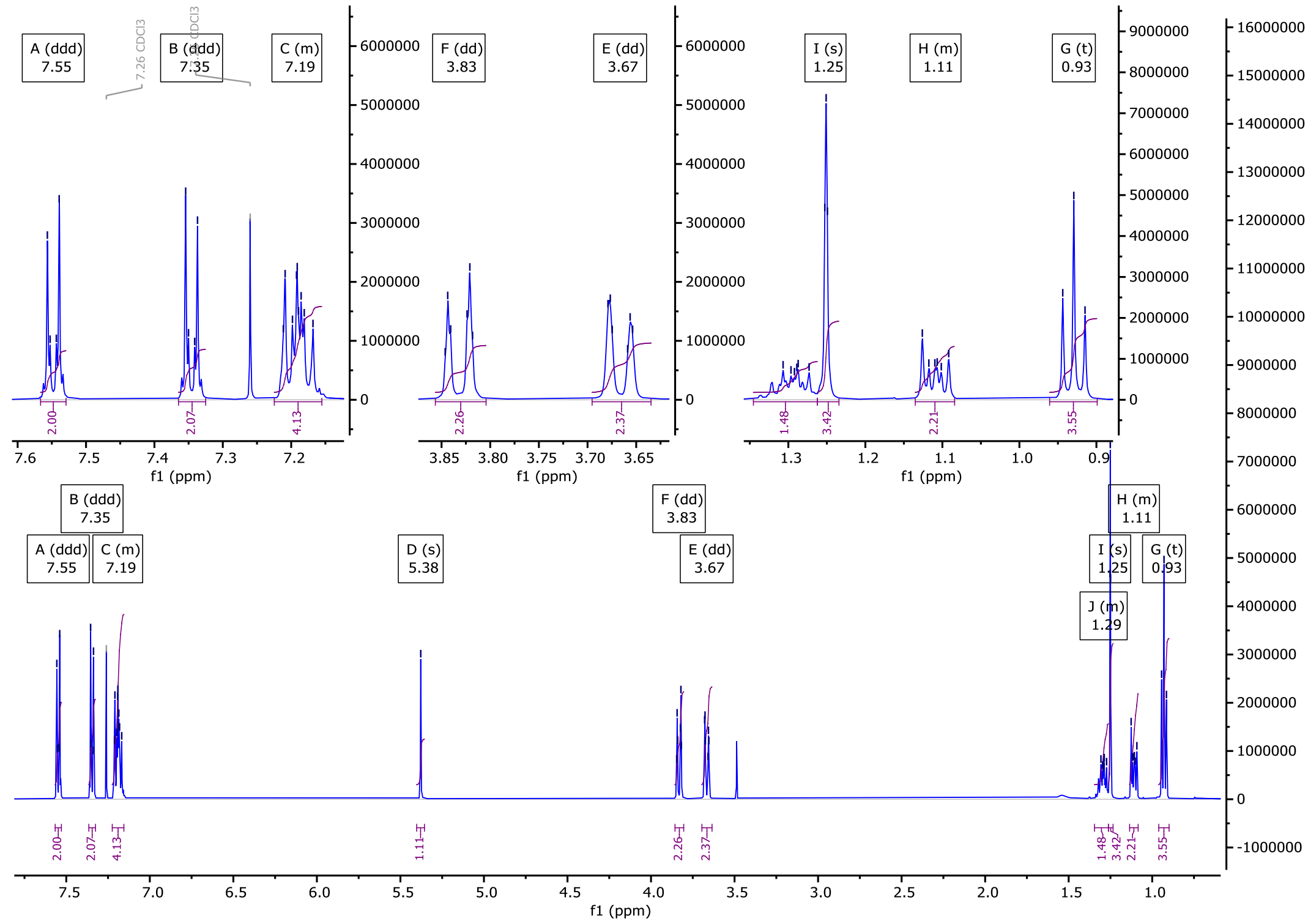

**Fig. S39.** $^1H$ NMR of **5-Me 6** in $CDCl_3$.

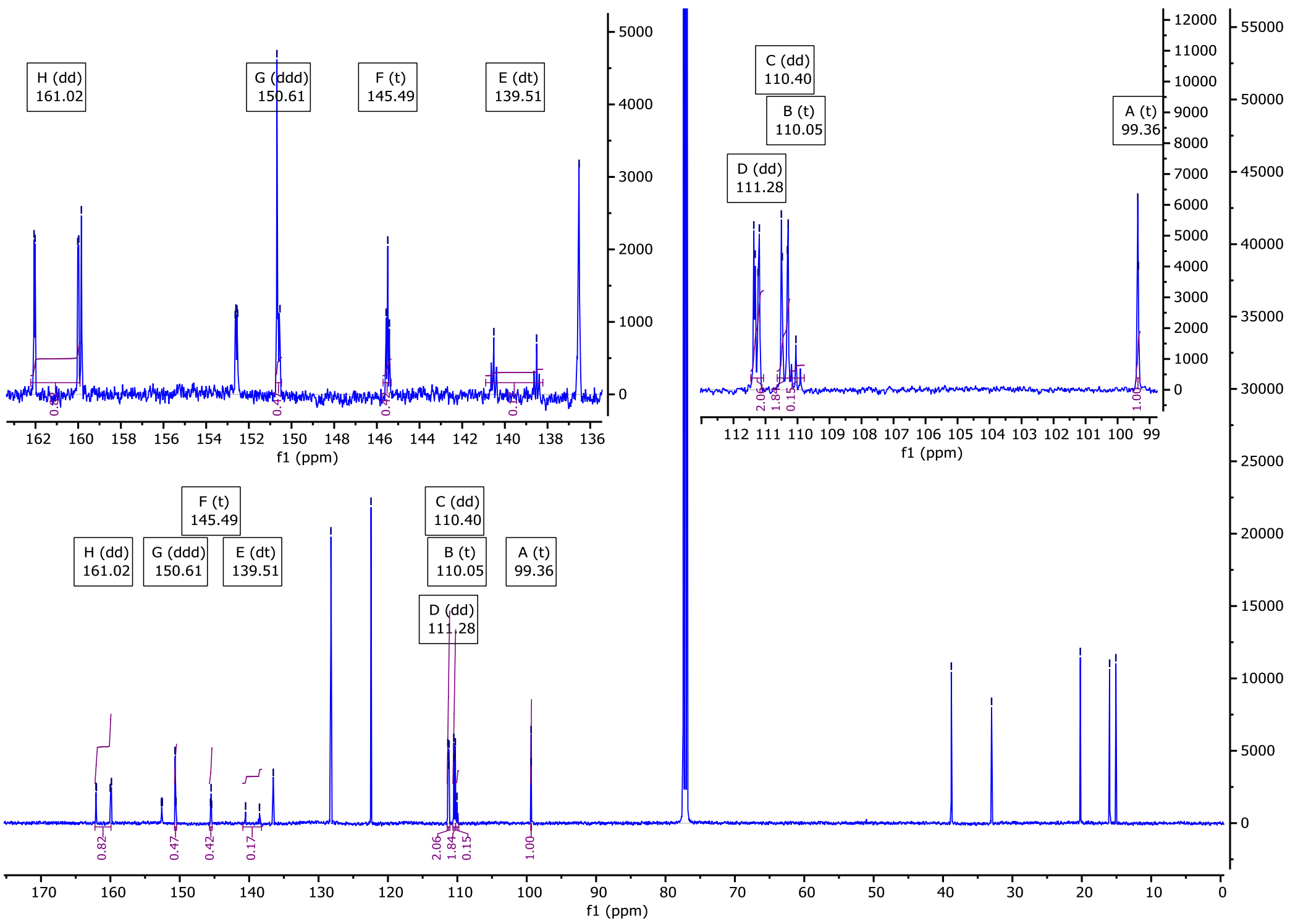

**Fig. S40.** $^{13}C\{^1H\}$ NMR of **5-Me 6** in $CDCl_3$.

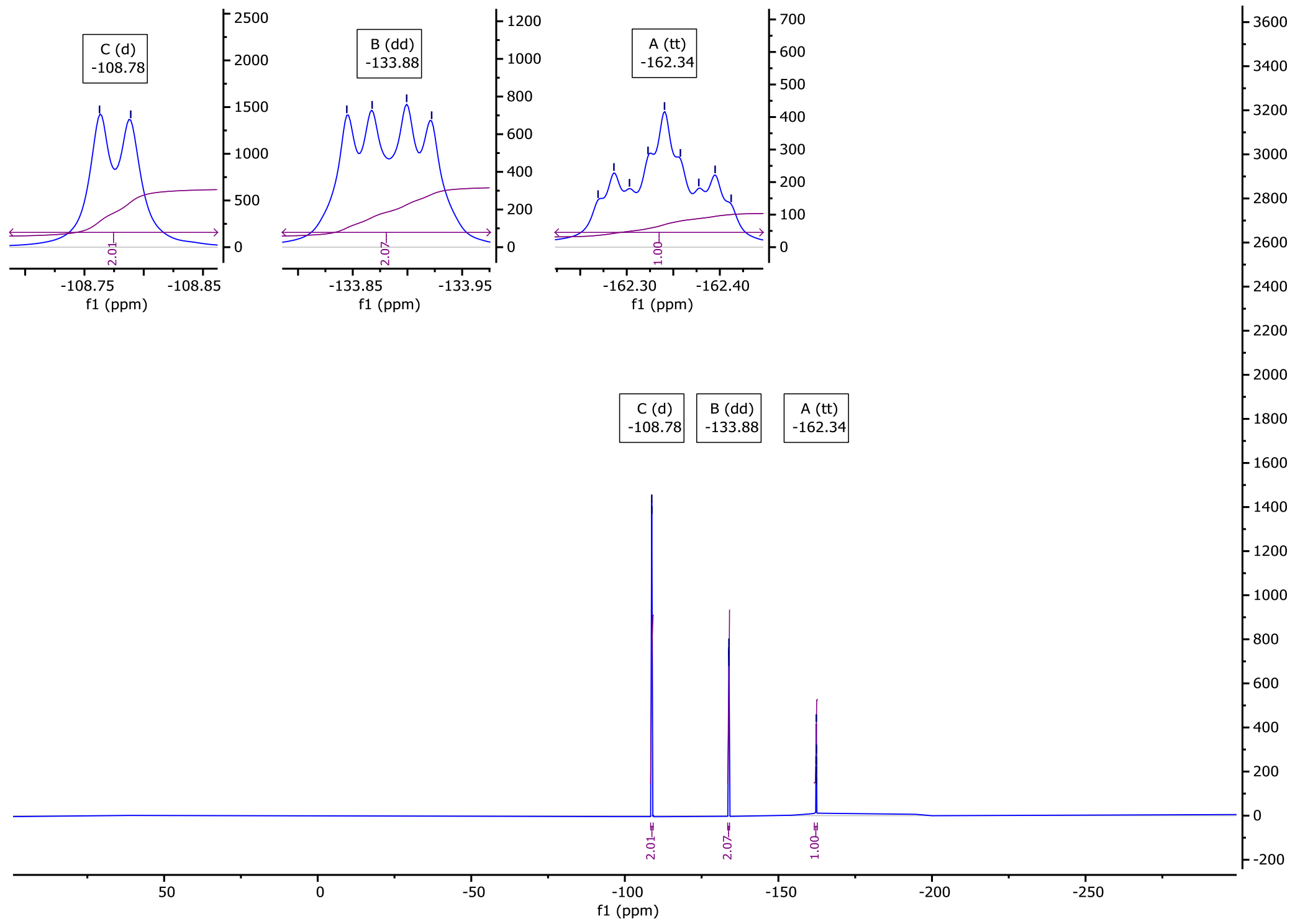

**Fig. S41.** $^{19}F$ NMR of **5-Me 6** in $CDCl_3$.

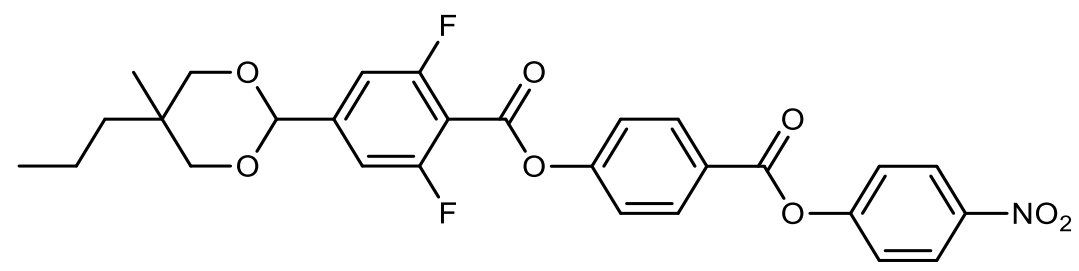

**5-Me 7**

*4-((4-nitrophenoxy)carbonyl)phenyl 2,6-difluoro-4-(5-methyl-5-propyl-1,3-dioxan-2-yl)benzoate*

| | |
|---|---|
| Yield: | (White crystalline solid) 175 mg, 65 %. |
| $R_F$ (DCM: hexanes [1:1]): | 0.34 |
| Re-crystallisation solvent: | MeOH |
| $^1$H NMR (501 MHz): | 8.37 – 8.28 (m, 4H, Ar-**H**)*, 7.48 – 7.41 (m, 4H, Ar-**H**)*, 7.22 (d, *J* = 9.8 Hz, 2H, Ar-**H**), 5.39 (s, 1H, Ar-C**H**-$O_2$), 3.84 (dd, *J* = 9.9, 1.4 Hz, 2H, 2x O-C$\mathbf{H_{eq}}H_{ax}$-C), 3.68 (dd, *J* = 10.5, 1.4 Hz, 2H, 2x O-C$\mathbf{H_{ax}}H_{eq}$-C), 1.34 – 1.27 (m, 2H, Me-C-C$\mathbf{H_2}$-$CH_2$), 1.26 (s, 3H, $CH_2)_2C(CH_2)$-**Me**), 1.15 – 1.09 (m, 2H, $CH_2$-C$\mathbf{H_2}$-$CH_3$), 0.93 (t, *J* = 7.2 Hz, 3H, $CH_2$-C$\mathbf{H_3}$).* Overlapping Signals. |
| $^{13}C\{^1H\}$ NMR (126 MHz): | 163.82, 163.62, 161.12 (dd, *J* = 258.5, 5.7 Hz), 159.29, 155.72 (d, *J* = 10.8 Hz), 154.91, 145.83 (t, *J* = 9.8 Hz), 145.64, 132.24, 132.19, 126.99, 126.33, 125.48, 122.79, 122.44, 122.26, 110.50 (dd, *J* = 23.5, 3.4 Hz), 109.58 (t, *J* = 17.0 Hz), 99.32, 38.80, 33.02, 20.23, 16.00, 15.10. |
| $^{19}$F NMR (376 MHz): | -108.45 (d, $J_{F-H}$ = 9.7 Hz, 2F, Ar-**F**). |

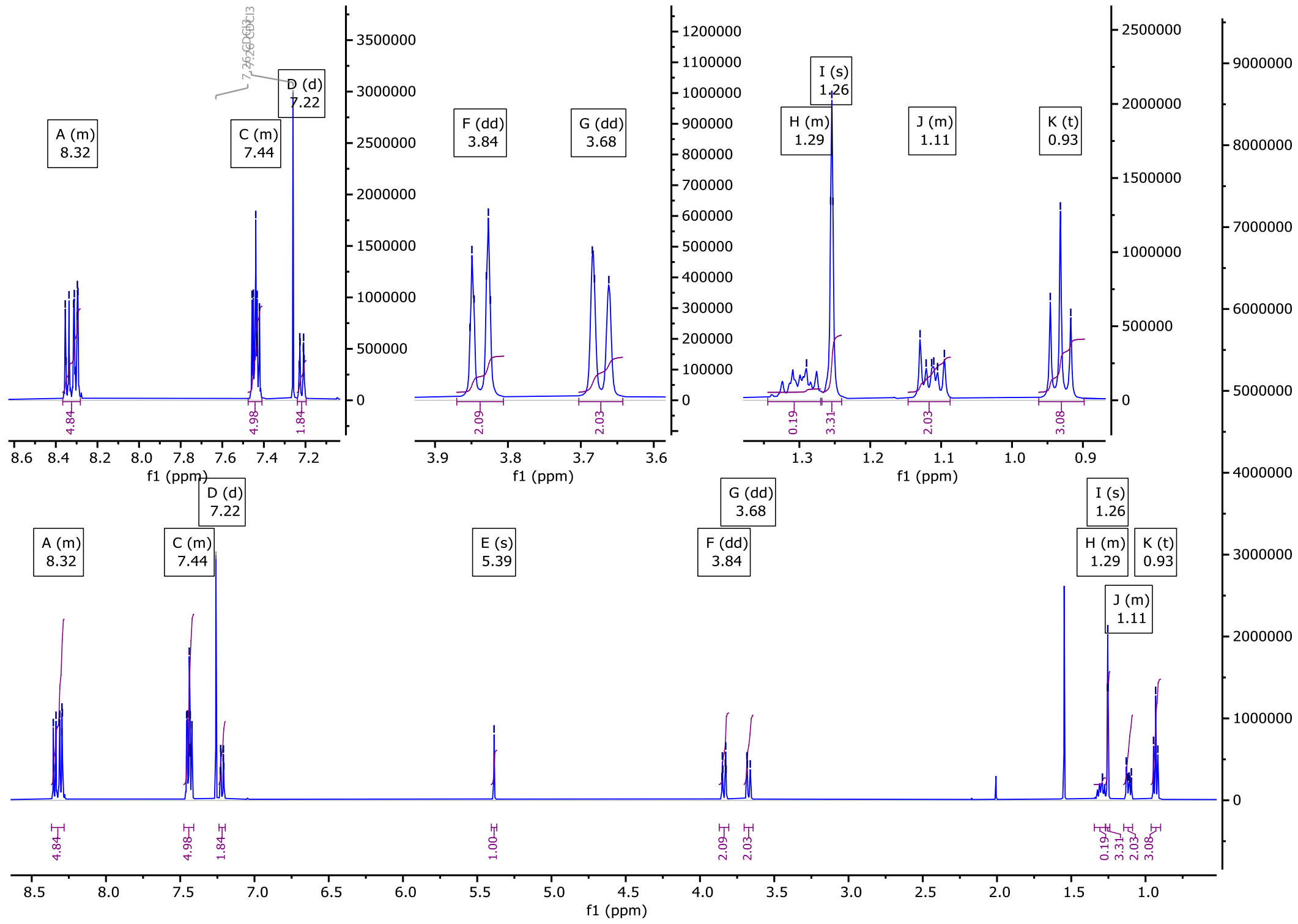

**Fig. S42.** $^{1}H$ NMR of **5-Me 7** in $CDCl_3$.

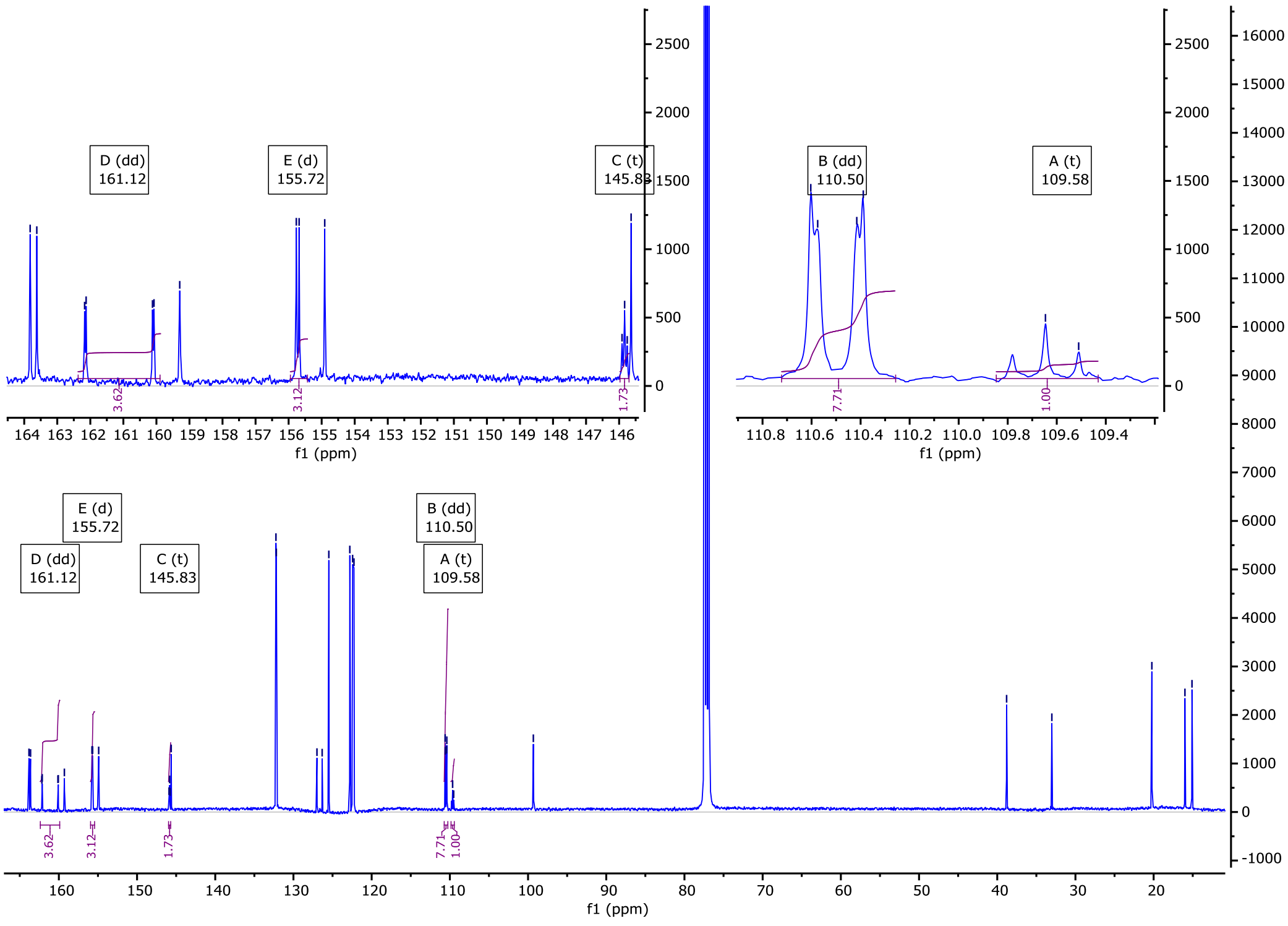

**Fig. S43.** $^{13}C\{^{1}H\}$ NMR of **5-Me 7** in $CDCl_3$.

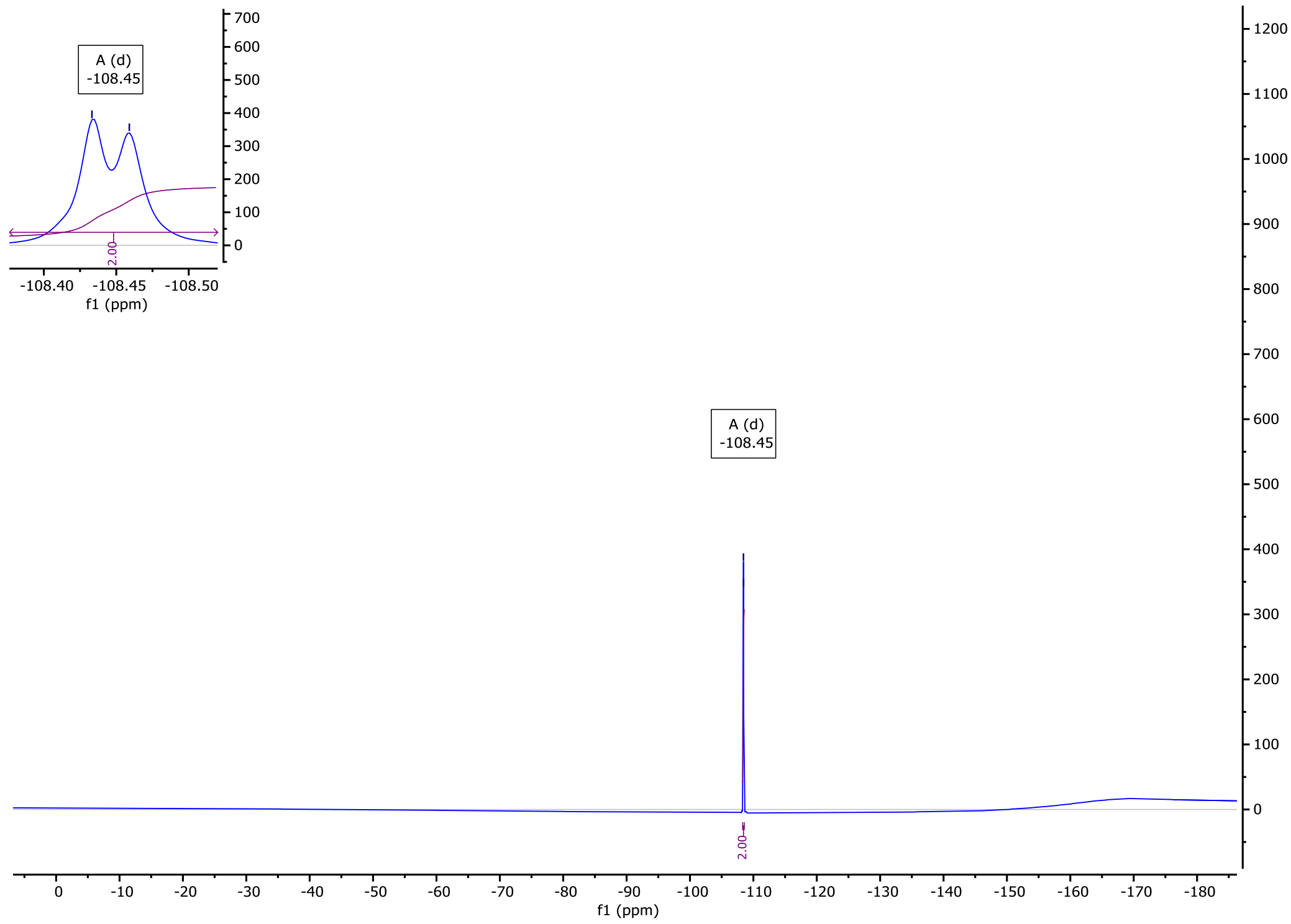

**Fig. S44.** $^{19}F$ NMR of **5-Me 7** in $CDCl_3$.

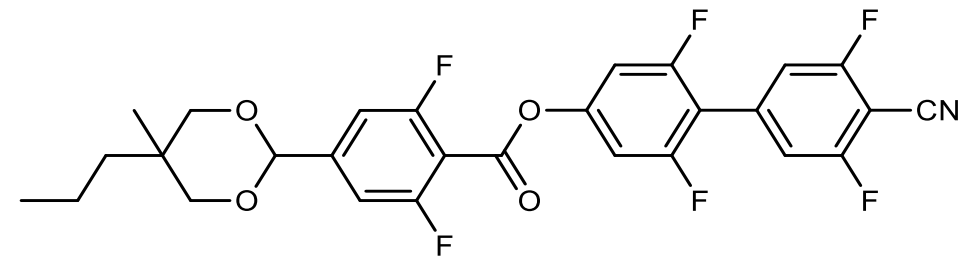

**5-Me 8**

*4'-cyano-2,3',5',6-tetrafluoro-[1,1'-biphenyl]-4-yl 2,6-difluoro-4-(5-methyl-5-propyl-1,3-dioxan-2-yl)benzoate*

Yield: (White crystalline solid) 225 mg, 82 %.

$R_F$ (DCM: hexanes [1:1]): 0.19

Re-crystallisation solvent: MeOH

$^1$H NMR (501 MHz): 7.24 – 7.19 (m, 4H, Ar-**H**), 7.06 (ddd, *J* = 8.1, 2.7, 2.6 Hz, 2H, Ar-**H**), 5.38 (s, 1H, , Ar-C**H**-$O_2$), 3.83 (dd, *J* = 9.8, 1.3 Hz, 2H, 2x O-C$\mathbf{H_{eq}}$$H_{ax}$-C), 3.67 (dd, *J* = 11.3, 1.6 Hz, 2H, 2x O-C$\mathbf{H_{ax}}$$H_{eq}$-C), 1.35 – 1.26 (m, 2H, Me-C-C$\mathbf{H_2}$-$CH_2$), 1.25 (s, 3H, $CH_2)_2C(CH_2)$-**Me**), 1.15 – 1.08 (m, 2H, $CH_2$-C$\mathbf{H_2}$-$CH_3$), 0.93 (t, *J* = 7.2 Hz, 3H, $CH_2$-C$\mathbf{H_3}$). * Overlapping Signals.

$^{13}$C{$^1$H} NMR (126 MHz): 162.73 (dd, *J* = 261.7, 4.6 Hz), 160.43 (dd, *J* = 259.5, 6.8 Hz), 158.85, 158.76 (dd, *J* = 252.2, 8.0 Hz), 151.88 (t, *J* = 14.4 Hz), 146.31 (t, *J* = 9.9 Hz), 136.72 (t, *J* = 10.6 Hz), 114.38 (dd, *J* = 21.1, 2.1 Hz), 113.18 (t, *J* = 16.7 Hz), 110.58 (dd, *J* = 23.3, 3.5 Hz), 109.18 – 108.79 (m)*, 107.38 – 106.76 (m), 99.23 (t, *J* = 2.3 Hz), 92.42 (t, *J* = 19.2 Hz), 38.79, 33.02, 20.22, 16.00, 15.09. * Overlapping Signals.

$^{19}$F NMR (376 MHz): -103.61 (d, $J_{F-H}$ = 8.7 Hz, 2F, Ar-**F**), -108.09 (d, $J_{F-H}$ = 9.9 Hz, 2F, Ar-**F**), -111.52 (d, $J_{F-H}$ = 9.0 Hz, 2F, Ar-**F**).

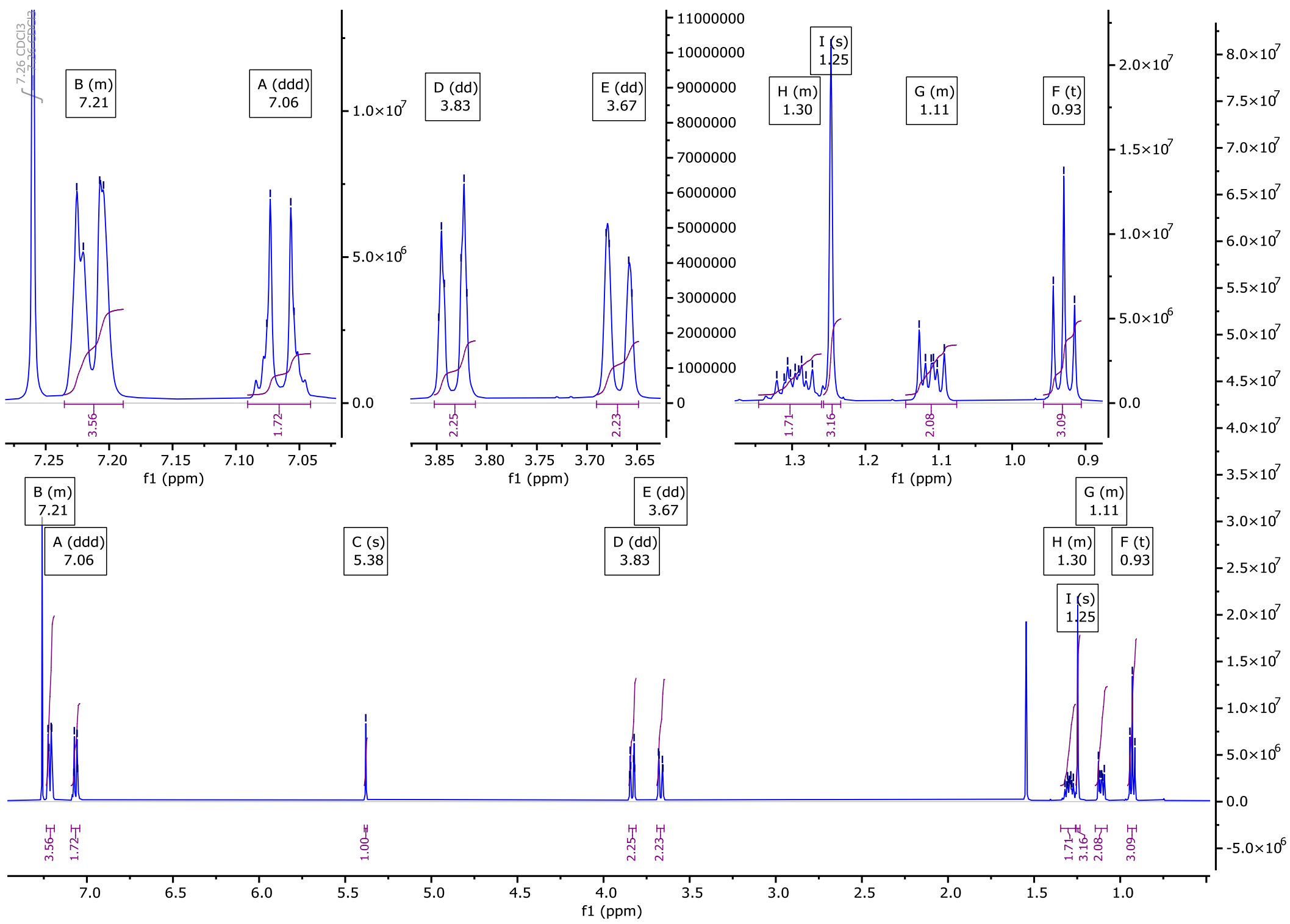

**Fig. S45.** $^{1}H$ NMR of **5-Me 8** in $CDCl_3$.

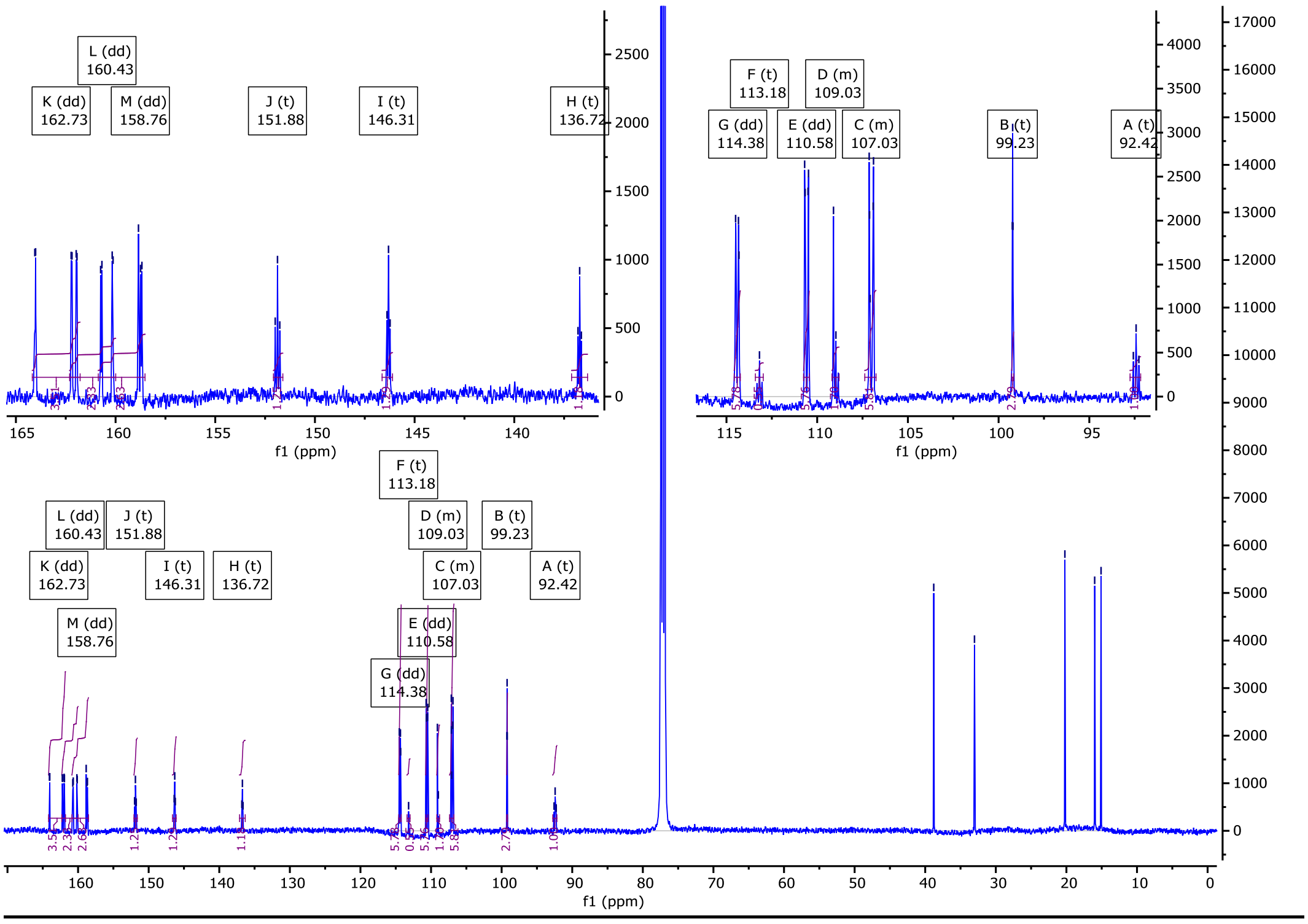

**Fig. S46.** $^{13}C\{^{1}H\}$ NMR of **5-Me 8** in $CDCl_3$.

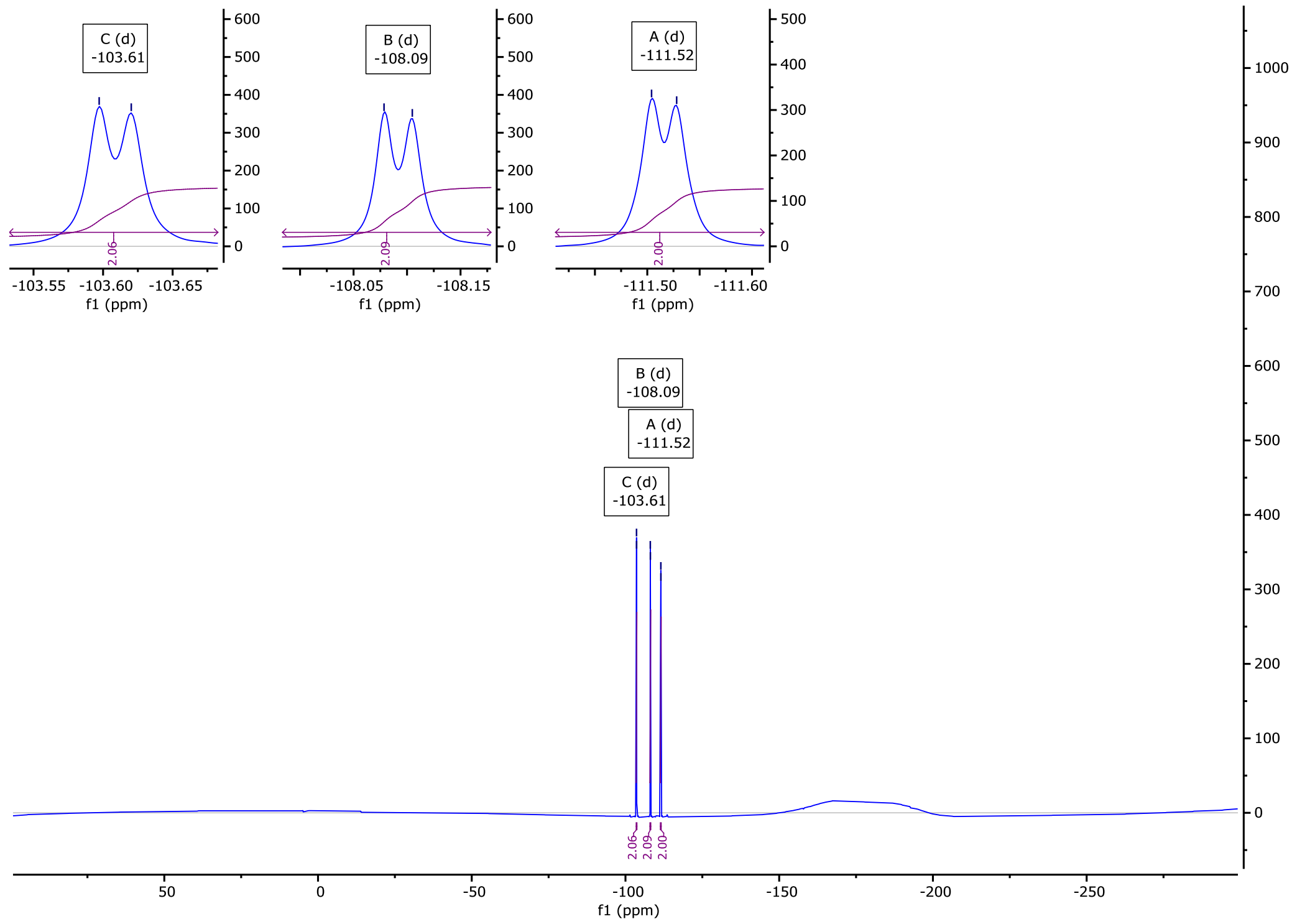

**Fig. S47.** $^{19}F$ NMR of **5-Me 8** in $CDCl_3$.

### 3.3 Preparation of compounds 1, 5-Me 1, 2-Me 1, and 2,5-Me 1.

The preparation of compounds **1**, **5-Me 1**, **2-Me 1**, and **2,5-Me 1** is outlined in **Scheme S2**. 2',3,5,6'-tetrafluoro-4'-formyl-[1,1'-biphenyl]-4-carbonitrile and 4'-acetyl-2',3,5,6'-tetrafluoro-[1,1'-biphenyl]-4-carbonitrile were synthesised using the general Suzuki coupling procedure (**3.3.1**) and compounds **1**, **5-Me 1**, **2-Me 1**, and **2, 5-Me 1** were prepared using the standard condensation reaction procedure (**3.3.2**). Structural characterisation data for **2-Me 1** and **2,5-Me 1** are not provided here due to them being inseparable mixtures of the *cis*- and *trans*-geometric isomers (see main manuscript for more details).

i) Pd(XPhos), $K_2CO_{3(aq)}$, THF, 70 °C, $N_{2(g)}$
ii) p-TsOH, Toluene, 110 °C, $N_{2(g)}$
R = H or Me

**Scheme S2:** The preparation of compounds **1**, **5-Me 1**, **2-Me 1**, and **2,5-Me 1**.

#### 3.3.1 General Suzuki coupling procedure

A reaction flask was charged with the appropriate aryl bromide (1.0 mol equiv.) and the appropriate pinacol ester benzonitrile (1.05 mol equiv.) which were dissolved in the appropriate amount of THF (to yield a final concentration circa 0.1 M) and 2 M $K_2CO_{3(aq)}$ (2.0 mol equiv.). The resultant solution was sparged with $N_{2(g)}$ for 30 minutes before reaction flask was then heated to 70 °C and the catalyst Pd XPhos G3 (0.01 mol equiv.) added in one portion. The reaction was monitored by TLC with the completion of the reaction being determined by the complete consumption of the bromo-substrate. The reaction was then cooled, the aqueous and organic layers separated with the organics being dried over $MgSO_4$. The organics were then passed through a silica plug before the filtrate was concentrated under reduced pressure and purified as indicated.

*2',3,5,6'-tetrafluoro-4'-formyl-[1,1'-biphenyl]-4-carbonitrile*

Quantities used: 4-bromo-2,5 diflurobenzaldehyde (45 mmol), 2,6-Difluoro-4-(4,4,5,5-tetramethyl-1,3,2-dioxaborolan-2-yl)benzonitrile (47 mmol), 150 mL THF, 30 mL 2M $K_2CO_3$. The reaction was performed according to the general Suzuki coupling procedure (3.3.1). The product was purified by flash chromatography over silica gel with a gradient of hexanes/DCM. Finally, the product re-crystallised from hexanes: toluene (10:1) to give the desired product as an off-white solid.

Yield: 8.5 g, 66 %
$R_f$ (DCM: Hexanes (1:1): 0.33

$^{1}$H NMR (400 MHz): 10.00 (t, $J$ = 1.7 Hz, 1H, CO-**H**), 7.63 – 7.54 (m, 2H, Ar-**H**), 7.26 – 7.21 (m, 2H, Ar-**H**).

$^{13}$C{$^{1}$H} NMR (101 MHz): 188.55 (t, $J$ = 1.9 Hz), 162.89 (dd, $J$ = 261.7, 4.8 Hz), 159.92 (dd, $J$ = 255.4, 5.5 Hz), 138.70 (t, $J$ = 7.8 Hz), 135.95 (t, $J$ = 10.6 Hz), 120.28 (tt, $J$ = 17.5, 2.2 Hz), 114.64 – 113.93 (m), 113.29 – 112.50 (m), 111.13 (dd, $J$ = 21.4, 3.4 Hz), 108.72, 108.40, 93.01 (t, $J$ = 19.2 Hz).

$^{19}$F NMR (376 MHz): -103.04 (d, $J_{F\text{-}H}$ = 9.0 Hz, 2F, Ar-**F**), -110.42 (d, $J_{F\text{-}H}$ = 7.8 Hz, 2F, Ar-**F**).

*4'-acetyl-2',3,5,6'-tetrafluoro-[1,1'-biphenyl]-4-carbonitrile*

Quantities used: 1-(4-bromo-3,5-difluorophenyl)ethan-1-one (4.2 mmol), 2,6-Difluoro-4-(4,4,5,5-tetramethyl-1,3,2-dioxaborolan-2-yl)benzonitrile (4.6 mmol), 30 mL THF, 5 mL 2M $K_2CO_3$. The reaction was performed according to the general Suzuki coupling procedure (3.3.1). The product was purified by flash chromatography over silica gel with a gradient of hexanes/DCM. Finally, the product re-crystallised from EtOH to give the desired product as a white solid.

Yield: 1.00 g, 83 %

$R_f$ (DCM: Hexanes (1:1): 0.40

$^{1}$H NMR (400 MHz): 7.62 (ddd, $J$ = 8.5, 3.0, 2.2 Hz, 2H, Ar-**H**), 7.23 (ddd, $J$ = 7.7, 1.3, 1.2 Hz, 2H, Ar-**H**), 2.64 (s, 3H, CO-C**H$_3$**).

$^{13}$C{$^{1}$H} NMR (101 MHz): 194.51, 162.87 (dd, $J$ = 261.6, 5.0 Hz), 159.56 (dd, $J$ = 253.8, 5.5 Hz), 139.76 (t, $J$ = 7.9 Hz), 136.17 (t, $J$ = 10.1 Hz), 114.26 (dd, $J$ = 21.3, 3.4 Hz), 112.02 (dd, $J$ = 26.4, 6.7 Hz), 111.24 (dd, $J$ = 22.0, 3.3 Hz), 108.79, 92.80 (t, $J$ = 19.3 Hz), 26.60.

$^{19}$F NMR (376 MHz): -103.22 (d, $J_{F\text{-}H}$ = 8.7 Hz, 2F, Ar-**F**), -111.44 (d, $J_{F\text{-}H}$ = 8.8 Hz, 2F, Ar-**F**).

### 3.3.2 General condensation reaction procedure

An oven dried flask, fitted with a dean-stark apparatus, was cooled under an atmosphere of dry nitrogen and charged with the appropriate benzaldehyde (1.0 mol equiv.), the relevant diol (1.1 mol equiv.), *p*-toluenesulfonic acid (cat.) and an appropriate volume of anhydrous toluene. The reaction mixture was heated to reflux, with any evolved water being removed at regular intervals until no further water was captured within the dean-stark apparatus (circa 48 h). The reaction was cooled to room temperature, concentred *in vacuo* and purified by flash chromatography using a gradient of hexanes:EtOAc. The chromatographed material was filtered through a 200 nm PTFE syringe filter, concentrated to dryness, and finally recrystalised from MeOH to give the reported yields.

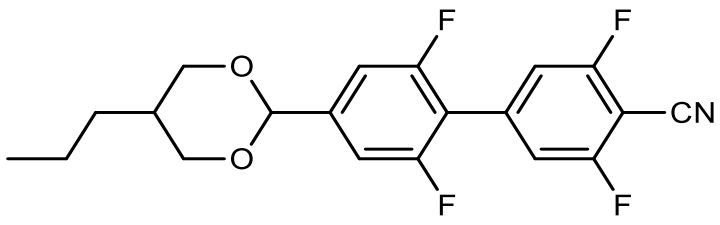

**1**

*4'-(5-ethyl-1,3-dioxan-2-yl)-2',3,5,6'-tetrafluoro-[1,1'-biphenyl]-4-carbonitrile*

Quantities used: 2',3,5,6'-tetrafluoro-4'-formyl-[1,1'-biphenyl]-4-carbonitrile (279 mg, 1.0 mmol), 2-propylpropane-1,3-diol (130 mg, 135 mL, 1.1 mmol), 20 mL toluene. The reaction was performed according to the general condensation reaction procedure (3.3.2).

| | |
|---|---|
| Yield: | (white needles) 197 mg, 52 % |
| $R_f$ (Hexanes: EtoAc (10:1): | 0.31 |
| $^{1}$H NMR (400 MHz) | 7.22 – 7.14 (m, 4H, Ar-**H**)*, 5.40 (s, 1H, Ar-C**H**-$O_2$), 4.26 (dd, *J* = 11.9, 4.7 Hz, 2H, O-C**H$_{ax}$**($H_{eq}$)-CH), 3.54 (t, *J* = 11.4 Hz, 2H, O-C**H$_{eq}$**($H_{ax}$)-CH), 2.12 – 1.97 ($m_{apparent}$, 1H, $(CH_2)_2$-C**H**-$CH_2$), 1.17 (p, *J* = 7.4 Hz, 2H, CH-C**H$_2$**-$CH_3$), 0.94 (t, *J* = 7.5 Hz, 3H, $CH_2$-C**H$_3$**). (*Overlapping Signals). |
| $^{13}$C{$^{1}$H} NMR (101 MHz): | 162.92 (dd, *J* = 261.0, 5.1 Hz), 159.38 (dd, *J* = 251.5, 6.3 Hz), 143.15 (t, *J* = 9.8 Hz), 137.39, 114.88 (t, *J* = 17.3 Hz), 114.36 (dd, *J* = 20.9, 2.8 Hz), 110.32 (dd, *J* = 21.7, 5.5 Hz), 109.17, 99.15 (t, *J* = 2.4 Hz), 92.17 (t, *J* = 19.2 Hz), 72.72, 34.05, 30.37, 19.67, 14.33. |
| $^{19}$F NMR (376 MHz): | -104.01 (d, $J_{F\text{-}H}$ = 9.0 Hz, 2F, Ar-**F**), -113.41 (d, $J_{F\text{-}H}$ = 9.4 Hz, 2F, Ar-**F**). |

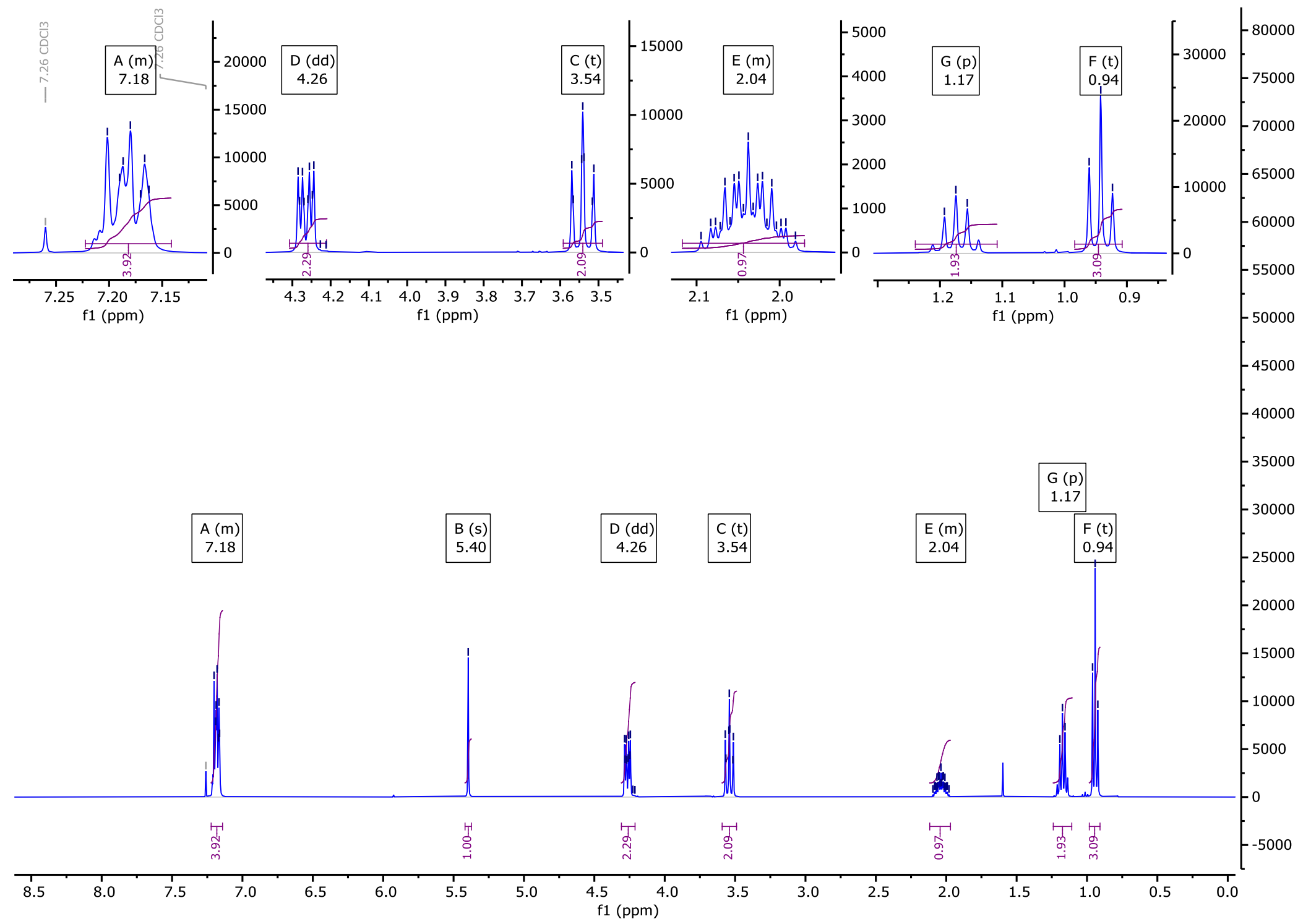

**Fig. S48.** $^{1}H$ NMR of **1** in $CDCl_3$.

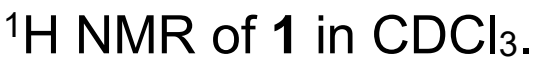

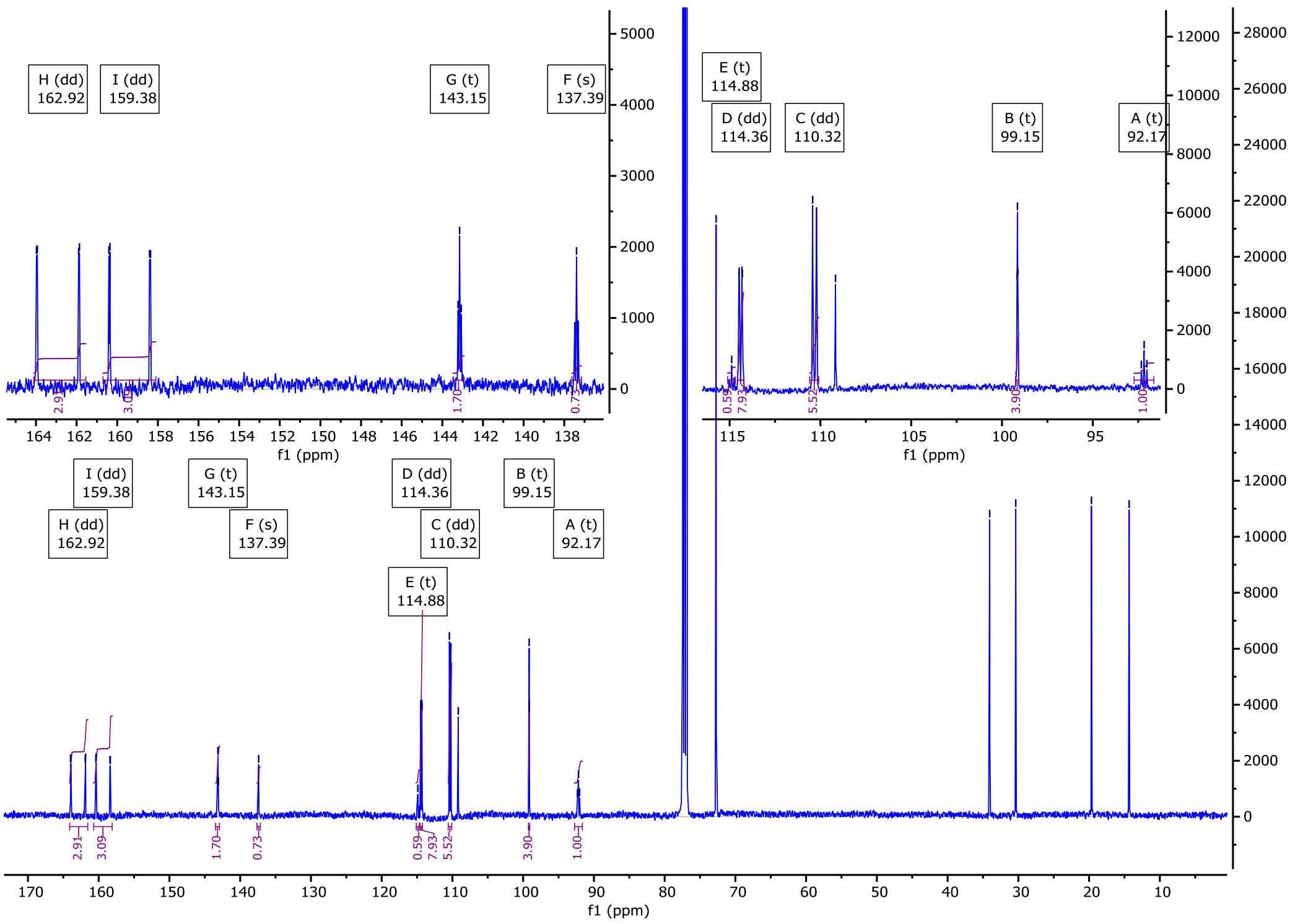

**Fig. S49.** $^{13}C\{^{1}H\}$ NMR of **1** in $CDCl_3$.

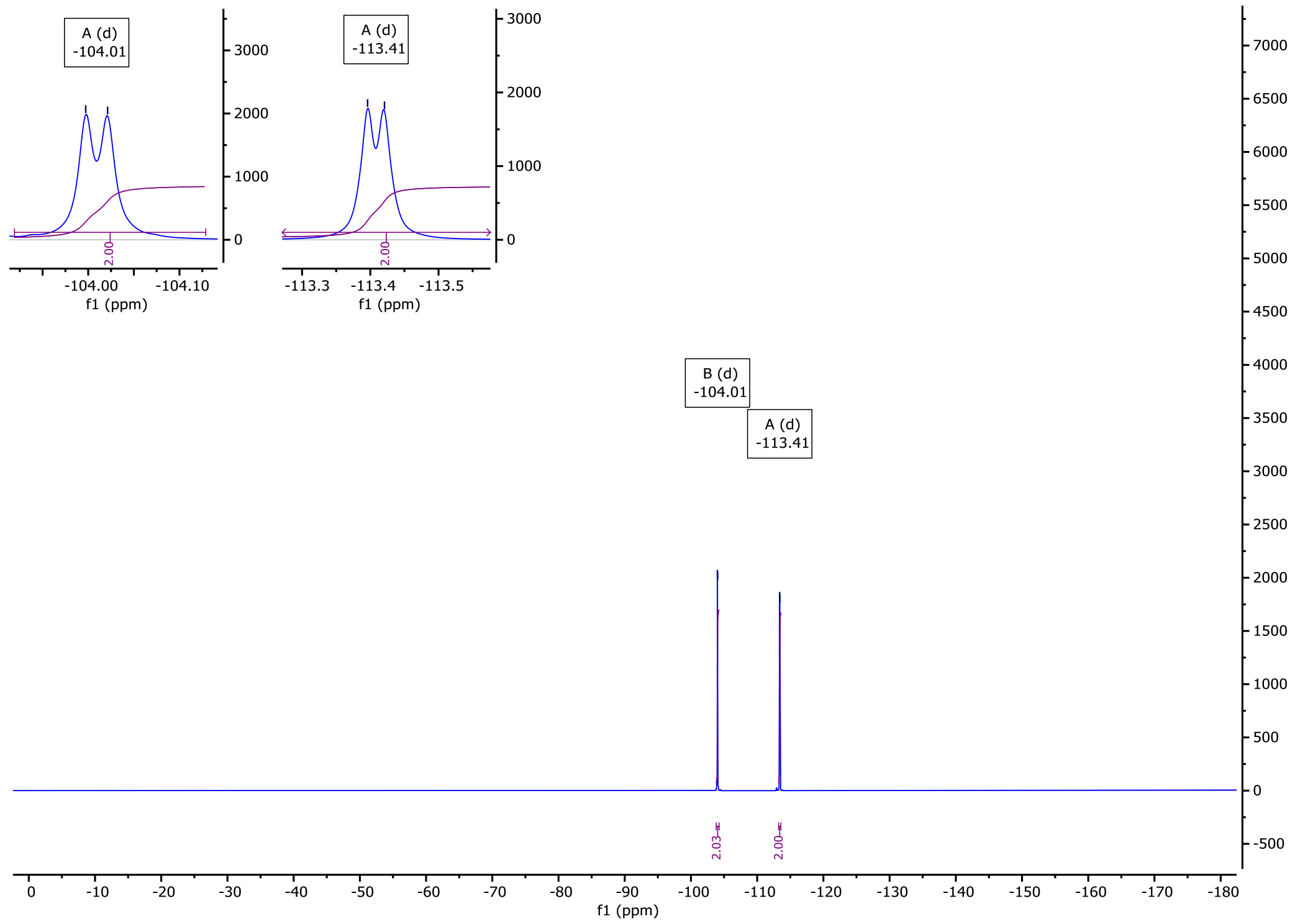

**Fig. S50.** $^{19}F$ NMR of **1** in $CDCl_3$.

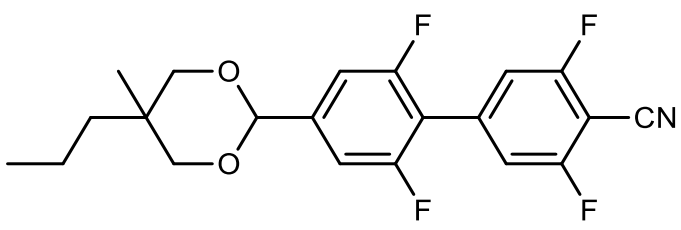

**5-Me 1**

*2',3,5,6'-tetrafluoro-4'-(5-methyl-5-propyl-1,3-dioxan-2-yl)-[1,1'-biphenyl]-4-carbonitrile*

Quantities used: 2',3,5,6'-tetrafluoro-4'-formyl-[1,1'-biphenyl]-4-carbonitrile (279 mg, 1.0 mmol), 2-methyl-2-propylpropane-1,3-diol (145 mg, 1.1 mmol), 20 mL toluene. The reaction was performed according to the general condensation reaction procedure (3.3.2).

Yield: (white needles) 224 mg, 57 %

$R_f$ (Hexanes: EtoAc (10:1): 0.35

$^{1}$H NMR (400 MHz): 7.25 – 7.14 (m, 4H, Ar-**H**)*, 5.38 (s, 1H, Ar-C**H**-$O_2$), 3.82 (d, *J* = 11.3 Hz, 2H, O-C$\mathbf{H_{ax}}$($H_{eq}$)-CH), 3.67 (d, *J* = 11.1 Hz, 2H, O-C$\mathbf{H_{eq}}$($H_{ax}$)-CH), 1.35 – 1.26 (m, 2H, $CH_2$-C$\mathbf{H_2}$-$CH_3$), 1.25 (s, 3H, $(CH_2)_2$C-C$\mathbf{H_3}$-(CH2)), 1.14 – 1.07 (m, 2H, $CH_2$-C$\mathbf{H_2}$-$CH_3$), 0.93 (t, *J* = 7.2 Hz, 3H, $CH_2$-C$\mathbf{H_3}$). (*Overlapping Signals).

$^{13}$C{$^{1}$H} NMR (101 MHz): 162.93 (dd, *J* = 261.0, 5.1 Hz), 159.43 (dd, *J* = 251.6, 6.3 Hz), 143.19 (t, *J* = 9.7 Hz), 137.40 (t, *J* = 10.6 Hz), 114.87 (t, *J* = 17.5 Hz), 114.39 (dd, *J* = 20.8, 3.4 Hz), 110.35 (dd, *J* = 21.7, 5.2 Hz), 99.53 (t, *J* = 2.4 Hz), 109.18, 99.55, 99.53, 99.51, 92.20 (t, *J* = 19.2 Hz), 38.82, 33.00, 20.23, 16.00, 15.10.

$^{19}$F NMR (376 MHz): -103.92 (d, $J_{F-H}$ = 9.0 Hz, 2F, Ar-**F**), -113.32 (d, $J_{F-H}$ = 9.3 Hz, 2F, Ar-**F**).

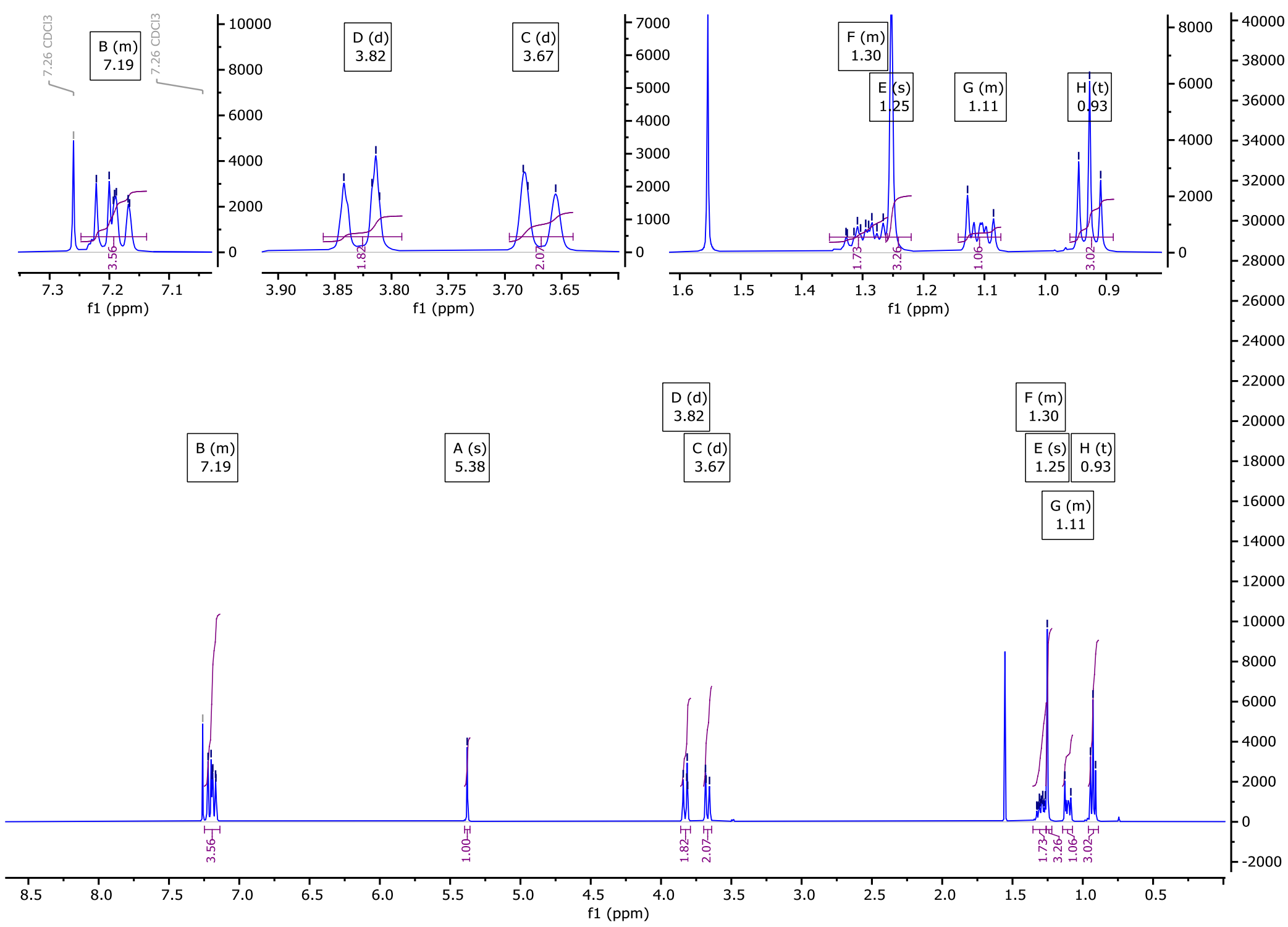

**Fig. S51.** $^{1}H$ NMR of **5-Me 1** in $CDCl_3$.

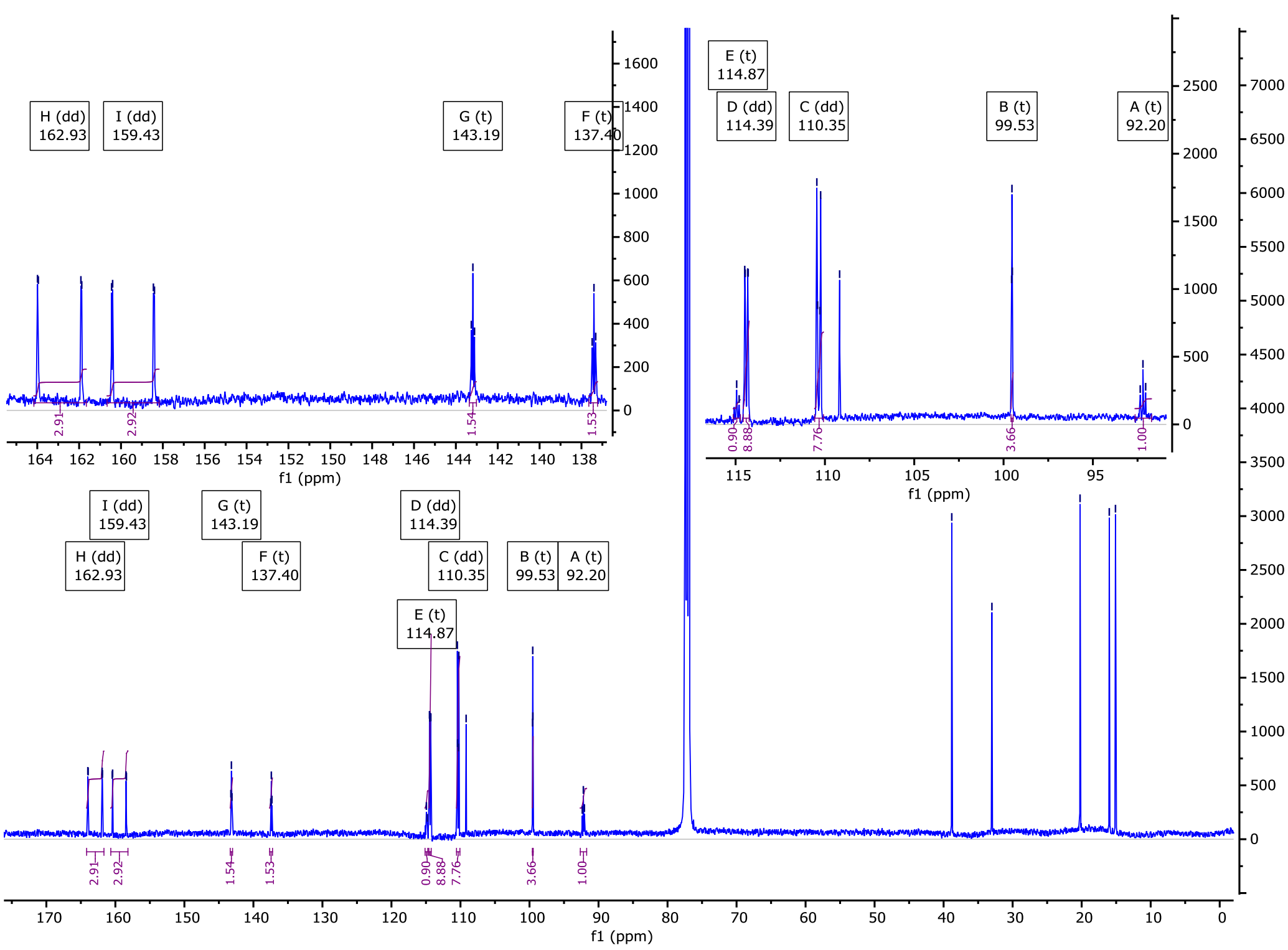

**Fig. S52.** $^{13}C\{^{1}H\}$ NMR of **5-Me 1** in $CDCl_3$.

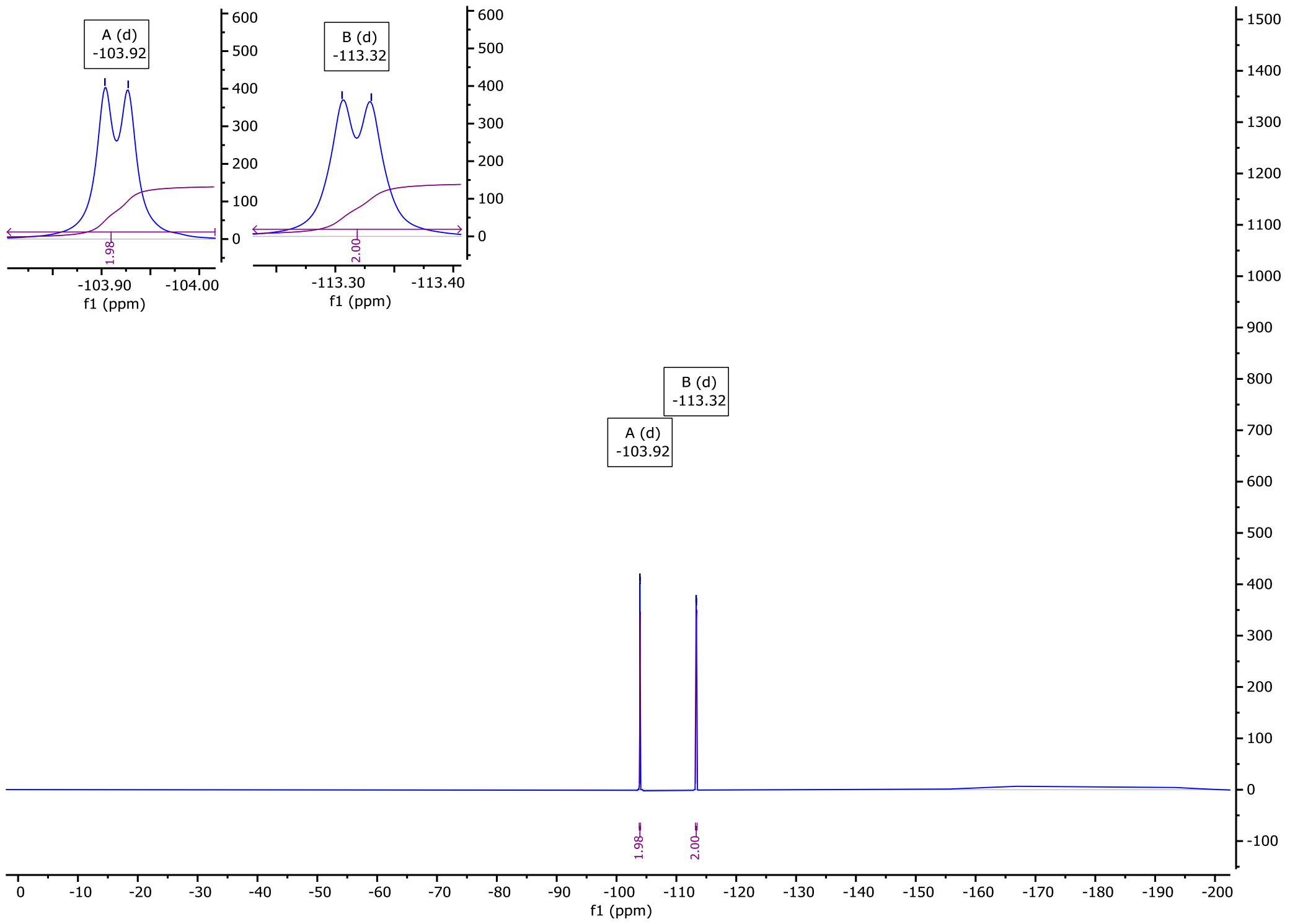

**Fig. S53.** $^{19}F$ NMR of **5-Me 1** in $CDCl_3$.

### 3.4 Preparation of compound 5-Me 1 (Si)

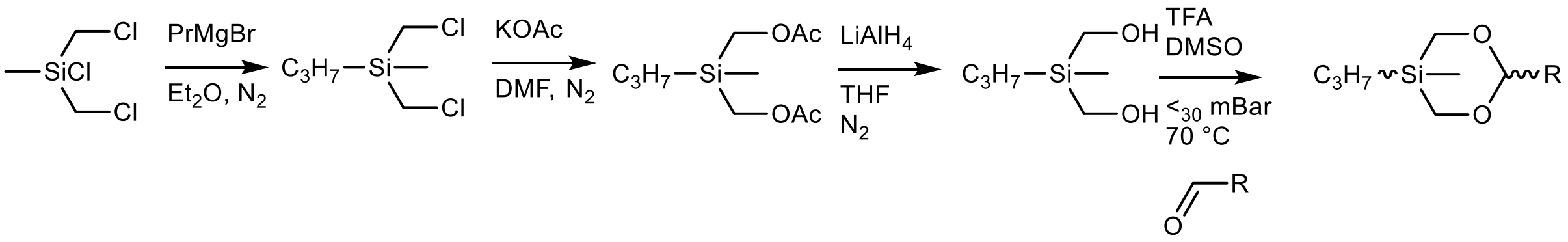

**Scheme S3** The preparation of compounds **5-Me 1 (Si)**

Alkylation of bis(chloromethyl)methylchlorosilane with propyl magnesium bromide proceeded in quantitative yield. Treatment with potassium acetate in DMF afforded the bis acetate, which was subsequently reduced with lithium aluminium hydride in THF to afford bis(hydroxymethyl)propylmethylsilane. Acid catalysed acetalization with an aldehyde *vide infra*) under modified Dean-Stark conditions [22] affords the 5-methyl-1,3,5-dioxasilinane as an inseparable mixture of isomers.

#### 3.4.1 Preparation of bis(chloromethyl)(methyl)(propyl)silane

In an oven dried flask under an atmosphere of dry nitrogen gas, propyl magnesium chloride (200 mmol, 1.4 eqv., 100 ml, 2M in $Et_2O$) was added dropwise to a solution of methyl bis(chloromethyl)chlorosilane (25 g, 140 mmol, 1 eqv.) in anhydrous THF (100 ml). The suspension was stirred for 2h at ambient temperature before quenching by careful addition of saturated aqueous ammonium chloride (200 ml). Diethyl ether (100 ml) was added; the organic layer was separated and retained. The aqueous layer was washed with diethyl ether (3x 50 ml) and discarded. The combined organics were washed with saturated aqueous NaCl (150 ml), dried over $MgSO_4$, and concentrated *in vacuo*. The crude material was filtered over a plug of silica gel, eluting with petroleum ether 40-60, and concentrated *in vacuo* to yield the title compound as a colourless oil.

*bis(chloromethyl)(methyl)(propyl)silane*

Yield: 26g (>99%)

[1]H NMR: 2.68 (s, 4H, $J_{H\text{-}Si}$ = 12 Hz, Si-C$\mathbf{H_2}$-Cl), 1.25-1.13 (m, 2H, Si-C$H_2$-C$\mathbf{H_2}$-C$H_3$), 0.76 (t, 3H, $J$ = 7.3 Hz, Si-C$H_2$-C$H_2$-C$\mathbf{H_3}$), 0.61-0.52 (m, 2H, Si-C$\mathbf{H_2}$-C$H_2$-C$H_3$), 0.00 (s, 3H, $J_{H\text{-}Si}$ = 6.7 Hz, Si-C$\mathbf{H_3}$)

Spectral data in keeping with literature values [23].

### 3.4.2 Preparation of (methyl(propyl)silanediyl)bis(methylene) diacetate

A flask was charged with propyl methyl bis(chloromethyl)silane (22.7 g, 122 mmol), potassium acetate (44 g, 500 mmol), and DMF (125 ml). The suspension was vigorously stirred whilst heating under reflux at a temperature of 125 °C for 18 h. The suspension was cooled and diluted with water (800 ml). The product was extracted into hexane (6 x 50 ml), and the aqueous set aside. The combined organic extracts were washed with water (1x 100 ml), aqueous LiCl (1 x 50 ml, 5% w/v), and saturated aqueous NaCl (100 ml). The organics were dried over $MgSO_4$ and concentrated *in vacuo*. The crude material was passed through a plug of silica gel, eluting with petroleum ether 40-60; removal of volatiles afforded the title compound as a colourless liquid.

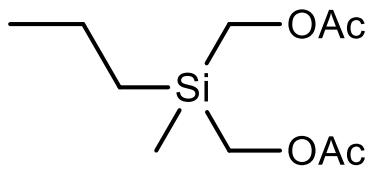

*(methyl(propyl)silanediyl)bis(methylene) diacetate*

Yield: 28 g (99%)

[1]H NMR: 3.74 (s, 4H Si-C$\mathbf{H_2}$-OAc), 1.92 (s, 6H, $J_{H\text{-}Si}$ = 6.0 Hz, OC(=O)-C$\mathbf{H_3}$), 1.35 – 1.19 (m, 2H, Si-C$H_2$-C$\mathbf{H_2}$-C$H_3$), 0.85 (t, 3H, $J$ = 7.2 Hz, Si-C$H_2$-C$H_2$-C$\mathbf{H_3}$), 0.62 – 0.51 (m, 2H, Si-C$\mathbf{H_2}$-C$H_2$-C$H_3$), 0.00 (s, 3H, $J_{H\text{-}Si}$ = 6.9 Hz, Si-C$\mathbf{H_3}$).

Spectral data in keeping with literature values [23]

### 3.4.3 Preparation of (methyl(propyl)silanediyl)dimethanol.

An oven dried flask under an atmosphere of dry nitrogen was charged with a solution of $LiAlH_4$ in THF (2M, 100 ml, 200 mmol) and diluted to 500 ml. Neat (methyl(propyl)silanediyl)bis(methylene) diacetate (19.4 g, 83 mmol) was added slowly. Once the addition was complete, the reaction was heated under reflux for 1 h. The reaction was then cooled, EtOAc (100 ml) was added slowly followed by water (100 ml) and 4M HCl (200 ml). The organic layer was separated and retained; the aqueous layer was washed with diethyl ether (4x 50 ml) and set aside. The combined organics were washed with saturated aqueous NaCl (100 ml), dried over $MgSO_4$, and concentrated *in vacuo*. The crude material was filtered over a plug of silica gel, eluting with dichloromethane, to afford the title compound as a pale yellow liquid.

*(methyl(propyl)silanediyl)dimethanol*

Yield: 9.7 g (79%)

$^1$H NMR: 3.35-3.34 (m [AB'], 4H Si-C$\mathbf{H_2}$-OH), 2.85 (s, 2H, Si-O**H**), 1.37 – 1.25 (m, 2H, Si-$CH_2$-C$\mathbf{H_2}$-$CH_3$), 0.88 (t, 3H, $J$ = 7.2 Hz, Si-$CH_2$-$CH_2$-C$\mathbf{H_3}$), 0.60 – 0.54 (m, 2H, Si-C$\mathbf{H_2}$-$CH_2$-$CH_3$), 0.00 (s, 3H, $J_{H\text{-}Si}$ = 6.6 Hz, Si-C$\mathbf{H_3}$).

Spectral data in keeping with literature values [23].

## 4 Supplemental References